\documentstyle[12pt,epsfig,axodraw]{article}

\def\ar{\rightarrow}
\def\a0{\bar\alpha_0}

\def\as{\alpha_{\mbox{\tiny S}}}
\def\b0{\beta_0}
\def\cN{{\cal N}}
 
\def\Ecm{E_{\mbox{\scriptsize cm}}}
\def\ee{e^+e^-}
\def\tee{$\ee$}

\def\lms{\Lambda^{(5)}_{\overline{\mbox{\tiny MS}}}}

\def\st{\sigma_{\mbox{\scriptsize tot}}} \def\ycut{y_{\mbox{\tiny cut}}}
 \def\frac#1#2{ {{#1} \over {#2} }}

\def\VEV#1{\left\langle #1\right\rangle}
\def\cO#1{{\cal{O}}\left(#1\right)}
\def\beq{\begin{equation}}
\def\beeq{\begin{eqnarray}}
\def\eeq{\end{equation}}
\def\eeeq{\end{eqnarray}}
\def\Ord{\buildrel{\scriptscriptstyle <}\over{\scriptscriptstyle\sim}}

\def\Geneva{{\sc Geneva}}
\def\Jade{{\sc Jade}}
\def\Durham{{\sc Durham}}
\def\LDurham{{\sc Durham/Lu}}
\def\AODurham{{\sc Angular-Ordered Durham}}
\def\AODurhamshort{{\sc Ang-Ord Durham}}
\def\AOLuclus{{\sc Angular-Ordered Luclus}}
\def\Luclus{{\sc Luclus}}
\def\Pyclus{{\sc Pyclus}}
\def\Cambridge{{\sc Cambridge}}
\def\LCambridge{{\sc Cambridge/Lu}}
\def\Diclus{{\sc Diclus}}
\def\Arclus{{\sc Arclus}}
\def\jetset{{\sc Jetset}}
\def\pythia{{\sc Pythia}}
\def\ariadne{{\sc Ariadne}}
\def\herwig{{\small HERWIG}}
\def\eerad{{\sc EERAD}}
\def\debrecen{{\sc DEBRECEN}}
\def\minuit{{\sc MINUIT}}
\def\bfp{{\mathbf{p}}}

\def\sensibly{noticeably}
\def\nosensibly{}
\def\fig#1{fig.\ \ref{#1}}
\def\tpt{$p_\perp$}
\def\pl #1 #2 #3 {{\it Phys.\ Lett.} {\bf#1} (#2) #3}
\def\np #1 #2 #3 {{\it Nucl.\ Phys.} {\bf#1} (#2) #3}
\def\zp #1 #2 #3 {{\it Z.\ Phys.} {\bf#1} (#2) #3}
\def\pr #1 #2 #3 {{\it Phys.\ Rev.} {\bf#1} (#2) #3}
\def\prep #1 #2 #3 {{\it Phys.\ Rep.} {\bf#1} (#2) #3}
\def\prl #1 #2 #3 {{\it Phys.\ Rev.\ Lett.} {\bf#1} (#2) #3}
\def\mpl #1 #2 #3 {{\it Mod.\ Phys.\ Lett.} {\bf#1} (#2) #3}
\def\rmp #1 #2 #3 {{\it Rev.\ Mod.\ Phys.} {\bf#1} (#2) #3}
\def\cpc #1 #2 #3{{\it Comp.\ Phys.\ Commun.\ }{\bf #1} (#2) #3}
\def\jp #1 #2 #3{{\it J.\ Phys.\ }{\bf #1} (#2) #3}
\def\jhep #1 #2 #3 {{\it J.\ High Energy Phys.} {\bf #1} (#2) #3}
\def\hepph #1 {{\tt hep-ph/#1}}
 
{\end{list}}
\newcounter{enumct}

\newlength{\abstwidth}
\begin{document}
 
\sloppy
 
\pagestyle{empty}
\setcounter{page}{0}
 
\begin{flushright}
RAL-TR-98-003  \\
LU--TP 98--7 \\
hep-ph/9804296\\
April 1998 
\end{flushright}
 
\vspace{\fill}
 
\begin{center}
{\LARGE\bf New and Old Jet Clustering Algorithms\\[5mm]
for Electron-Positron Events}\\[10mm]
{\Large Stefano Moretti} \\[3mm]
{\it Rutherford Appleton Laboratory,
Chilton, Didcot, Oxon OX11 0QX, UK} \\[1mm]
{E-mail: {\tt moretti@v2.rl.ac.uk}}\\[5mm]
{\Large Leif L\"onnblad, Torbj\"orn Sj\"ostrand} \\[3mm]
{\it Department of Theoretical Physics, Lund University, Lund, Sweden} \\[1mm]
{E-mail: {\tt leif@thep.lu.se}, {\tt torbjorn@thep.lu.se}}\\[5mm]
\end{center}
 
\vspace{\fill}
 
\begin{center}
{\bf Abstract}\\[2ex]
\begin{minipage}{\abstwidth}
Over the years, many jet clustering algorithms have been proposed
for the analysis of hadronic final states in $e^+e^-$ annihilations.
These have somewhat different emphasis and are therefore more
or less suited for various applications. We here review some of
the most used 
and compare them from a theoretical and
experimental point of view. 
\end{minipage}
\end{center}
 
\vspace{\fill}
 
\clearpage
\pagestyle{plain}
\setcounter{page}{1}
 
\section{Introduction}
\label{sec_intro}

Clustering algorithms have come to be an indispensable tool in the
study of multi-hadronic events. They take the large
number of particles produced in high-energy scatterings and cluster them 
into a small number of `jets'. Such a simplified characterization of 
the event should help focus on the main properties of the 
underlying dynamics.
In particular, the reconstructed jets should reflect the  
partonic picture, and thus allow a separation of perturbative and 
non-perturbative QCD physics aspects. Of course, such a separation 
can never be perfect, since there will always be smearing effects
that cannot be compensated, and since there is not even a well-defined 
transition from perturbative to non-perturbative QCD. 

Jet finders can be applied to a variety of tasks.  The number of
well-separated jets found in an event sample allows a determination of
an $\as$ value. The distribution in angles between jets can be used as
test of the fundamental properties of QCD, such as the gluon spin and
the QCD color factors. The flow of particles around jet directions
probes soft physics, both perturbative and non-perturbative.  The
clustering of jets may help to identify massive particles, such as
$W^\pm$ and $t$, or to search for new ones.

The essential ingredients of  jet clustering algorithms are basically the same
independently of the phenomenological applications. Nonetheless,
the kinematics and dynamics of, e.g.,  
$e^+e^-$, $ep$ and $p\overline{p}$ collisions are sufficiently 
different that computational methods have to be modified accordingly 
(see, e.g., \cite{ellis,ktclus}). 
We will in this paper concentrate on algorithms for electron-positron 
annihilations, where there are no spectator jets and thus
schemes can be made especially simple.

Over the years, several algorithms have been proposed for the study of
$e^+e^-$ events. Recently, advances in the understanding of soft
perturbative physics lead to the introduction of further ones
\cite{camjet}. This has made it even more difficult for a user to
understand differences and to know which algorithm to use where. The
purpose of the current paper is to review several of the existing jet
finders and compare them in various ways. Neither the choice of
algorithms nor the selection of comparisons is exhaustive, but it
should still help give some useful hints.  We also introduce a few new
hybrid algorithms to better understand the results.
%
%
By using several event generators, we cross-check our findings. In a
sense, our study is an update of the corresponding one carried out in
Ref.~\cite{BKSS}, in view of the new algorithms that have been
proposed since then \cite{camjet,diclus} and of the advent of LEP2.

The conclusions might seem disappointing at first glance: while some
algorithms fare markedly less well than the better ones, there 
is not {\sl one} single best choice that sticks out in {\sl all}
 phenomenological
contexts we have studied. However, this should be of no surprise. 
In fact, given the varied use, there need not exist one algorithm that is
optimal everywhere. Instead, we will show that, depending on the tasks
assigned to the algorithm and on the physics dominion where this is applied, 
it is often possible to clearly individuate the most suitable to use.

In the following Section we review the historical evolution of
clustering algorithms and describe some of the more familiar ones.
Sections \ref{sec_pertcomparison} and \ref{sec_nonpertcomparison} 
contain comparisons between algorithms, for next-to-leading-order 
(NLO) and resummed perturbative QCD results, jet rates, jet energy
and angle reconstruction,
$W^\pm$ mass reconstruction, and so on. Finally Section 
\ref{sec_summary} contains a summary and outlook. 

\section{Clustering algorithms}
\label{sec_algorithms}

The first studies of jet structure in $e^+e^-$ annihilations 
were undertaken to establish the spin $1/2$ nature of quarks 
\cite{MarkI}. It was then only necessary to define a common
event axis for two back-to-back jets, and for this purpose
{\sl event measures} such as thrust \cite{thrust} or sphericity 
\cite{sphericity} are quite sufficient. 

With the search for and study of gluon jets at PETRA it became 
necessary to define and analyze three-jet structures. It is
possible to generalize thrust to triplicity \cite{triplicity}, 
in which the longitudinal momentum sum is maximized with respect 
to three jet axes. The maximization procedure can be rather 
time-consuming, however, in view of the large number of 
possibilities to subdivide particles into three groups.
Since three jets span a plane (neglecting initial-state 
QED radiation), special tricks are possible: if all particles
are projected onto an event plane, they can be ordered in 
angle such that only contiguous ranges of particles need be 
considered as candidate jets. The tri-jettiness measure 
\cite{trijettiness} uses the sphericity tensor to define the 
event plane, and thereafter finds the subdivision into three 
jets by minimizing the sum of squared transverse momenta, where 
each $p_{\perp}$ (or, equivalently, $k_{\perp}$)
is defined relative to the jet axis the particle is assigned to. 

Such special-purpose algorithms have the disadvantage that, first, one
procedure is needed to determine the number of jets in an event and,
thereafter, another to find the jet axes. The algorithms may also be
less easily generalizable to an arbitrary number of jets, or very
time-consuming. The task is not hopeless: with some tricks and
approximations, thrust/triplicity can be extended to an arbitrary
number of jet axes \cite{Bab,God,Bac} and tri-jettiness to four jets
\cite{Wu}. However, alternatives were sought, and more generic jet
algorithms started to be formulated. Several ideas were proposed and
explored around 1980 \cite{Bab,Bac,Dor,Dau,Lan}.  Most were based on a
binary clustering, wherein the number of {\sl clusters}\footnote{Here
  and in the following, the word `cluster' refers to hadrons or
  calorimeter cells in the real experimental case, to partons in the
  theoretical perturbative calculations, and also to intermediate jets
  during the clustering procedure.} %
is reduced one at a time by combining the two most (in some sense)
nearby ones. The joining procedure is stopped by testing against some
criterion, and the final clusters are called jets. An alternative
technique, top-down rather than bottom-up, is that of the minimum
spanning tree, where a complete set of links are found and then
gradually removed to subdivide the event suitably \cite{Dor}.

The starting configuration for the binary joining normally had each 
final-state particle as a separate cluster, but some algorithms
contained a `preclustering' step \cite{Bac,Dau}. Here, very nearby 
particles are initially merged according to some simplified scheme, 
in order to speed up the procedures or to make them less sensitive to 
soft-particle production. The possibility of `reassignment' between 
clusters was also used to improve on the simple binary joining recipe
\cite{Bab,God,Bac}. Normally all particles were assigned to some jet 
but, in the spirit of the Sterman--Weinberg jet definition 
\cite{StermanWeinberg}, a few algorithms allowed some fraction 
of the total energy to be found outside the jet 
cones \cite{Dau}. 

The distance measure between clusters always contained an angular 
dependence, explicit or implicit, while the energy/absolute momentum 
entered in different ways or not at all. As one example, of some
interest to compare with later measures, we note the use of 
thrust/triplicity generalized to $n$-jet axes \cite{Bab,God,Bac}:
\begin{equation}
T_n = \frac{1}{E_{\mathrm{tot}}} \max \sum_{i=1}^n |\bfp_i| =
      \frac{1}{E_{\mathrm{tot}}} \max \sum \sqrt{E_i^2 - m_i^2} \approx
      1 - \frac{1}{2E_{\mathrm{tot}}} \min \sum \frac{m_i^2}{E_i} ~, 
\end{equation} 
where each $\bfp_i$ is obtained as the vector sum of the momenta of
the particles assigned to jet $i$ (of energy $E_i$ and mass $m_i$).
Thus a maximization of $T_n$ is almost the same as a minimization of
$\sum m_i^2$, except that more energetic jets also can have a larger
mass. Note that the relation $m_{\mathrm{jet}}^2 \propto
E_{\mathrm{jet}}$ is approximately respected by non-perturbative
iterative jet fragmentation models \cite{FF,AGIS}.

The algorithms thus were rather sophisticated. Seen from a
modern perspective, the main shortcoming is that in
those days they could only  be tested against generators producing a
fixed number of partons --- two, three or, at most, four --- based 
on a leading-order (LO) matrix-element (ME) description. Therefore a 
`correct' number of jets existed, and criteria were devised to find 
this number. Those criteria tended to be rather complex, at times
even contrived, and thus often over-shadowed the simplicity of the 
basic algorithm. However, there is probably no fundamental reason why 
not several of these algorithms could have been used successfully 
even today, at least for some  tasks.

The oldest algorithm still in use is the \Luclus\ one
\cite{luclus}, which again is based on a binary joining
scheme, with additional preclustering and reassignment steps.
There were two advances. One was the choice of transverse momentum
as distance measure, which is better adapted to the conventional
picture of non-perturbative jet fragmentation and thus allows
a cleaner separation of perturbative and non-perturbative aspects of the
QCD dynamics. The other was that
no attempt was made to define a correct number of jets, but 
rather a parameter was left free, with the explicit purpose to 
correspond to different `jet resolution powers'.

The \Jade\ algorithm \cite{jade} offered a further 
simplification, in that only the binary joining was retained,
without preclustering or reassignment. The choice of distance
measure was based on invariant mass, corresponding to what was
available in most ${\cal O}(\as^2)$ calculations
of the time \cite{secondorderme}. The \Jade\ algorithm was 
therefore optimal for $\as$ determination studies, and came to
set the standard. By contrast, it performs less well in the 
handling of event-by-event
hadronization corrections, i.e., in the matching of jet directions and
energies between the parton and hadron level \cite{BKSS}.

Advances in the understanding of the perturbative expansion showed
that soft-gluon emission does not exponentiate when ordered in
invariant mass, while it does if transverse momentum is used instead
\cite{exponentiation,durham}. This gave birth to the \Durham\ 
algorithm \cite{durham}. Alternatives such as the \Geneva\ one were
also proposed \cite{BKSS}.

Recently, further advances in the understanding of soft-gluon emission
has lead to the introduction of new algorithms based on the \Durham\ 
scheme, the \AODurham\ and the \Cambridge\ ones \cite{camjet}, which
modify the clustering procedure of the former in order to remedy some
of its shortcomings. Given the huge increase in the computing power of
modern computers, one can now reverse the historical trend towards
simplification without compromising the efficiency of the algorithm,
e.g., indulging in procedures more sophisticated than the simple
binary joining.
 
The \Diclus\ (also called \Arclus) algorithm
\cite{diclus} does not really fit into the above scheme, in that
it is not based on the binary joining of two clusters to one
but on the joining of three clusters to two. This is well matched
to the dipole picture of cascade evolution. Like in many other
algorithms, the distance measure is based on transverse momentum. 

The connection between perturbative QCD cascades and jet clustering
algorithms is not only limited to the \Diclus\ case. In general one
may describe clustering algorithms as an attempt to reconstruct a QCD
cascade backwards in time. In fact, when we in the following argue
that one clustering should be performed \emph{before} another it is
based on experience from how to formulate QCD cascades where color
coherence is correctly taken into account. Such QCD cascades have been
the basis of the enormous success modern event generators have had in
describing the detailed structure of \tee\ annihilation events.

\begin{figure}[t]
  \begin{center}
    
    \setlength{\unitlength}{0.07mm}
    \begin{picture}(2000,700)(-100,0)
      \put(1050,650){\makebox(0,0){$\kappa$}}
      \put(1650,100){\makebox(0,0){$y$}}
      \put(400,100){\vector(1,0){1200}}
      \put(1000,100){\vector(0,1){600}}
      \put(500,100){\line(1,1){500}}
      \put(1500,100){\line(-1,1){500}}
      \thicklines
      \put(850,650){\makebox(0,0)[r]{\ariadne}}
      \put(900,700){\vector(0,-1){100}}
      \put(600,400){\makebox(0,0)[r]{\jetset}}
      \put(650,450){\vector(1,-1){70}}
      \put(1400,400){\makebox(0,0)[l]{\jetset}}
      \put(1350,450){\vector(-1,-1){70}}
      \put(1020,150){\vector(1,0){100}}
      \put(980,150){\vector(-1,0){100}}
      \put(1000,50){\makebox(0,0)[c]{\herwig}}

    \end{picture}
    \caption{The ordering of emissions in \ariadne, \herwig, and \jetset\
      QCD showers in the plane of $\kappa$ and $y$ (logarithm of
      transverse momentum and rapidity of emitted gluons).}
    \label{fig:timeordering}
  \end{center}
\end{figure}
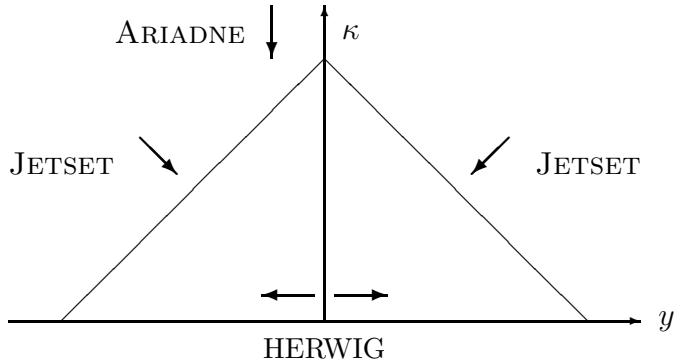

It should be noted, however, that the notion of time ordering in QCD
cascades is not unambiguous. Looking at the three most successful
coherent cascade implementations today, they all have different
ordering of emissions. \herwig\ orders emissions in angle while
\jetset\ orders in invariant mass with an additional angular
constraint to ensure coherence. Finally \ariadne\ orders the emissions
in transverse momentum. In fig.~\ref{fig:timeordering} we show the
approximate phase space available for gluon emission in an
\tee\ annihilation event, in the plane of logarithm of transverse
momentum ($\kappa$) and rapidity ($y$). The notion of time is
indicated for the three programs. In the \herwig\ and \jetset\ cases,
where emissions from the $q$ and $\bar{q}$ are treated separately,
there is one direction for each, while in the \ariadne\ case there is
only one direction for the ordering of emissions from the $q\bar{q}$
\textit{dipole}.

These three descriptions, although very different, are consistent with
perturbative QCD and it has not been possible to say that one in
better than another, although some experimental observables have been
suggested \cite{PhotonOrdering}. Common for all programs is that they
treat gluon emissions in a coherent way, and it may be easiest to look
at this in terms of angular ordering. In the following we present
three example diagrams of $\ee\rightarrow q_1+\bar{q}_2+g_3+g_4+...$.
In all cases we have drawn them as one Feynman diagram, but in general
all multi-gluon states are of course coherent sums of many diagrams.
It is clear that a good clustering algorithm in some sense should
cluster an event according to the dominating diagram for each given
partonic state. 

In the following Subsections we give a more detailed description of
several of the currently used algorithms. The order is not purely
historical, but is rather intended to allow a gradual introduction of
new concepts.

\subsection{\Jade}
\label{subsec_JADE}

The \Jade\ algorithm \cite{jade} may be viewed as the archetype 
of a binary joining scheme. 

In this class of methods, a 
distance measure $d_{ij}$ between two clusters
$i$ and $j$ is defined as a function of their respective four-momenta,
$p_{i,j} = (E_{i,j}, \bfp_{i,j})$. Since the measure is normally not Lorentz 
invariant, it is assumed that the analysis is performed in the hadronic 
rest frame of the event. To the extent this frame is not known, the 
lab frame is used instead, and the effects of initial-state QED 
radiation should then be included as a correction to the final 
physics results. 

The algorithm starts from a list of particles, that is considered
as the initial set of clusters. The two clusters with the smallest
relative distance are found and then merged into
one, provided their distance is below the desired minimum separation
$d_{\mathrm{cut}}$. The four-momentum of the new cluster $k$ is found 
from its constituents $i$ and $j$ by simple addition, e.g., $p_k = p_i + p_j$. 
The joining procedure is repeated, until all pairs of clusters have a 
separation above $d_{\mathrm{cut}}$. This final set of clusters is 
called jets.

In the \Jade\ algorithm the distance measure is given by
\begin{equation}
d_{ij}^2 = 2 E_i E_j (1 - \cos\theta_{ij}),  
\end{equation}
where $\theta_{ij}$ is the opening angle between the momentum 
vectors of the two clusters. As written here, $d_{ij}$ has dimensions
of mass. The scaled expression 
\begin{equation}\label{yJ}
y_{ij} = \frac{d_{ij}^2}{E_{\mathrm{vis}}^2} = 
\frac{2 E_i E_j (1 - \cos\theta_{ij})}{E_{\mathrm{vis}}^2}  
\end{equation}
is more often quoted. The visible energy $E_{\mathrm{vis}}$ would
agree with the centre-of-mass (CM) energy for a perfect detector but, to the
extent that some particles are lost or mismeasured, normalization
to $E_{\mathrm{vis}}$ gives some cancellation of errors between
numerator and denominator. In the following we will usually give the 
$y$-expression, but note that a translation between the two alternative 
forms is always possible. This also applies to the cut-off scale
$y_{\mathrm{cut}} = d_{\mathrm{cut}}^2/E_{\mathrm{vis}}^2$.

Whether the dimensional or scaled dimensionless form is preferable
is normally a matter of application and physics point of view. 
The $\as$ evolution with energy, and all other comparisons of jet
rates at different energies, are best done in terms of scaled 
variables $y$. The transition between perturbative and non-perturbative 
physics, on the other hand, is expected to occur at some fixed 
dimensional scale  of the order of 1 GeV. Such a hypothesis is 
supported, e.g., by the observable scaling violations of fragmentation 
functions in jets defined by a fixed $\ycut$. Therefore, we expect 
the `true' partonic multiplicity of an event to increase with
energy, tracing the increase of the hadronic multiplicity, while
the jet rate above a given $y$ drops, tracing the running of $\as$. 
 
The $d_{ij}$ measure above is closely related to the invariant mass
\begin{equation}
m_{ij}^2 = (p_i + p_j)^2 = m_i^2 + m_j^2 + 2 (E_i E_j - 
|\bfp_i| |\bfp_j| \cos\theta_{ij}),
\label{Jade_E_dist}
\end{equation}
and the use of the correct mass as distance measure defines the 
so-called E variant of the \Jade\ scheme. Given its Lorentz 
invariant character, mass would have been a logical choice, had it
not suffered from instability problems. The reason is well
understood: in general, particles tend to cluster closer in
invariant mass in the region of small momenta. The clustering process
therefore tends to start in the center of the event, and only
subsequently spreads outwards to encompass also the fast particles.
Rather than clustering slow particles around the fast ones (where the
latter na\"{\i}vely should best represent the jet directions), the 
invariant mass measure tends to cluster fast particles around 
the slow ones.

The $d_{ij}$ and $m_{ij}$ measures coincide when $m_i = m_j = 0$.
For non-vanishing cluster masses $d_{ij}$ normally drops below  
$m_{ij}$, and the difference between the two measures increases
with increasing net momentum of the pair. This tends to favor 
clustering of fast particles somewhat, and thus makes the standard
\Jade\ algorithm more stable than the one based on true invariant 
mass. 

There would seem to be a mismatch in comparisons between fixed-order 
perturbation theory based on the correct invariant mass expression
\cite{secondorderme} 
and experimental analyses based on the $d$ measure. However, the
perturbative results are normally presented in terms of massless 
outgoing partons, so the $m$ and $d$ measures agree on the parton 
level. A definition of hadronic cluster separation as if clusters were 
massless therefore better matches the partonic picture, and should
give smaller hadronization corrections. When performing 
NLO perturbative calculations, it is of course then 
of decisive importance to impose the same kind of clustering scheme
as will be used on the hadron level.

Further variants of the \Jade\ scheme have been introduced 
\cite{BKSS}. In the p alternative, the energy of a cluster $k$ is 
defined to be $E_k = |\bfp_k|$, so that the cluster is explicitly 
made massless, at the expense of violating energy conservation
when pairing two clusters. In the E0 scheme, massless clusters
are instead obtained by momentum violation, defining
$\bfp_k = E_k (\bfp_i + \bfp_j) / |\bfp_i + \bfp_j|$. In this paper 
we stay with the standard scheme, however.

\subsection{\Durham}
\label{subsec_Durham}

The \Durham\ algorithm \cite{durham} can be obtained from the \Jade\ 
one by a simple replacement of the distance measure from mass to
transverse momentum. In scaled variables
\begin{equation}\label{yD}
y_{ij} =  
\frac{2 \min(E_i^2, E_j^2) (1 - \cos\theta_{ij})}{E_{\mathrm{vis}}^2},
\end{equation}
i.e., with $E_i E_j \to \min(E_i^2, E_j^2)$. Some special features
should be noted. Firstly, strictly speaking, the measure is transverse
energy rather than transverse momentum, just like the \Jade\
measure is based on energies.  Secondly, the transverse momentum is
defined asymmetrically, as the $p_{\perp}$ of the lower-energy one
with respect to a reference direction given by the higher-energy one.
And, thirdly, the angular dependence only agrees with that of the
transverse momentum $p_{\perp} = E \sin\theta$ for small angles, where
$\sin^2 \theta \approx 2(1 - \cos\theta)$. The reason for retaining
the same angular dependence as in \Jade\ is obvious enough: the
correct $p_{\perp}$ would vanish for two back-to-back particles and
thus allow unreasonable jet assignments.  The \Durham\ algorithm
has eventually taken over the \Jade\ r\^ole of standard jet
finder. There are two main reasons for preferring \Durham.

Firstly,
fixed-order perturbative corrections are quite sizeable for the \Jade\
algorithm. This is particularly true for the case of the NLO
 ones to the three-jet rate $f_{3}(\ycut)$ 
(see later on, in Sect.~\ref{subsec_fixedorder} for its definition)
\cite{Yellow,jadef3}. 
The importance of this aspect is evident if one considers that 
$f_3(\ycut)$ provides a direct measurement of $\as$. Such behaviours
can be seen by noticing the large renormalization scale
dependence of $f_3(\ycut)$ at NLO, indicating that higher order
corrections are not yet negligible in the perturbative expansion.
Since it was (and still is) unthinkable with present computational
technology to attempt the evaluation of next-to-next-to-leading order
(NNLO) terms, the path to be necessarily followed in order to reduce
the
scale dependence of $f_3(\ycut)$ was to define new clustering
algorithms having smaller perturbative corrections. Secondly, the jet
fractions obtained in the \Jade\ scheme
do not show the usual Sudakov exponentiation of multiple soft-gluon
emission \cite{exponentiation}, despite having an expansion of the
form $\as\ln^2\ycut$ at small values of the resolution parameter.

\begin{figure}[!t]
\begin{center}
\vskip-3.0cm
\begin{picture}(455,200)
\SetScale{.9}
\SetWidth{1.2}
\SetOffset(0,0)

\Line(250,100)(300,100)
\Gluon(300,100)(400,120){2}{4}
\Line(300,100)(470,80)

\Line(250,100)(200,100)
\Gluon(100,120)(200,100){2}{4}
\Line(200,100)(30,80)

\Text(10,70)[]{$q_1$}
\Text(442.5,70)[]{$\bar q_2$}
\Text(375,120)[]{$g_3$}
\Text(75,120)[]{$g_4$}

\Text(225,90)[]{$\times$}

\end{picture}
\vskip-1.5cm
\caption{The seagull diagram
  with $E_3,E_4\ll E_1,E_2$ and $\theta_{14},\theta_{23}\ll\theta_{34}$.}
\label{fig_seagull}
\end{center}
\end{figure}
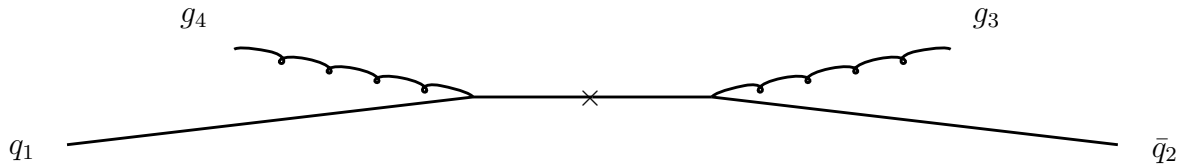

The source of such misbehaviors at both large (i.e., in fixed-order
calculations) and small (i.e., in the resummation of leading logarithms) 
$\ycut$ values is indeed the same, namely, the large rate of soft
gluons radiated in the hard scattering process and the way they are dealt
with in the clustering procedure. The problem can be exemplified by referring
to one of the possible configurations in which two soft gluons $g_3$ and
$g_4$ can be
emitted by two leading (i.e., highly energetic)
back-to-back quarks $q_1$ and $\bar q_2$. Let us imagine the first gluon
to be radiated in one of the two hemispheres defined by the plane
transverse to the axis of the two quarks and 
the second one on the opposite side (i.e., the `seagull diagram' of 
Ref.~\cite{camjet}: see Fig.~\ref{fig_seagull}). By adopting as
measure $y_{ij}$ the expression given in eq.~(\ref{yJ}),  the iterative
algorithm would combine the two gluons with each other first, so that the
net results is a `ghost jet' in a direction along which no original parton 
can be found. Such behaviours end up representing a serious challenge in 
perturbative calculations. On the one hand, the problems encountered by
fixed-order QCD in  cancelling divergences are amplified 
by the clustering of two soft particles, so that in general one naturally 
expects larger higher order terms.
On the other hand, such unnatural clustering induces a redistribution
of the partons in the final state that spoils the exponentiation
properties of large logarithms $\ln\ycut$ for $\ycut\ar0$   
\cite{partition}.

The simple modification \cite{yuri} given in eq.~(\ref{yD})  
is enough to cure the two above mentioned problems. This is clear if one
considers that, by adopting the \Durham\ measure, in the seagull diagram
configuration one of the soft gluons will always be combined first with 
the nearby high-energy quark, unless the angle that it forms
with the other gluon is smaller than that with respect to the leading
parton.
As a consequence, the stability of the fixed-order results is greatly
improved and the factorization of large leading and next-to-leading
logarithms guaranteed.

\subsection{\Luclus}
\label{subsec_LUCLUS}

Historically, \Luclus\ \cite{luclus} has not been used for 
$\as$ determinations and related QCD studies, but is instead widely 
used for other jet topics, such as search for new particles in
invariant mass distributions. In main properties it is similar to 
the \Durham\ algorithm, but with several differences.

Firstly, the transverse-momentum-based distance measure is
\begin{equation}\label{yL}
y_{ij} =  \frac{2 |\bfp_i|^2 |\bfp_j|^2 (1 - \cos\theta_{ij})}%
{(|\bfp_i| + |\bfp_j|)^2 E_{\mathrm{vis}}^2}.
\end{equation}
Geometrically, in the small-angle approximation, this can be viewed as 
the transverse momentum of either particle with respect to a reference 
direction given by the vector sum of the two momenta.

Apart from the difference between $|\bfp|$ and $E$, the step from the
\Durham\ to the \Luclus\ distance measure is given by the replacement
$\min(E_i, E_j) \to E_i E_j/(E_i + E_j)$.  Clearly the two expressions
agree when either of $i$ or $j$ is much softer than the other, so all
the soft-gluon exponentiation properties of the \Durham\ measure carry
over to the \Luclus\ one. In the other extreme, when $E_i = E_j$, the
two $y$-expressions differ by a factor of 4 (that is, by a factor of 2
at $p_{\perp}$ level).

The usage of $|\bfp|$ rather than $E$ in \Luclus\ is based on
non-perturbative physics considerations, specifically on the
properties of string fragmentation \cite{AGIS}. Here, primary
particles are given a Gaussian transverse momentum spectrum with
respect to the string direction, typically around 400 MeV, common for
all particle species. Secondary decays give a final mean $p_{\perp}$
that is around 100~MeV lower for pions than for Kaons or protons, but
this is still smaller than the mass difference between the particles.
Therefore, a jet is a set of particles with limited $p_{\perp}$ with
respect to the common jet direction, and using $E_{\perp}$ only
introduces unnecessary smearing. From a perturbative point of view,
arguments could be raised for the use of energy (see the discussion on
the \Jade\ algorithm above).  Also note that, in the string model, the
$p_{\perp}$ width of a non-perturbative jet is independent of
longitudinal momentum, to first approximation. This concept is
preserved by the symmetric way in which \Luclus\ defines
$p_{\perp}$. The asymmetric $p_{\perp}$ definition of \Durham\ is
more appropriate if high-energy particles are better lined up in
$p_{\perp}$ with the true jet axis than low-energy ones. This may
occur when multi-partonic states are considered, see discussion below,
so the matter is not quite clearcut.

The original \Luclus\ routine differs from the others presented here 
in that it contains preclustering and reassignment steps. These options
can both be switched off, individually, but the reassignment step was a 
part of the basic philosophy at the time the algorithm was written. 
The preclustering one, on the other hand, was purely intended to speed 
up the algorithm without affecting the final results significantly. 
The amount of preclustering can be varied, with much preclustering
giving a faster algorithm at the expense of some residual effects
of the preclustering step. Speed was an important consideration at the 
time the algorithm was originally formulated, but is normally no issue 
with modern workstations. Today users should therefore feel free to
switch off this step entirely.

First consider the reassignment aspect.
When two clusters are merged, some particles belonging to the new 
and bigger cluster may actually be closer to another cluster. A simple 
example is once again the seagull diagram of Fig.~\ref{fig_seagull},
with the quark-gluon opening angles not necessarily small.
With the \Luclus\ $p_{\perp}$ measure, it can happen (in fact, more easily 
than with
the \Durham\ one, see Fig.~\ref{fig_measure} later on) that the two soft 
particles are first combined to one cluster and thereafter this cluster is 
merged with one of the hard particles. One of the soft particles is that way
combined with the hard particle it is furthest away from. The `natural' 
subdivision would have been with one hard and one nearby soft particle 
in each final cluster. That is, a procedure that is good for going from 
four to three clusters and from three to two clusters may be less good 
for the combined operation of going from four to two clusters. 
The problem is that simple binary joining algorithms do not allow
previous assignments to be corrected in the light of new information.  

Hence the reassignment: after each joining of two clusters, each
particle in the event is reassigned to its nearest cluster. For
particle $i$, this means that the distance $d_{ij}$ to all clusters
$j$ in the event has to be evaluated and compared. After all particles
have been considered, and only then, are cluster momenta recalculated
to take into account any reassignments. To save time, the assignment
procedure is not iterated until a stable configuration is reached.
(Again, the time cost of these iterations could be acceptable today
but it was not at the time the algorithm was written.)  All particles
are reassigned after each binary joining step, however, and not only
those of the new cluster. Therefore an iteration is effectively taking
place in parallel with the cluster joining. Only at the very end, when
all $d_{ij} > d_{\mathrm{cut}}$, is the reassignment procedure
iterated to convergence --- still with the possibility to continue the
cluster joining if some $d_{ij}$ should drop below $d_{\mathrm{cut}}$
due to the reassignment.

The \Luclus\ algorithm was conceived mainly based on non-perturbative
considerations. The reassignment procedure is completely deterministic,
however, and can therefore also be applied to any perturbative 
calculation, just like the simple binary joining. The price is that
analytic calculations become more difficult to survey.
A reassignment cannot occur after the first binary joining of an
event, though, but only after the second. It therefore does not affect
leading or NLO results, but only NNLO and higher orders. 
 
In the preclustering step the original large number of particles
are put together in a smaller number of clusters. This is done as follows. 
The particle with the highest momentum is found, and thereafter all 
particles within a distance $d_{ij} < d_{\mathrm{init}}$ from 
it. Here it is intended that $d_{\mathrm{init}} \ll d_{\mathrm{cut}}$
for preclustering to give negligible effects. 
Together these very nearby particles
are allowed to form a single cluster. For the remaining particles, 
not assigned to this cluster, the procedure is iterated, until all 
particles have been used up. Particles in the central momentum region, 
$|\bfp| < 2d_{\mathrm{init}}$ are treated separately: if their 
vectorial momentum sum is above $2d_{\mathrm{init}}$ they are allowed 
to form one cluster, otherwise they are left unassigned in the initial 
configuration and only appear in the first reassignment step.

The value of $d_{\mathrm{init}}$, as long as reasonably small, should
have no physical importance, in that the same final cluster configuration 
will be found as if each particle initially is assumed to be a cluster by 
itself. That is, the particles clustered at this step are so nearby 
that they almost inevitably must enter the same jet. 
`Mistakes' in the preclustering can however
 be corrected by the reassignment procedure
in later steps of the iteration. Therefore reassignment may be seen as 
a prerequisite and guarantee for successful preclustering. 

In this respect, we would like to give a word of caution, about the 
actual meaning of `reasonably small'. The value chosen for 
$d_{\mathrm{init}}$ should depend on the $d_{\mathrm{cut}}$-range considered 
in the analysis. For example, the default value of 0.25 GeV is clearly 
inappropriate for $\ycut=d_{\mathrm{cut}}^2/s\approx 0.0001$, as some 
residual effects of preclustering are then visible (see 
Sects.~\ref{subsec_tubemodel} and \ref{subsec_jetrates} later on).
A scaling, e.g., like $d_{\mathrm{init}} = d_{\mathrm{cut}}/10$  
would have removed them (we have explicitly verified this in our numerical 
simulations). Though we recommend the mentioned scaling, should the 
preclustering step be retained, we have decided to keep the default 
value of 0.25 GeV in order to illustrate the consequences of a fixed
$d_{\mathrm{init}}$ for small $\ycut$.

{}From a perturbative physics point of view, the $d_{\mathrm{init}}$
parameter plays a r\^ole very similar to that, e.g., of the $y_0$ parameter
in the phase space slicing method of handling higher-order corrections
to MEs (see, e.g., Sect.~4.8 of \cite{LEP2QCDgen}).
Below $y_0$ the cancellation of real and virtual corrections is
carried out analytically in an approximate treatment of phase space,
while between $y_0$ and $y_{\mathrm{cut}}$ the addition of contributions 
is performed numerically with full kinematics. Hence $y_0$ should be 
picked as small as computer resources allow, and always much smaller 
than the physical $y_{\mathrm{cut}}$ parameter.

In this paper we will focus our attention on four possible options
of the \Luclus\ algorithm, namely the default one, and it stripped off
either preclustering or reassignment or both. We will call the latter
the \Durham\ scheme with the \Luclus\ measure (with the acronym
DL), as this effectively differs from the algorithm introduced in 
Sect.~\ref{subsec_Durham} only in the choice of the distance measure.

\Luclus\ has always been distributed as part of \jetset.
With the merger of \jetset\ and \pythia\ the routine has 
been renamed \Pyclus, but we will here refer to it by its original 
name. 

\subsection{\Geneva}
\label{subsec_Geneva}

The \Geneva\ algorithm \cite{BKSS} is based on pure binary
joining, with a dimensionless distance measure
\begin{equation}\label{yG}
y_{ij} = \frac{8}{9} \frac{E_i E_j (1 - \cos\theta_{ij})}%
{(E_i + E_j)^2}.
\end{equation}
Unlike the other algorithms studied, the measure depends only on the
energies of the particles to be combined, and not on the energy of the
event. The energy factor $E_i E_j/(E_i + E_j)^2 \approx \min(E_i, E_j)
/ \max(E_i, E_j)$ favors the clustering of soft particles to the
hardest ones and disfavors the combination of soft particles with each
other.  The soft-gluon problems of the \Jade\ algorithm are thus
avoided, indeed in a more effective way than in the \Durham\
scheme. In fact, a soft gluon will only be combined with another soft
gluon if the angle between them is {\sl much} smaller than the angle
between the former and the nearby high-energy particle. As a
consequence, it turns out that the \Geneva\ scheme exhibits a more
reduced scale dependence as compared to the \Durham\ algorithm in
the three-jet rate at NLO \cite{BKSS}. Indeed, in
Ref.~\cite{genevabest} it was shown that such property remains true
also in the case of the NLO four-jet rates $f_4(\ycut)$.  Furthermore,
it has been pointed out \cite{genevabest} that the \Geneva\
algorithm is particularly sensitive to the number of light flavors,
this rendering it most suitable for the study of New Physics effects.
For example, in the context of four-jet events in $e^+e^-$
annihilations \cite{genevabest}, where the existence of the so-called
`light gluino' events has been advocated in the past years
\cite{window}. As for the exponentiation properties of large
logarithms at small values of the resolution parameter, these have not
thoroughly been studied yet for this scheme. However, a simple 
example\footnote{For which we are indebted to Yuri Dokshitzer.},
should help understanding that the \Geneva\ algorithm can manifest severe 
misassignment problems in the soft regime. It suffices to consider a 
$q_1\bar q_2 g_3g_4$ configuration (e.g., with the antiquark and the two 
gluons in the same hemisphere), with the two gluons produced via a 
triple-gluon vertex and ordered in energy, such that $E_2\gg E_3\gg E_4$.  
In the region where $\theta_{34}<\theta_{23}$ the gluon
$g_4$ can be assigned by the \Geneva\ algorithm to 
the antiquark $\bar q_2$ rather than to other gluon $g_3$, since 
$(E_2+E_4)^2\gg(E_3+E_4)^2$ in the denominator of eq.~(\ref{yG}). 
Hence, one expects the $C_F$ factor instead of the correct $C_A$ one
to describe the radiation intensity of such an event. This induces a 
breakdown  of the correct exponentiation picture \cite{exponentiation,durham}:
see eqs.~(\ref{Fx})--(\ref{Fy}) in Subsect.~\ref{subsec_resummed}.
By contrast, a transverse momentum measure, such as the \Durham\ (\ref{yD})
or \Luclus\ (\ref{yL}) ones, would have more naturally assigned $g_4$ to $g_3$, 
as in the limit $\theta_{34}\ll\theta_{23}$ the full infrared (i.e., both
soft and collinear, driven by the gluon propagator) singularity sets on, 
which renders the triple-gluon diagram dominant in the kinematics above. 

The \Geneva\ algorithm has had little phenomenological impact so
far. One reason is that it is rather sensitive to hadronization
effects, as already pointed out in Ref.~\cite{BKSS}.  For instance,
compare a single original large-energy hadron with a system of hadrons
of the same total energy and collinear within the hadronization
$p_{\perp}$ spread of a few hundred MeV. Then the former can collect
particles further away in angle at the early stages of clustering.
Therefore clustering of gluon jets, which start out with a lower
energy and tend to fragment softer than quark jets, is disfavored.
This will introduce systematic biases, e.g., in jet energy
distributions, that can only be unfolded given a detailed
understanding of the hadronization process. In addition, the \Geneva\ 
algorithm is more sensitive to measurement errors since the measure
contains the energy of individual particles also in the denominator,
where other algorithms have instead the total visible energy, which is
more precisely measured.

\subsection{\AODurham}
\label{subsec_Angular}

This algorithm maintains the same measure (\ref{yD}) of the \Durham\ 
scheme, while modifying the clustering procedure. It was introduced in
Ref.~\cite{camjet} to obviate one of the flaws of that algorithm:
namely, its tendency of inducing `junk-jet' formation at small values
of the resolution parameter.

\begin{figure}[!t]
\begin{center}
\vskip-3.0cm
\begin{picture}(455,200)
\SetScale{.9}
\SetWidth{1.2}
\SetOffset(-10,0)

\Line(250,100)(50,100)
\Line(250,100)(300,100)
\Gluon(300,100)(315,115){2}{2}
\Line(300,100)(350,95)
\Line(350,95)(470,80)
\Gluon(350,95)(420,100){2}{4}
\Text(30,90)[]{$q_1$}
\Text(442,70)[]{$\bar q_2$}
\Text(395,92)[]{$g_3$}
\Text(295,112.5)[]{$g_4$}

\Text(225,90)[]{$\times$}

\end{picture}
\vskip-1.5cm
\caption{Parton branching with `unresolved', soft, large-angle gluon emission
$g_4$. Here, one has the following configuration:
$E_2\gg E_3\gg E_4$, and $\theta_{23}\ll \theta_{24}\approx\theta_{34}$.}
\label{fig_soft_unresolved}
\end{center}
\end{figure}
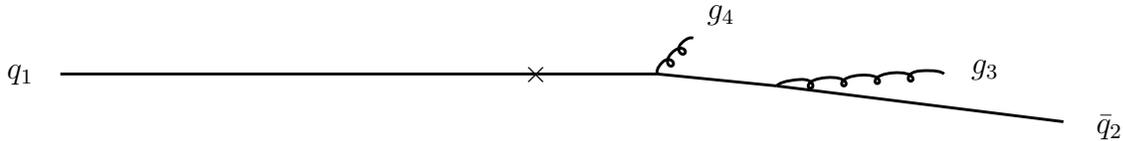

The problem is as follows.
Let us imagine the configuration in Fig.~\ref{fig_soft_unresolved},
with two back-to-back high-energy partons (the quarks $q_1$ and $\bar
q_2$) plus a double gluon emission ($g_3$ and $g_4$) \cite{camjet} in
the same hemisphere defined by the plane transverse to the direction
dictated by the two quarks, one of the gluons being at large angle and
soft ($g_4$) and the other ($g_3$) collinear to the nearby leading
particle on the same side ($\bar q_2$).  Then, according to the
clustering procedure adopted in the \Durham\ scheme, one usually
starts from the softest particle (i.e., one of the two gluons: here
$g_4$) and merges this with the nearest in {angle}, to minimize the
$p_{\perp}$-measure. Thus, such a particle gets clustered not with one
of the leading partons (i.e., $\bar q_2$ here) but, typically, with
the softest among the particles which happen to lie on the same side
(i.e., to the other gluon, $g_3$, in our example). This is contrary to 
our picture of the large-angle $g_4$ being emitted coherently by the
$\bar q_2$ and $g_3$, so that most of the recoil to the $g_4$ should
have been taken by the more energetic $\bar q_2$. Such a procedure
gets iterated in the case in which more particles are
involved (e.g., radiated in between the least, $g_4$, and the most
energetic, $\bar q_2$, ones in one hemisphere).  Since at each stage
the new pseudo-particle acquires more and more four-momentum, in the
end the latter will have a $p_{\perp}$ relative to the leading
particle in the same hemisphere larger than the resolution scale
adopted. This way, a third jet is eventually resolved.
A good algorithm should then be designed so that the starting
configuration remains classified as a two-jet final state down to the
smallest possible values of the resolution $\ycut$, at which the third
(junk-)jet separates. However, one should notice that if one
hemisphere of an event is significantly broadened by multiple
soft-gluon emission, where the gluons together carry away
non-negligible energy and $p_{\perp}$, it would be reasonable to argue
that the event could be legitimately recognised as a three-jet one.
Clearly, in this as in many other cases, it is the status of
our present computation technology and of its list of priorities that
induces the choice of strategy to be adopted. In fact, the latter needs
to be neither unique nor definitive.
On the other hand, as well demonstrated in Ref.~\cite{stan} (see, e.g., 
Fig.~1 there) and as we shall further see below, the remedy adopted
by the \AODurham\ (and \Cambridge, too) scheme in order to alleviate the
above mechanism appears more than adequate for present investigations.

In Ref.~\cite{camjet} it was shown that a simple modification of the
\Durham\ algorithm suffices to delay the onset of junk-jet formation,
which results mainly from a non-optimal sequence of clustering, rather
than from a poor definition of the test variable (as was the case for
the \Jade\ algorithm in the seagull diagram).  The key to reduce the
severity of the problem resides in distinguishing between the variable
$v_{ij}\equiv 2(1-\cos\theta_{ij})$, used to decide which pair of
objects to test first, and the variable $y_{ij}$ to be compared with
the resolution parameter $\ycut$. The algorithm then operates as
follows. One considers first the pair of objects $(ij)$ with the
smallest value of $v_{ij}$ (in Ref.~\cite{camjet}, this procedure was
referred to as `angular-ordering').  If $y_{ij}<\ycut$, they are
combined. Otherwise the pair with the next smallest value of $v_{ij}$
is considered, and so on until either a $y_{ij}<\ycut$ is found or, if
not, clustering has finished and all remaining objects are defined as
jets.

Coming back to the example configuration described before, but with
the new clustering procedure,
 one should expect the collinear quark $\bar q_2$ and gluon $g_3$
to be paired first,
with the soft, large-angle gluon $g_4$ eventually joining the new cluster.
In case more radiation is present around the leading quark,
the procedure
always iterates so that the pairing always starts amongst the particles
collinear to the leading quark $\bar q_2$, with the soft, large-angle gluon 
$g_4$ entering
the clustering procedure only at the very end. Indeed, this way, the
original configuration will more likely be recognised as a two-jet
one in the \AODurham\ than in the original \Durham. 
We will exemplify this in Sect.~\ref{subsec_jetrates}.

By generalizing the procedure to the full hard scattering process, one
indeed realizes that, at a given $\ycut$, the two-jet fraction at NLO
as given by the \AODurham\ is larger than in the original \Durham\ 
scheme, as illustrated in Ref.~\cite{camjet}.  Conversely, the
three-jet rate at the same order is smaller. Thus, since it is not
unreasonable to argue that jet algorithms having smaller NLO terms may
also have smaller higher-order corrections, one would imagine the
scale dependence of the three-jet rate for the \AODurham\ to be
reduced as compared to that of the \Durham\ algorithm.
%
%
This was shown explicitly again in Ref.~\cite{camjet} (see also
Sect.~\ref{subsec_fixedorder} later on).  Given the phenomenological
relevance of $f_3(\ycut)$, this should represent an improvement from
the point of view of the accuracy achievable, e.g., in $\as$
determinations, given that the theoretical error should diminish
accordingly. As for the exponentiation properties at small $\ycut$,
these remain unspoilt in the new algorithm, as discussed in
Ref.~\cite{camjet}.

Before closing this Section, we should remind the reader that the
\AODurham\ should be intended as an intermediate step between the
\Durham\ and \Cambridge\ schemes, rather than a new one. Indeed, we
will recall in the next Section another shortcoming of the \Durham\ 
algorithm which carries over into the \AODurham\ one and which can
have a strong impact in jet-rate studies. Nonetheless, for purposes of
comparison, we will present results for the \AODurham\ on the same
footing as the other algorithms.

\subsection{\Cambridge}
\label{subsec_Cambridge}

The \Cambridge\ algorithm was defined and its properties 
discussed in Ref.~\cite{camjet}.
It implements the same distance measure as the \Durham\ scheme, 
while further 
modifying the clustering procedure of the \AODurham\ one.
As a matter of fact, the sole introduction of $v_{ij}$ is not enough to
remedy the problem of `misclustering' of the \Durham\ algorithm, that
is,  the tendency of soft `resolved' particles 
of attracting wide-angle radiation \cite{camjet}.

Let us imagine that the soft, large-angle gluon $i=4$ 
in the previous example has 
been eventually resolved at a certain (possibly low) scale $\ycut$
(i.e., $y_{4j}>\ycut$). 
Clearly, the very same ability that it  had of attracting radiation 
(because of its softness)
as unresolved parton (i.e., when $y_{4j}<\ycut$) survives above the new
$\ycut$.
In particular, if further wide-angle (with respect to the leading quark in
the same hemisphere of the resolved gluon: i.e.,
$\bar q_2$) radiation occurs, say, two additional
gluons $g_5$ and $g_6$ 
one of which ($g_5$) happens to lie in angle a little closer to 
$g_4$ than to the other, then such a gluon will be erroneously assigned
(to $g_4$ rather than to $g_6$), when $\theta_{45}>\theta_{4\hat{23}}$
(thus assuming that $\bar q_2$ and $g_3$ have already been clustered).

In order to cure this problem, the \Cambridge\ scheme implements the
sterilization (i.e., the removal from the table of particles
participating in the clustering) of the softest particle in a resolved
pair, a procedure called `soft-freezing'. In our example, once $g_4$
has been removed from the sequence of clustering (when
$y_{4\hat{23}}>\ycut$), then the unwanted pairing of $g_4$ and $g_5$
(yielding $y_{45}<\ycut$ if the gluons are soft enough) is
successfully prevented.

\begin{figure}[!t]
\begin{center}
\vskip-1.7cm
\begin{picture}(455,200)
\SetScale{.9}
\SetWidth{1.2}
\SetOffset(-10,0)

\Gluon(180,150)(175,125){2}{2}
\Gluon(175,125)(115,130){2}{4}
\Gluon(200,100)(175,125){2}{3}
\Line(200,100)(50,70)
\Line(250,100)(200,100)
\Line(250,100)(300,100)
\Gluon(300,100)(325,115){2}{2}
\Line(300,100)(350,95)
\Line(350,95)(470,80)
\Gluon(350,95)(420,100){2}{4}
\Text(27.5,60)[]{$q_1$}
\Text(442,70)[]{$\bar q_2$}
\Text(395,92)[]{$g_3$}
\Text(305,112.5)[]{$g_4$}
\Text(170,145)[]{$g_5$}
\Text(90,117.5)[]{$g_6$}

\Text(225,90)[]{$\times$}

\end{picture}
\vskip-1.5cm
\caption{Parton branching with `resolved', soft, large-angle gluon emission
$g_4$. In this case, 
$\theta_{45}>\theta_{4\hat{23}}$, where $\hat{23}$ symbolizes
the cluster formed by the merging of $\bar q_2$ and $g_3$.}
\label{fig_soft_resolved}
\end{center}
\end{figure}

Note, however, that the diagram in \fig{fig_soft_resolved} is not the
only important one for the described parton configuration. For
example, we should also consider the diagram where $g_5$ is attached
to $\bar{q}_2$ (to the left of $g_4$),
in which case the freeze-out of $g_4$ could prevent $g_5$ from being
correctly clustered.
The relative importance of the two, as well as of the others appearing
at the same order, is clearly a function of the dynamics 
of the final state.  Numerical results will finally establish the
effectiveness and/or the limitations of the above approach.

As a matter of fact, the mentioned misclustering phenomenon could well
manifest itself in studies of high multi-jet rates (as in
Fig.~\ref{fig_soft_resolved}) as well as in those of the internal jet
sub-structure when examining the history of the mergings (e.g., in the
`would-be' cluster $\hat{45}$, artificially over-populated with
gluons).  In particular, it was shown in Ref.~\cite{camjet}, that such an
additional step plugged onto the \AODurham\ algorithm allows one to
increase the final event multiplicity (e.g., that of the NLO
$f_3(\ycut)$ rate) while not deteriorating the scale dependence of the
results. In addition, as for the \AODurham\ scheme,
the properties of factorization/exponentiation of large $\ln\ycut$
terms at small $\ycut$'s remain completely unaffected.
 
Altogether, as shown in Ref.~\cite{camjet}, the \Cambridge\ scheme
came at the time of those studies to represent the most suitable
choice when dealing with phenomenological studies involving infrared
(i.e., soft and collinear) configurations of hadrons/partons, also in
view of its performances with respect to the size of the hadronization
corrections (see Ref.~\cite{camjet} and
Sect.~\ref{sec_nonpertcomparison}).  In our forthcoming studies we
will allow for a variation of the basic \Cambridge\ scheme. Namely, we
will also adopt the \Luclus\ measure along with the \Durham\ one, and
we will label the corresponding algorithm as CL (with the same
clustering as the original \Cambridge, though).

\subsection{\Diclus}
\label{subsec_DICLUS}

The \Diclus\ algorithm is very different from the other ones
considered in this paper, in that each step clusters three jets
into two, rather than two into one. Although unconventional in this
respect, it is not unnatural. If one considers, e.g., the Lund string
fragmentation model, a hadron is produced in the color field between
two partons rather than stemming from one individual parton.
Especially soft hadrons between jets can never unambiguously be
assigned to one jet. Also on the perturbative level, gluons are
emitted coherently by neighboring partons. The partonic cascade
can therefore be formulated in terms of color-dipole radiation of gluons 
from pairs of color-connected partons, as in the \ariadne\ program
\cite{ariadne}. In a conventional parton cascade, this coherence is
instead formulated in terms of angular ordering.

Just as a conventional binary clustering algorithm can be viewed as an
attempt to reconstruct backwards a parton shower step by step, the
\Diclus\ algorithm tries to reconstruct a dipole cascade\footnote{This
  similarity with the \ariadne\ cascade is the reason the algorithm
  was originally called \Arclus.}. The ordering variable in \ariadne\ 
is a Lorentz-invariant transverse momentum measure defined for an
emitted parton $i$ with respect to the two emitting partons $j$ and
$k$ as
\begin{equation}\label{yAR}
  p^2_{\perp i(jk)}=\frac{(s_{ji}-(m_i+m_j)^2)(s_{ik}-(m_i+m_k)^2)}
  {s_{ijk}}\label{LL:invpt},
\end{equation}
where $s_{ij}$($s_{ijk}$) is the squared invariant mass of two (three)
partons. When reconstructing the dipole cascade backwards in time with
\Diclus, the same measure is used and the clustering procedure is as
follows.
\begin{itemize}
\item For each cluster $i$, find the two other clusters $j$ and $k$
  for which $p^2_{\perp i(jk)}$ is minimized.
\item Take the combination $i,j,k$ which gives the minimum $p^2_{\perp
    i(jk)}$, and if this is below a cutoff, remove cluster $i$ and
  distribute its energy and momentum among $j$ and $k$.
\end{itemize}
These steps are repeated until no $p^2_{\perp i(jk)}$ is below cutoff.

The joining is performed in the rest-frame of the three clusters, which
are replaced by two massless, back-to-back ones aligned with the one
of the original clusters with the largest energy ({\tt mode=1}).
Alternatively, the new clusters are placed in the plane of the
original ones with an angle
$\psi=\frac{E_k^2}{E_j^2+E_k^2}(\pi-\theta_{jk})$ from the highest
energy cluster $j$ ($k$ is the second highest energy cluster and
$\theta_{jk}$ is the angle between $j$ and $k$) ({\tt mode=0}). These
two options correspond exactly to the two ways of distributing
transverse recoil in an emission in \ariadne. Mode 1 is more similar
to the binary algorithms and is the one mostly used in this paper.

In fact, for most cases, the measure in eq.~(\ref{LL:invpt}) is closer
to the transverse mass of cluster 2 rather than its transverse
momentum, except when two clusters are almost at rest w.r.t.\ 
each other. Removing the subtraction of masses in eq.~(\ref{LL:invpt})
gives a measure which is closer to transverse mass everywhere:
\begin{equation}\label{mtAR}
  m^2_{\perp i(jk)}=\frac{s_{ji}s_{ik}}{s_{ijk}}\label{LL:invmt}.
\end{equation}
This measure is used in {\tt mode=2}, which is otherwise the same as
{\tt mode=1}. Since the jets in \Diclus\ are always massless, the only
difference will be in the initial clustering of massive hadrons, but
it turns out that this actually makes some difference in the
reconstruction of jet energies and angles below.

There is no straight-forward translation between the distance measure
in \Diclus\ and the ones used in binary clustering algorithms, since
it depends on three clusters rather than two. However, in the limit of
small measures, the $p_{\perp i(jk)}$ is equal to the \Luclus\ 
measure taken between the two softest clusters in the rest frame of
the clusters $i,j,k$. Also, if the cluster $i$ is much softer than $j$
and $k$, and much closer to, e.g., $j$ than $k$, $p_{\perp i(jk)}$
is again equal to the \Luclus\ $d_{ij}$.

To better understand how \Diclus\ works we look at the example
diagrams above. In \fig{fig_seagull}, the first step would be to
cluster $g_3$ into $\bar{q}_2$ and $g_4$ (or $g_4$ into $q_1$ and
$g_3$). In most cases the anti-quark would be given the major part or
the gluons transverse momentum, thus \Diclus\ resembles the other
transverse momentum based algorithms for this case. However, in some
cases the neighboring gluon will get a large fraction of the
transverse momentum, especially if the invariant mass of $g_3$ and
$g_4$ is smaller than that of $g_3$ and $\bar{q}_2$. This may happen
even if the angle between $g_3$ and $g_4$ is larger than that between
$g_3$ and $\bar{q}_2$, so the problem present in algorithms based on
invariant mass measures is not solved completely.

In \fig{fig_soft_unresolved}, assuming $g_4$ has smaller \tpt\ than
$g_3$, the first step would be to cluster $g_4$ into $q_1$ and $g_3$,
giving extra \tpt\ to $g_3$ possibly pushing it above the cutoff.
This is a good description of how this parton configuration would have
been generated in a dipole cascade. However, parton cascades in
general only agrees completely with perturbative QCD in the limit of
strong ordering of emissions where recoils do not matter, but, as we
have discussed above, it makes some difference here and it would
certainly be more reasonable to say that $g_4$ was radiated by $g_3$
and $\bar{q}_2$ coherently. Finally in \fig{fig_soft_resolved},
assuming now that $g_3$ and $\bar{q}_2$ have already been joined,
\Diclus\ could very well cluster $g_5$ into $g_6$ and $g_4$, which is
how it would have been produced in a dipole cascade, although if $g_5$
is soft, it would be more reasonable to have it produced from a dipole
between $g_6$ and the ($g_4g_3\bar{q}_2$) system, where the latter
acts coherently as one color charge.

Since \Diclus\ clusters three particles into two, it is not directly
possible to say which final state hadron belongs to which jet. It is,
however, possible to assign each particle to a jet after the jet
directions have been found, simply by finding the two jets $j$ and $k$
for each particle $i$ for which $p^2_{\perp i(jk)}$ is smallest and
then assigning particle $i$ to the jet which is closest in angle in the
rest frame of $ijk$. In this way it is also possible to redefine the
jet directions and energies by summing the momentum of the particles
assigned to each jet.  This reclustering is used below and is then
labeled `{reclustered}'.

In the remainder of the paper, in order to avoid any confusion with
the \Durham\ algorithm, we will sometimes label the three modes of the
\Arclus/\Diclus\ scheme by AR0, AR1 and AR2, for the {\tt mode=0,1,2}
cases respectively.

\section{Perturbative comparisons}
\label{sec_pertcomparison}

In this Section we will compare the performances of the
jet clustering algorithms introduced in Sect.~\ref{sec_intro} 
with respect to several quantities calculable in perturbative QCD
which are relevant to hadronic studies in electron-positron annihilations. 
It is subdivided in two Subsections. In the first 
we deal with fixed-order results whereas in the second we present
resummed perturbative quantities. The treatment in 
Subsect.~\ref{subsec_fixedorder} is mainly numerical, whereas
in Subsect.~\ref{subsec_resummed} is analytical.

To produce the results in the first case, we have made 
use of the `QCD parton generator'
\eerad\ \cite{EERAD}\footnote{An up-to-date list and a description
  of similar codes publicly available can be found in
  Ref.~\cite{QCDgenerators}.}.  Such programs calculate NLO corrections
to arbitrary infrared-safe two- and three-jet quantities, through the
order ${\cal O}(\as^2)$ in QCD perturbation theory.  Although they
resort to Monte Carlo (MC) multidimensional integration techniques,
they differ substantially from the QCD-based `Monte Carlo event
generators' that we will introduce later on (in
Sect.~\ref{sec_nonpertcomparison}).  For a start, the former compute
the  exact ${\cal O}(\as^2)$ ME result, rather than implementing only 
the infrared
QCD dynamics in the usual ${\cal O}(\alpha_s)$ ME + Parton Shower (PS)
modeling \cite{book}.  In addition, the phase-space configurations
generated are not necessarily positive definite, so that a
probabilistic interpretation is not possible.  Finally, these programs
only consider partonic states and no treatment of the hadronization
process is given. This kind of generators thus
represents a complementary tool for QCD analyses to the
phenomenological MCs which will be described and used in
Sect.~\ref{sec_nonpertcomparison}.

One final remark, before we start our investigations in pQCD. That is,
although we look here at some individual properties,  we remind the reader
that when choosing an algorithm for a particular measurement, one may
have to consider many different aspects altogether. When, e.g., measuring
$\as$ from the three-jet rate, it is not enough to find the
algorithm with smallest scale dependence, especially if this behavior
is found at a larger resolution scale where the three-jet rate is
lower and thus giving a larger statistical error in the measurement.
The goal must be to minimize the total error which may include both
the statistical error as well as systematical errors due to detector
unfolding, hadronization corrections, scale dependencies, etc.
This is however well beyond the scope of our study.

\subsection{Fixed-order perturbative results} 
\label{subsec_fixedorder}

In this Subsection we study the 
$y$-dependent three-jet fraction\footnote{Here and in the following
Subsection, in order to 
simplify the notation, we shall use $y$ to represent $\ycut$ and
refer to the various jet clustering algorithms/schemes by using their initials 
only. In addition, we acknowledge our abuse in referring to the latter
both as algorithms and as schemes, since the last term was originally
intended to identify the composition law of four-momenta when pairing two 
clusters (see Sect.~\ref{subsec_JADE}). This is in fact a well admitted
habit which we believe will not generate confusion in our discussion.}
$f_3(y)$, defined through the relation
\beq
\label{f3}
f_3(y) =     \left( \frac{\as}{2\pi} \right)    A(y)
           + \left( \frac{\as}{2\pi} \right)^2 (B(y)-2A(y)) + ... ,
\eeq
having implicitly assumed the choice $\mu=Q$ of the renormalization scale
(in the $\overline{\mbox{MS}}$ scheme). In eq.~(\ref{f3}),
$\as$ represents the strong coupling constant whereas $A(y)$ and
$B(y)$ are the so-called leading and next-to-leading `coefficient functions'  
of the three-jet rate, respectively. The terms of order $\cO{\as^2}$
involving $A(y)$ take account of the normalization to $\st$ rather
than to $\sigma_{0}$, which we assume throughout the paper. In fact,
we define the $n$-jet fraction
$f_n(y)$ as
\beq
\label{fn}
f_n(y)=\frac{\sigma_n(y)}
                  {\sum_m \sigma_m(y)}
            =\frac{\sigma_n(y)}
                  {\st(y)},
\eeq
where $\sigma_n(y)$ is the $n$-jet production cross section at a given $y$.
If $\st$ identifies the {\sl total} hadronic cross section
$\st=\sigma_{0}(1+\as/\pi+ ... )$, $\sigma_{0}$ being the
lowest-order Born one, then the constraint $\sum_n f_n(y)=1$ applies.
For $n=3$, eq.~(\ref{f3}) represents the three-jet fraction
in NLO approximation in perturbative-QCD (pQCD). 

Out of the thirteen jet clustering algorithms that we originally chose
for our study, we focus here our attention on the D, A, C, DL, CL, AR0
and AR1 schemes. We neglect considering the others for the following
reasons.  On the one hand, the J and G schemes have already been
documented extensively in the specialized literature
\cite{BKSS,Yellow} and, on the other hand, we would expect them to
have little phenomenological applications in the future, at least in
QCD studies in the infrared dominion. In fact, as already recalled,
the former does not allow for factorization properties of large
logarithms $\ln y$ at small $y$-values whereas for the latter these
have not been proven to hold yet. Indeed, they both share the feature
of being based on an obsolete invariant mass measure, whose flaws go
beyond the realm of perturbative QCD, as it is reflected by the more
fundamental r\^ole played by the transverse momentum in setting the
scale of jet evolution, as the argument of the running coupling, and
in defining the boundary between perturbative and non-perturbative
physics \cite{book}. Furthermore, of the four possible options of the
\Luclus\ scheme introduced previously, we only consider here the
simplest one (which we labeled DL), which implements neither the
preclustering nor the reassignment steps. Anyhow, because of the
kinematic simplicity of the partonic final states entering in the NLO
three-jet rates,
the differences among the four implementations turn out to be very
marginal.  Firstly, the reassignment option is inactive in three-jet
rates until the NNLO, see Sect.~\ref{subsec_LUCLUS}. Secondly, the
preclustering procedure can be incorporated easily with imperceptible
effects, so that the cancellations between the loop- and the
bremsstrahlung-diagrams still take place effectively without
deteriorating the accuracy of the results. (If this is not the case,
the $d_{\mathrm{init}}$ parameter has not been set appropriately, see
Sect.~\ref{subsec_LUCLUS}.)  Thus, the claim made in the literature,
that the \Luclus\ algorithm is not suitable for perturbative
calculations (see, e.g., Ref.~\cite{BKSS}), does not apply in the
present context: i.e., in {\sl numerical} computations of NLO
three-jet observables. Also, although analytical calculations with the
original \Luclus\ scheme may be prohibitively difficult, one can
certainly say that, without preclustering and reassignment (here, DL
scheme), \Luclus\ remains a reasonable option to adopt, especially in
view of some of the results that we will present in the following. Besides,
its properties with respect to the Sudakov exponentiation of
soft-gluon emission in the resummation procedure of large $\ln y$
logarithms are on the same footing as the D, A and C schemes
\cite{camjet} (see next Section). For \Diclus\ we note that the
measures in eqs.~(\ref{yAR}) and (\ref{mtAR}) are equivalent for
massless partons, so that in the following AR2 coincides with AR1 as
they have the same recoil assignment (see Subsect.~\ref{subsec_DICLUS}).

\begin{figure}[htb]
\begin{center}
~\epsfig{file=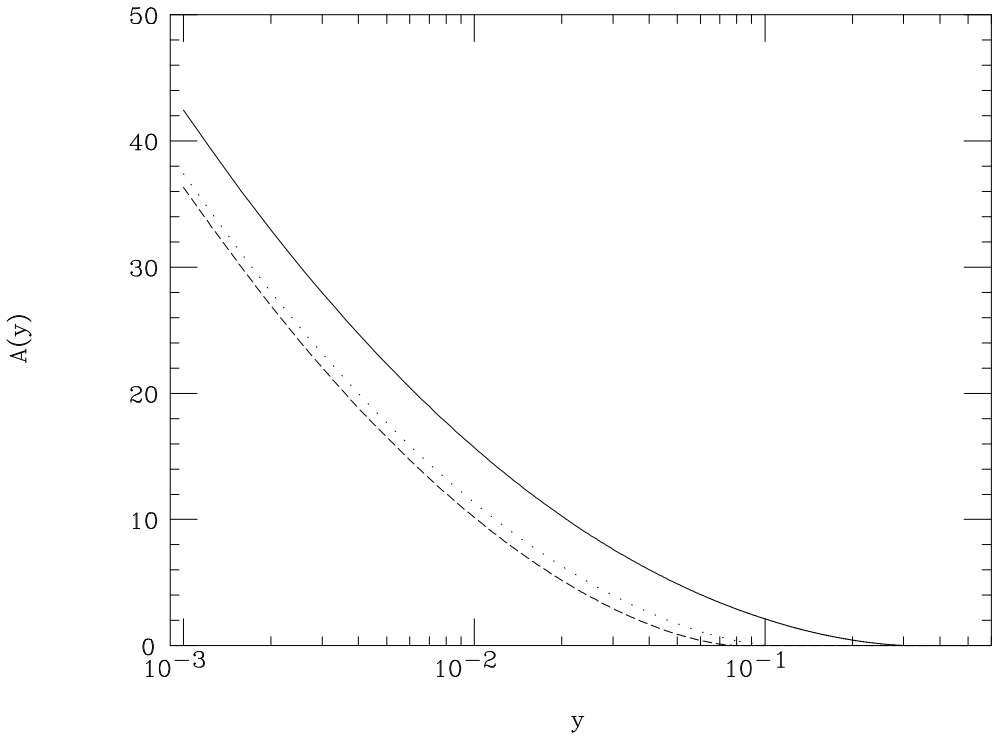,height=16cm,width=12cm,angle=90}
\caption{The parton level $A(y)$ function entering in the three-jet
fraction (\ref{f3}) at LO and NLO in the D, A and C schemes (solid line), 
DL and CL schemes (dashed line), AR0 and AR1 schemes (dotted line).}
\label{fig_afunction_bis}
\end{center}
\end{figure}

Fig.~\ref{fig_afunction_bis} shows the $A(y)$ function, over the range
$ 0.001 \leq y \Ord 0.1$, for the selected algorithms. Notice that
in several cases the curves coincide. In particular, it 
occurs for the D, A and C
schemes, the DL and CL algorithms and the AR0
and AR1 ones, respectively. This is evident if one considers that (apart from
the C and CL options)
the various schemes, within each of the three subsets, differ only in the
clustering procedure of unresolved particles, 
which clearly does not affect the LO three-jet rates.
As for the C and CL algorithms, one should consider that, for $n=3$ partons,
kinematical constraints impose that, on the one hand, the two closest
particles are also those for which $y$ is minimal and, on the other hand,
the identification of the softest of the three partons as a jet implies that
the remaining two particles are naturally the most energetic and far
apart. This ultimately means that the $A(y)$ function is the same also
for the schemes implementing the soft-freezing step.

The pattern of the curves in Fig.~\ref{fig_afunction_bis} is easily understood
in terms of the measures of the algorithms. For given values of angle
and energies (or three-momenta) of the parton pair $(ij)$, $\theta_{ij}$ and  
$E_{i,j}$ (or $|\bfp_{i,j}|$), respectively, the value of $y_{ij}$ is 
generally larger in the D, A and C schemes, as compared
to the DL and CL ones: see eqs.~(\ref{yD}) and (\ref{yL}). 
Therefore, over an identical portion of phase space, more three-parton
events will be accepted as three-jet ones in the D, A and C algorithms
than in the DL and CL ones (see also Fig.~\ref{fig_measure} below),
this ultimately increasing the value of $A(y)$. The comparison of the
two measures (\ref{yD}) and (\ref{yL}) to that of the \Diclus\ schemes
(\ref{yAR}) is clearly less straightforward, as already discussed.
For a fixed $y$, the latter is on average larger than that of the DL measure
but smaller than that of the D one, as can be deduced from 
Fig.~\ref{fig_afunction_bis}.

\begin{figure}[htb]
\begin{center}
~\epsfig{file=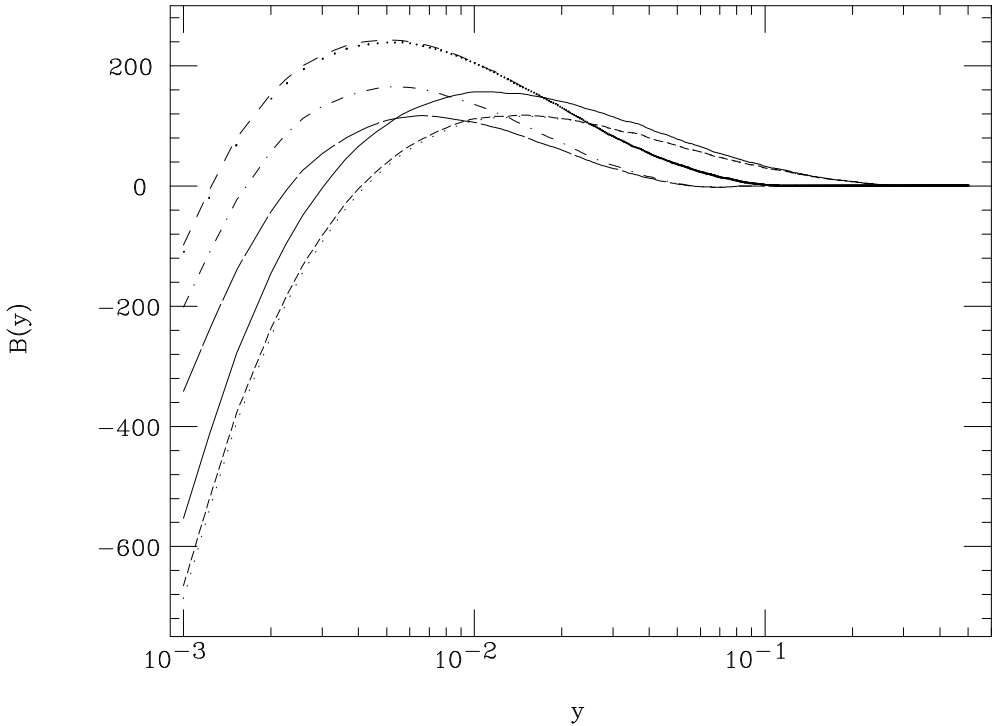,height=16cm,width=12cm,angle=90}
\caption{The parton level $B(y)$ function entering in the three-jet
fraction (\ref{f3}) at NLO in the D scheme (solid line),
A scheme (short-dashed line),
C scheme (dotted line), DL scheme (dot-dashed line),
CL scheme (long-dashed line),
AR0 scheme (dashed line) and AR1 scheme (fine-dotted line).}
\label{fig_bfunction_bis}
\end{center}
\end{figure}

In Fig.~\ref{fig_bfunction_bis} we present the NLO $B(y)$ function.
Because of the different recombination procedures of the schemes,
the various curves now all separate. The interplay between the D, A and C
rates has been analysed in great detail in Ref.~\cite{camjet}, so we do not
repeat those discussions in this paper. We only spot here the correspondence 
existing between, on the one hand, the D and C algorithms and, on the other 
hand, the DL and CL ones. In the sense that, the former two differ from each 
other in the same way as the latter two do: i.e., in the angular-ordering
and soft-freezing procedure recommended in Ref.~\cite{camjet}. Indeed,
the relative behaviors (of D vs. C and of DL vs. CL) are qualitatively similar.
Further notice the tendency of the DL, CL, AR0 and AR1 schemes
of yielding at small $y$'s a $B(y)$ consistently 
higher than that due to the other three algorithms, and vice versa at
large values of the resolution parameter. 
In other terms, they emphasize the real four-parton component with one 
unresolved pair more than the virtual three-parton at small $y$-values,
and vice versa as $y$ increases. However,
we remind that the peaking of $B(y)$ at different
$y$ values in itself does not have to say much, since the definition of 
$y$ is not the same. The simplest measure of the difference in cancellations
between real and virtual contributions is instead the maximum value of $B(y)$.


We are now going to carefully investigate the
interplay of the $A(y)$ and $B(y)$ functions in 
the expression of the three-jet fraction since,
as recalled in Ref.~\cite{camjet}, from the point of view
of perturbative studies, a `good' jet clustering algorithm should allow
for a reduced $\mu$-scale dependence of the fixed-order results, where $\mu$ 
is the subtraction (energy) scale regulating the infrared cancellations. 
As a matter of fact, the $\cO{\as^3}$ corrections are guaranteed
to cancel the $\mu$-dependence of the $\cO{\as^2}$ three-jet fraction
up to the order $\cO{\as^4}$, so that the smaller the variations with $\mu$
the lower the higher order corrections are expected to be. The
$\mu$-scale dependence is introduced in eq.~(\ref{f3}) by means of the 
two substitutions
\beq
\label{scaledep}
\as\rightarrow \as(\mu),\qquad\qquad\qquad
B(y)\rightarrow B(y)-A(y)\beta_0\ln(\frac{Q}{\mu}),
\eeq
where $\beta_0=11-2N_f/3$ is the first coefficient of the QCD $\beta$-function
and $N_f$ is the number of fermionic (colored)
flavors active at the energy scale $\mu$. 

A problem arises when studying the scale dependence of $f_3(y)$
for algorithms based on different measures, as for the same $y$
the three-jet fraction at NLO can be significantly different.
A more consistent procedure was outlined in 
Ref.~\cite{BKSS}: that is, to compare the NLO scale dependence of the various
schemes not at the same $y$-value, rather at the same $A(y)$, 
the three-jet fraction at LO. As can be viewed from 
Fig.~\ref{fig_bfunction_bis}, two possible
 combinations of $y$'s are the following;
$y_{\mathrm\tiny{D,A,C(DL,CL)[AR0,AR1]}}=0.010(0.005)[0.006]$ and
$y_{\mathrm\tiny{D,A,C(DL,CL)[AR0,AR1]}}=0.050(0.021)[0.025]$.
Such  values are typically in the three-jet region and, in addition, they are 
rather large, as compared to the minimum $y=0.001$ considered
so far, so that they can guarantee the full applicability of the perturbation
theory. 

\begin{figure}[htb]
\begin{center}
~\epsfig{file=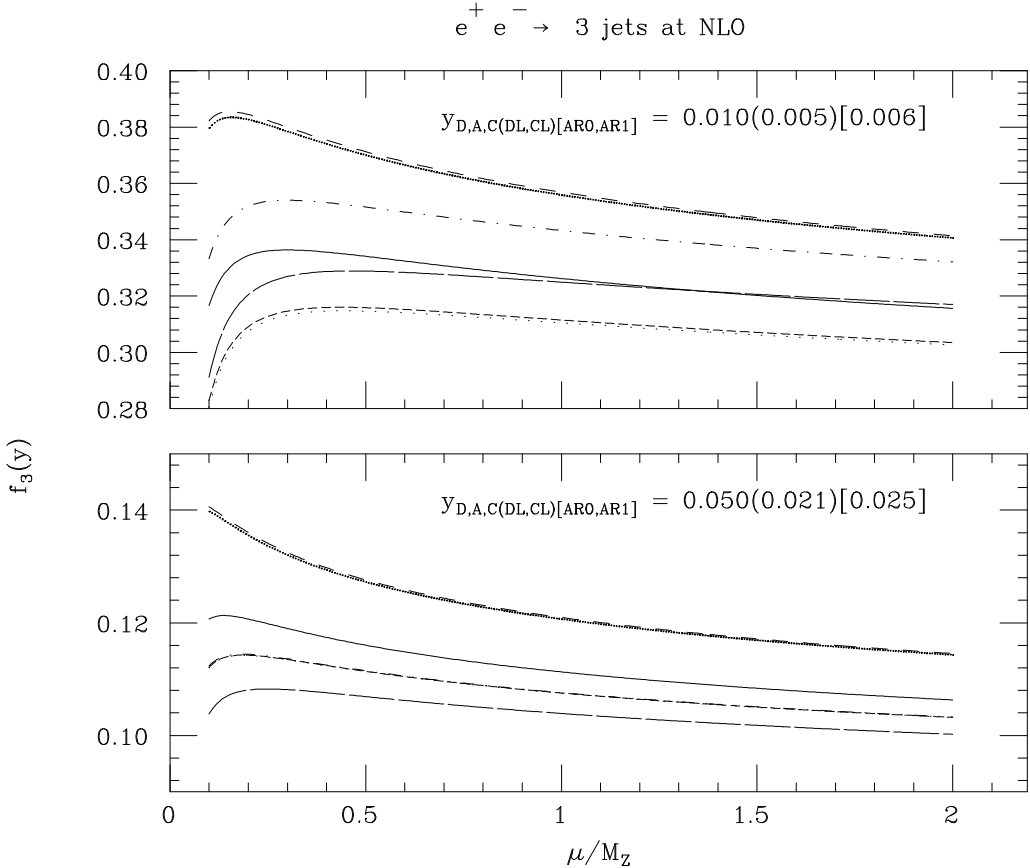,height=16cm,width=12cm,angle=90}
\caption{The parton level three-jet fraction at NLO as a function of the
adimensional renormalization scale $\mu/M_Z$, for
$y_{\mathrm{\tiny{D,A,C(DL,CL)[AR0,AR1]}}}=0.010(0.005)[0.006]$ 
(upper plot) and
$y_{\mathrm{\tiny{D,A,C(DL,CL)[AR0,AR1]}}}=0.050(0.021)[0.025]$
(lower plot), in the 
the D scheme (solid line),
A scheme (short-dashed line),
C scheme (dotted line), DL scheme (dot-dashed line),
CL scheme (long-dashed line),
AR0 scheme (dashed line) and AR1 scheme (fine-dotted line).}
\label{fig_scale_bis}
\end{center}
\end{figure}

Fig.~\ref{fig_scale_bis} shows (again for the seven selected algorithms)
the value of $f_3(y)$ plotted against the adimensional scale parameter
$\mu/Q\equiv \mu/M_Z$, over the range between $1/10$ and
$2$, for the two mentioned combinations of the jet resolution parameter.
Note that for the 
strong coupling constant  we have used the two-loop expression,
with $N_f=5$ and
$\Lambda^{(5)}_{\overline{\mbox{\tiny{MS}}}}=250$
MeV, yielding $\as(M_Z)=0.120$, with $Q=M_Z$ as CM energy
at LEP1. 

Although the structure of the QCD perturbative expansion does not prescribe
which value should be adopted for the scale $\mu$, an obvious requirement
is that it should be of the order of the energy scale involved in the problem:
i.e., the CM energy $Q$ (see 
Ref.~\cite{subtraction} for detailed discussions). Indeed, this choice 
prevents the appearance of large terms of the form $(\as\ln(\mu/Q))^n$
in the QCD perturbative series. Furthermore, 
the physical scale of gluon emissions that actually give rise to three-jet 
configurations are to be found down to the energy scale
${\sqrt{y}}M_Z$, not above $M_Z$. In other terms, one should avoid 
 building up large logarithmic terms related to the
mismatch between $\mu\ge M_Z$ and the physical process scale ${\sqrt{y}}M_Z$.
Therefore, as a sensible range over which to estimate
the effects of the uncalculated ${\cal O}(\as^3)~+~...$
corrections one should adopt a reduced interval just below the value
$\mu/M_Z = 1$.
If one does so, then a remarkable feature of Fig.~\ref{fig_scale_bis}
is that the DL and CL algorithms show a \sensibly\ reduced scale dependence
as compared to the D and C ones, respectively,
at low and especially at high $y$-values. 
Furthermore, among these two algorithms, it is the CL one
that in general performs  better, on the same footing as the C 
algorithm does as compared to the D one.

For example, at the low(high) $y$-values considered, 
the differences between the maximum and minimum values of $f_3(y)$ between
$\mu/M_Z=1/2$ and $\mu/M_Z=1$ are 
$2.4(3.6)\%$ for the DL scheme and 
$1.2(2.9)\%$ for the CL one, respectively. 
In the case of the D and C algorithms, one has
$2.4(4.3)\%$ and
$1.3(3.6)\%$, correspondingly.
The numbers for the AR0 and AR1 algorithms are \sensibly\ larger:
$4.0(5.5)\%$ and
$4.0(5.5)\%$ at small(large) $y$-values, respectively.
To help the reader in disentangling the features of 
Fig.~\ref{fig_scale_bis}, we have reproduced some of the data points of
the figure in Tab.~\ref{tab_scale_bis}.

\begin{table}[htb]
\begin{center}
\begin{tabular}{|c|c|c|c|c|c|c|c|}
\hline
\multicolumn{8}{|c|}
{\rule[0cm]{0cm}{0cm}
$f_3(y)$}
\\ \hline\hline

\multicolumn{8}{|c|}
{\rule[0cm]{0cm}{0cm}
$y_{\mathrm\tiny{D,A,C(DL,CL)[AR0,AR1]}}=0.010(0.005)[0.006]$}
\\ \hline
\rule[0cm]{0cm}{0cm}
$\mu/M_Z$ & D & A & C & DL & CL & AR0 & AR1 \\ \hline
\rule[0cm]{0cm}{0cm}
$0.25$ & $0.336$ & $0.312$ & $0.311$ & $0.353$ & $0.325$ & $0.383$ & $0.381$ \\
$0.50$ & $0.334$ & $0.316$ & $0.315$ & $0.352$ & $0.329$ & $0.371$ & $0.370$ \\
$0.75$ & $0.330$ & $0.314$ & $0.313$ & $0.347$ & $0.327$ & $0.363$ & $0.362$ \\
$1.00$ & $0.326$ & $0.311$ & $0.310$ & $0.343$ & $0.325$ & $0.357$ & $0.356$ \\
$1.25$ & $0.323$ & $0.309$ & $0.308$ & $0.340$ & $0.323$ & $0.352$ & $0.351$ \\
$1.50$ & $0.320$ & $0.307$ & $0.306$ & $0.337$ & $0.320$ & $0.348$ & $0.347$ \\
\hline\hline

\multicolumn{8}{|c|}
{\rule[0cm]{0cm}{0cm}
$y_{\mathrm\tiny{D,A,C(DL,CL)[AR0,AR1]}}=0.050(0.021)[0.025]$}
\\ \hline
\rule[0cm]{0cm}{0cm}
$\mu/M_Z$ & D & A & C & DL & CL & AR0 & AR1 \\ \hline
\rule[0cm]{0cm}{0cm}
$0.25$ & $0.120$ & $0.114$ & $0.114$ & $0.114$ & $0.108$ & $0.134$ & $0.134$ \\
$0.50$ & $0.116$ & $0.111$ & $0.111$ & $0.111$ & $0.107$ & $0.128$ & $0.127$ \\
$0.75$ & $0.113$ & $0.109$ & $0.109$ & $0.109$ & $0.105$ & $0.124$ & $0.123$ \\
$1.00$ & $0.111$ & $0.108$ & $0.108$ & $0.108$ & $0.104$ & $0.121$ & $0.121$ \\
$1.25$ & $0.110$ & $0.106$ & $0.106$ & $0.106$ & $0.103$ & $0.119$ & $0.119$ \\ 
$1.50$ & $0.108$ & $0.105$ & $0.105$ & $0.105$ & $0.102$ & $0.117$ & $0.117$ \\
\hline\hline

\multicolumn{8}{|c|}
{\rule[0cm]{0cm}{0cm}
$e^+e^-\ar 3$ jets at NLO}
\\ \hline

\end{tabular}
\caption{The parton level three-jet fraction at NLO in correspondence
of  selected values
of the adimensional renormalization scale $\mu/M_Z$, for 
$y_{\mathrm\tiny{D,A,C(DL,CL)[AR0,AR1]}}=0.010(0.005)[0.006]$ (upper section) 
and 
$y_{\mathrm\tiny{D,A,C(DL,CL)[AR0,AR1]}}=0.050(0.021)[0.025]$ (lower section) 
in the D, A, C, DL, CL, AR0 and AR1 schemes.}
\label{tab_scale_bis}
\end{center}
\end{table}

The improvement in switching from CL to DL can be traced back to the
implementation of the angular-ordering and soft-freezing procedures, as one 
of their side effects is to reduce the 
three-jet fraction: compare to eq.~(\ref{f3}), where the $B(y)$ term enters
with a positive sign (the leading piece proportional to
$A(y)$ is the dominant one also at NLO). 
As pointed out in Ref.~\cite{camjet} and also discussed earlier on, 
the reduced scale dependence and
the smaller NLO corrections to the three-jet rate are intimately related.

\begin{figure}[htb]
\begin{center}
\centerline{}
\begin{minipage}[b]{.5\linewidth}
\centerline{\Durham\ measure: 
$2~\mbox{min}\left(\frac{x_i^2}{4},\frac{x_j^2}{4}\right)$\quad}
\centering\epsfig{file=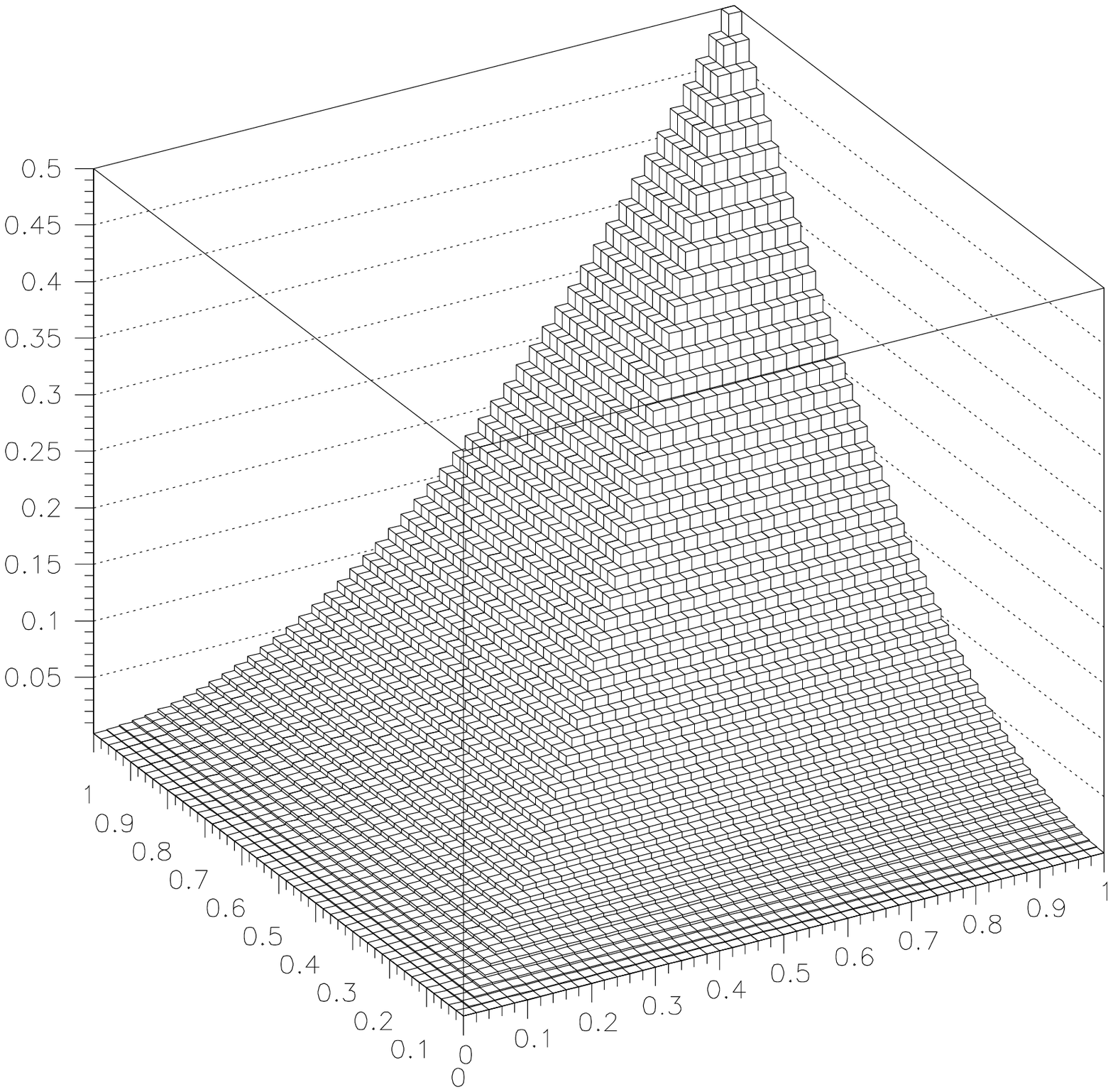,angle=0,height=8cm,width=\linewidth}
\vskip-2.0cm
\centerline{
$x_i$
\qquad\qquad\qquad\qquad\qquad\qquad\quad
$x_j$}
\end{minipage}\hfil
\begin{minipage}[b]{.5\linewidth}
\centerline{\Luclus\ measure: 
$2\left(\frac{x_i}{2}\right)^2 \left(\frac{x_j}{2}\right)^2/ 
\left(\frac{x_i}{2}+\frac{x_j}{2}\right)^2$}
\centering\epsfig{file=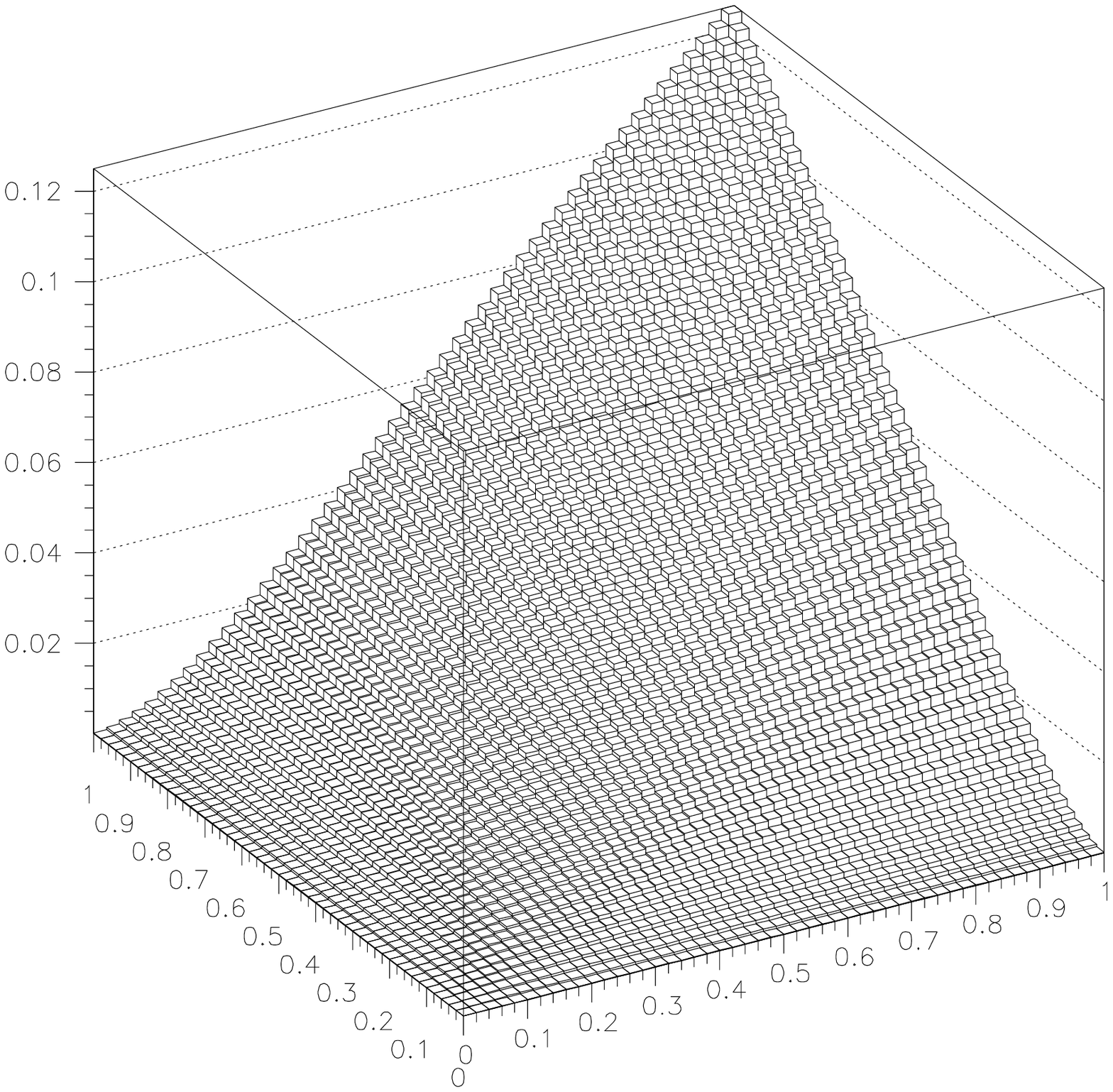,angle=0,height=8cm,width=\linewidth}
\vskip-2.0cm
\centerline{
$x_i$
\qquad\qquad\qquad\qquad\qquad\qquad\quad
$x_j$}
\end{minipage}\hfil
\centerline{}
\caption{The dependence of the \Durham\ (left-hand side) 
and \Luclus\ (right-hand side) measure as a function of the reduced
energies of the parton pair $(ij)$.}
\label{fig_measure}
\end{center}
\end{figure}

The differences between the CL(DL) algorithm and the C(D) one can only
be ascribed to the choice of the measure, as the clustering procedure
is the same for both schemes. From the numbers in Tab.~\ref{tab_scale_bis},
it is clear that the reason for the improved performance goes beyond the
relative importance of the LO and NLO pieces in the three-jet rate, as
in some cases the DL and CL rates are above the D and C ones, respectively:
e.g., at low $y$, where nonetheless the behaviour of the two measures 
is almost identical. Indeed, one could associate the better performances of
the \Luclus\ measure (\ref{yL}) as compared to the \Durham\ one
(\ref{yD}) to the fact that the first describes a {\em smooth}
function of its arguments over all the available phase space whereas
the second does not. This can be appreciated in
Fig.~\ref{fig_measure}, where the shape of the expression (here $Q$
plays the r\^ole of ${E_{\mathrm{vis}}} $ in
Sect.~\ref{sec_algorithms}) \beq
\label{measureD}
\frac{y_{ij}^D}{1-\cos\theta_{ij}}=
2\frac{\mbox{min}(E_i^2,E_j^2)}{Q^2}
\equiv 
2~\mbox{min}\left(\frac{x_i^2}{4},\frac{x_j^2}{4}\right)
\eeq 
for the \Durham\ measure is compared to that of the \Luclus\ one
\beq
\label{measureL}
\frac{y_{ij}^L}{1-\cos\theta_{ij}}=
2\frac{(|p_i|^2|p_j|^2)}{(|p_i|+|p_j|)^2 Q^2}
\equiv 2\frac{\left(\frac{x_i}{2}\right)^2 \left(\frac{x_j}{2}\right)^2} 
             {\left(\frac{x_i}{2}+\frac{x_j}{2}\right)^2},
\eeq
as bi-dimensional function
of the reduced energies $x_i=2E_i/Q$ and $x_j=2E_j/Q$. For simplicity, we assume
the two cluster $i$ and $j$ to be massless (i.e., $E_{i,j}=|{\bfp}_{i,j}|$)
and drop the angular dependence
$(1-\cos\theta_{ij})$.

It is well known that the presence of `edges' at the border of 
the phase space defined by a jet algorithm is a source of
misbehaviors in higher-order perturbation theory, as they ultimately 
generate infrared divergences (integrable, though) inside the physical region 
\cite{CatWeb}. For example, one can refer to the so-called 
`infrared instability' of the jet energy profile $(d E_T/dr)$
in (iterative) cone algorithms typically used in hadron-hadron collisions,
with $r$ being the Lorentz-invariant opening angle of the cone defined in terms
of pseudorapidity and azimuth (see, e.g., Ref.~\cite{Mike} for definitions
and details).
In fact, such a shape shows an edge in $\cO{\as^3}$ perturbation theory
at the cone radius $r=R$.
Although a resummation of logarithms $\pm\ln(|r-R|)$ to all orders cures
the problem (as the edge eventually becomes a `shoulder' !), a lesson
to be learned is that it is clearly desirable to avoid observables
with discontinuities 
when comparing with fixed-order predictions. (Similar conclusions
can be drawn for the $C$-parameter in $e^+e^-$ scatterings 
\cite{CatWeb,lastCatWeb}.)

Although not quite the same context, it is not unreasonable to expect
that the \Durham\ measure might reveal sooner or later problems
similar to those discussed, given its behavior along the trajectory
$x_i=x_j$. In this respect, the original \Luclus\ measure should
represent a `safer' observable.  Indeed, the more sensitive scale
dependence of the D and C algorithms, as compared to the DL and CL
ones, respectively, could well be a first notice of possible problems
in higher order pQCD.

Also the measure in \Diclus\ is a smooth function of its
arguments. However, as discussed in section \ref{subsec_DICLUS},
\Diclus\ still have some problems with the seagull diagram giving
larger three-jet rates and also larger scale dependence.

\begin{figure}[htb]
\begin{center}
~\epsfig{file=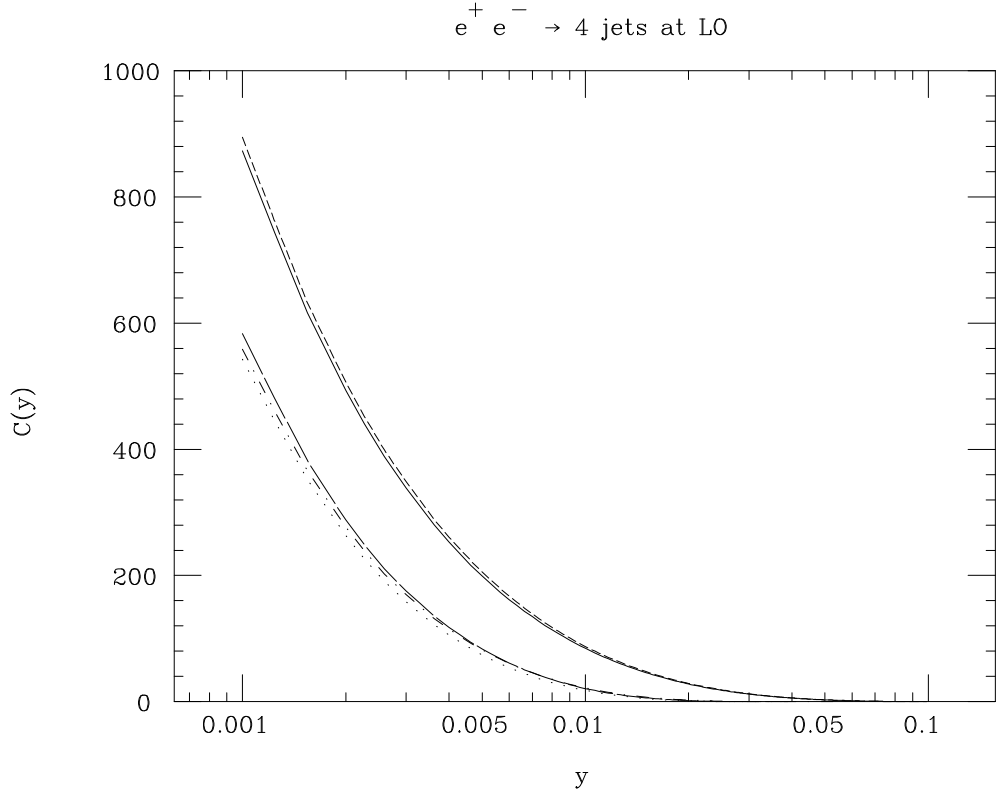,height=16cm,width=12cm,angle=90}
\caption{The parton level $C(y)$ function entering in the four-jet
fraction (\ref{f4}) at LO in the D and A schemes (solid line),
C scheme (short-dashed line),
DL scheme (dotted line), 
CL scheme (long-dashed line),
AR0 and AR1 schemes (dot-dashed line).}
\label{fig_cfunction_bis}
\end{center}
\end{figure}

For completeness, we also present the rates for the 
four-jet fraction at LO. The analytical expression reads as follows
(neglecting the $\mu$-scale dependence in $\as$)
\beq
\label{f4}
f_4(y) = \left( \frac{\as}{2\pi} \right)^2 C(y) + ... , \eeq where
$C(y)$ is the corresponding coefficient function introduced on the
same footing as $A(y)$ in the three-jet rates at LO. Its behavior is
shown in Fig.~\ref{fig_cfunction_bis}, for the D, A, C, DL, CL, AR0
and AR1 schemes. Note that D and A, on the one hand, and AR0 and AR1,
on the other hand, coincide as no clustering between unresolved parton
can take place at LO. Once again, we leave aside any comments about
the D, A and C rates, for which we refer the reader to
Ref.~\cite{camjet}. As for the DL and CL algorithms, notice the
increase of the LO rates due to the soft freezing mechanism, like
between the D and C schemes. In this case the increase is larger,
$7.4\%$ against $2.5\%$ at the minimum $y=0.001$. The absolute value
(of DL vs. D and of CL vs. C) is \sensibly\ smaller, though: by
approximately $35-38\%$ (at $y=0.001$). Such a difference can be
explained in terms of the \Luclus\ and \Durham\ measure, as was done
while commenting Fig.~\ref{fig_afunction_bis}.  The \Diclus\ curve
falls between the two \Luclus\ ones. Thus, like in the case of the
$A(y)$ function, for a fixed $y$, the $y_{ij}$ value of the former is
on average larger than the L but smaller than the D one.

\begin{table}[htb]
\begin{center}
\begin{tabular}{|c|c|c|c|c|c|c|c|}
\hline
\rule[0cm]{0cm}{0cm}
$F$ & Algorithm & $y$-range & $k_0$ & $k_1$ & $k_2$ & $k_3$ & $k_4$
                      \\ \hline\hline
\rule[0cm]{0cm}{0cm}
$A$ & DL, CL   & $0.001-0.06$ &
    $19.049$ & $-18.991$ & $5.891$ & $-0.619$ & $0.0314$ \\
$A$ & AR0, AR1 & $0.001-0.10$ &
    $11.436$ & $-12.879$ & $4.369$ & $-0.461$ & $0.0257$ \\
\hline
\rule[0cm]{0cm}{0cm}
$B$ & DL   & $0.001-0.06$ &
    $691.886$ & $-556.092$ & $117.215$ & $1.290$ & $-1.349$ \\
$B$ & CL   & $0.001-0.06$ &
    $622.657$ & $-504.153$ & $107.150$ & $1.388$ & $-1.341$  \\
$B$ & AR0  & $0.001-0.10$ &
    $55.037$ & $10.161$ & $-63.296$ & $27.529$ & $-2.757$  \\
$B$ & AR1  & $0.001-0.10$ &
    $154.687$ & $-99.096$ & $-20.233$ & $20.332$ & $-2.335$  \\
\hline
\rule[0cm]{0cm}{0cm}
$C$ & DL         & $0.001-0.03$ &
    $-172.821$ & $108.274$ & $-5.551$ & $-7.255$ & $1.152$  \\
$C$ & CL         & $0.001-0.03$ &
    $-239.022$ & $176.112$ & $-30.242$ & $-3.593$ & $0.981$  \\
$C$ & AR0, AR1   & $0.001-0.10$ &
    $-99.895$ & $50.846$ & $9.853$ & $-8.880$ & $1.214$  \\
\hline
\end{tabular}
\caption{Parametrization of the three- and four-jet QCD functions
$A(y)$, $B(y)$ and $C(y)$ as polynomials $\sum_n k_n(\ln(1/y))^n$, for the
DL, L, AR0 and AR1 algorithms. The range of validity in $y$
is given for each case.}
\label{tab_kparam_largey}
\end{center}
\end{table}

We conclude this Section by presenting a polynomial fit of
the form 
\beq
\label{fit}
F(y)=\sum_{n=0}^{4} k_n\left( \ln\frac{1}{y} \right)^n
\eeq
to the $F=A$, $B$ and $C$ functions, 
as already done in various instances in previous literature
(see, e.g., Refs.~\cite{camjet,BKSS}), now in the case 
of the DL, CL, AR0 and AR1 schemes. The lower limit
of our parameterization is $y=0.001$ for all three auxiliary functions.
We extend the fits up to values where the four schemes yield sizable
rates exploitable in  phenomenological studies (see, e.g.,
Ref.~\cite{OPALnj}). Typically, for the DL and CL algorithms these
are around $y=0.06(0.03)$ for $A$ and $B$($C$), whereas for AR0 and
AR1 a common value to the three function is $y=0.1$. 
The values of the coefficients $k_n$, with $n=0, ... 4$,
are given in Tab.~\ref{tab_kparam_largey}. Those for the D, A and C
algorithms were given in Ref.~\cite{camjet}.

As a summary of our fixed-order studies, though limited to the
NLO $\cO{\as^2}$ rates, several conclusions can be drawn.
\begin{enumerate}
\item The angular-ordering
%
%
  and soft-freezing
%
%
  procedures advocated in Ref.~\cite{camjet} represent a {\sl genuine
    improvement} in fixed-order perturbative studies in the infrared
  dominion, provided these are plugged onto a $p_{\perp}$-based
  algorithm. (In fact, one should recall their `inefficiency' when
  implemented within the \Jade\ scheme \cite{camjet}, based on a mass
  measure.)  This can be appreciated by noticing a\nosensibly\ reduced
  scale dependence of the NLO three-jet rates in both the C and CL
  schemes, as compared to the D and DL ones, respectively.
\item Of the two $p_{\perp}$-based binary measures considered here,
  the \Luclus\ one yields better performances than the \Durham\ one in
  terms of the stability of the NLO results against variation of the
  subtraction scale $\mu$. This is presumably related to its
  definition in terms of the energies of the partons involved, which
  does not contain discontinuities or edges at the border of the phase
  space selected by the resolution parameter, contrary to the case of
  the \Durham\ $p_{\perp}$-measure. Thus, an algorithm based on both
  angular-ordering and soft-freezing and exploiting the \Luclus\ 
  measure represents an alternative option to the \Cambridge\ scheme
  to be adopted in the kind of studies performed here.
\item However, as for multi-parton studies in higher order pQCD, the
  exploitation of the original \Luclus\ scheme should be limited to
  the adoption of its measure, not to the implementation of the
  preclustering and rearrangement steps recommended in the original
  version. On the one hand, these would introduce a considerable
  complication in both the numerical and (especially) analytical
  calculations.  On the other hand, they would spoil the well known
  factorization properties of $p_{\perp}$-based algorithms, in
  resumming to all order in perturbation theory terms of the form $\ln
  y$ at small values of the parameter $y$.  Indeed, one should notice
  that such properties are applicable to the case of the described DL
  and CL schemes, as the \Luclus\ measure reduces to the \Durham\ one
  in the soft limit. We will address this point specifically in the
  next Section. In addition, we will also show that the sole adoption
  of the \Luclus\ measure (i.e., the DL scheme) is not enough to
  reduce \sensibly\ the size of the hadronization corrections of the D
  scheme (see Sect.~\ref{subsec_jetrates}), so that CL is an option to
  be generally preferred to the DL one, as was the case for the C
  algorithm as compared to the D one \cite{camjet}.
\item Finally, the two \Diclus\ schemes based on the clustering of
  three-particles into two, do not show any substantial improvement,
  as compared to the conventional ones, which join two particles into
  one, at least in fixed-order perturbative studies of three-jet rates
  at NLO.
\end{enumerate}

Before closing, we should mention that further analyses on the same
footing as those described in this Section are under way, for the case
of the four-jet rate at NLO: from QCD, using the program \debrecen\ 
\cite{DEBRECEN}\footnote{In fact, the long-awaited $\cO{\as^3}$
  corrections to the four-jet rate have recently become available
  \cite{a3} and have been implemented in the mentioned code.}, and
from $W^+W^-$ decays, exploiting the code used in Ref.~\cite{Ezio}. An
account of the results in these contexts will be given in a future
publication \cite{preparation}.

\subsection{Resummed perturbative results}
\label{subsec_resummed}

In this Section we introduce a quantity which makes use of the results
present in the previous Section and which is very useful in
investigating the interplay between perturbative and non-perturbative
effects \cite{camjet}. This is the mean number of jets, defined as
\begin{equation}\label{n_jets}
n_{\mbox{\tiny jets}}\equiv\cN(y)=\sum_{n=1}^N n F_n(y),
\end{equation}
where $F_n(y)$ is nothing else than the $n$-jet fraction introduced in
eq.~(\ref{fn}) in theoretical terms (thus, $F_n(y)={f}_n(y)$ and $N=4$
in Sect.~\ref{subsec_fixedorder}), i.e., as a ratio of cross sections.
{}From the experimental point of view ($F_n(y)=\tilde{f}_n(y)$ and 
$N\ar\infty$), the corresponding quantity is defined as a ratio of 
numbers of events, i.e., 
\beq
\label{tfn}
\tilde{f}_n(y)=\frac{N_n(y)}
                    {\sum_m N_m(y)}
            =\frac{N_n(y)}
                  {N_{\mbox{\tiny tot}}(y)},
\eeq
where $N_n(y)$ is the amount of $n$-jet events and $N_{\mbox{\tiny tot}}(y)$
the total hadronic sample. (This is the definition that 
we will adopt in computing
the $n_{\mbox{\tiny jets}}$ rates using the MC programs, 
in Sect.~\ref{subsec_jetrates}.)

The mean number of jets can be
studied as a function of the jet resolution parameter $y$,
down to arbitrarily low values, at fixed energy. Furthermore, its 
perturbative
behavior at very low values of $y$ can be computed
including resummation of leading and next-to-leading
logarithmic terms to all orders in perturbation theory \cite{CDFW1}.
The corresponding predictions (particularly accurate at small $y$'s) 
can then be matched with the fixed-order
results (especially reliable at large $y$'s) of the previous Section 
over an appropriate interval in $y$, to give reliable pQCD estimates
throughout the whole range of $y$. Furthermore, these results are quite
stable against variation of the scale $\mu$ while being particularly
sensitive to $\Lambda^{(5)}_{\overline{\mbox{\tiny{MS}}}}$,
 making the jet multiplicity
$n_{\mbox{\tiny jets}}$ a particularly good quantity for the determination
of $\as$.  Non-perturbative
contributions to $n_{\mbox{\tiny jets}}$ can then be estimated
by comparing the perturbative results with those of 
MC event generators. This will be done in 
Sect.~\ref{subsec_jetrates}.

We first compute the resummed predictions for the DL and CL
algorithms.  In doing so, we make use of the results and formulae
presented in Ref.~\cite{CDFW1} for the case of the D scheme. Those
were obtained in the case of multiple soft-gluon emission at small
values of the resolution parameter. As we have already stressed in the
Introduction, since in the soft limit in which either of $i$ or $j$ is
much softer than the other the two measures (\ref{yD}) and (\ref{yL})
coincide, all the soft-gluon exponentiation properties of the \Durham\ 
algorithm carry over to the \Luclus\ one, provided no unnatural
partition of the phase space is introduced by preclustering and/or
reassignment. We do not perform the same fit for the \Diclus\ algorithms
for two reasons. First, neither the AR0 or AR1 schemes have proven
themselves being particularly suitable in pQCD studies, because of the
larger scale dependence of their NLO rates. Second, the calculation of
the resummed predictions would presumably be more complicated than in
the case of the D and L schemes and would deserve an all new paper on
its own.

Recalling that through second-order in $\as$ the two-jet fraction reads as 
\beq
\label{f2}
f_2(y) = 1 - \left( \frac{\as}{2\pi} \right)    A(y)
           + \left( \frac{\as}{2\pi} \right)^2 (2A(y)-B(y)-C(y)) + ...~,
\eeq
and using the expressions (\ref{f3}) and (\ref{f4}) of 
Sect.~\ref{subsec_fixedorder},
one easily finds that in ${\cal O}(\as^2)$ pQCD the mean number of jets
is 
\beq
\label{Nypert}
\cN(y) = 2 + \left( \frac{\as}{2\pi} \right)    A(y)
           + \left( \frac{\as}{2\pi} \right)^2
             \left( B(y)+2C(y)-2A(y) \right) + ...
\eeq

The behavior of the first-order coefficient $A(y)$ at small $y$ is
given by
\beq\label{Fx}
A(y)= C_F\left(\ln^2 y + 3\ln y + r(y)\right),
\eeq
with the non-logarithmic contribution  being \cite{durham,partition}
\beq\label{ry}
r(y) = 6\ln 2 + \frac{5}{2} - \frac{\pi^2}{6}
+ 4\left(\ln(1+\sqrt 2)-2\sqrt 2\right)\sqrt y - 3.7 y\ln y +\cO y,
\eeq
whereas for the second-order coefficient one has
\beq\label{Fy}
F(y)\equiv B(y)+2C(y)-2A(y)= C_F \left[\frac{1}{12}C_A\ln^4 y
- \frac{1}{9}(C_A-N_f)\ln^3 y + \cO{\ln^2 y} \right].
\eeq
In eqs.~(\ref{Fx}) and (\ref{Fy}) the two quantities
$C_F=4/3$ and $C_A=3$ are the Casimir operators of the fundamental and adjoint
representations of the QCD gauge group $SU(N_C)$, these quantifying 
the strength of the $q\ar qg$ and $g\ar gg$ splittings, respectively, 
with $N_C=3$ the number of colors, whereas the number of flavors is $N_f=5$.

Note that, as long as terms of order $\ln^2 y$ are neglected, the above 
expressions are identical for the two versions of the \Luclus\ based
algorithms DL and CL (and so are they for the D, A or C schemes). 
In order to introduce the algorithm-dependent ${\cal O}(\ln^2y)$ coefficients
we adopt the same procedure as in Ref.~\cite{camjet}. That is, we perform
a fit of the form (\ref{fit}), restricted to the interval, say, 
$0.001<y<0.01$, with the coefficients $k_3$ and $k_4$ fixed at the values 
prescribed by Eq.~(\ref{Fy}).  The quantities $k_0$, $k_1$, $k_2$ 
are instead treated as free parameters. The numerical 
results are given in Tab.~\ref{tab_kparam_smally}. Therefore, one can simply
use the fits in Tab.~\ref{tab_kparam_smally} for, say,
$y<0.005$ and those in Tab.~\ref{tab_kparam_largey} for $y>0.005$. (Indeed,
over the region $0.001<y<0.01$, the transition between the two 
parameterization is very smooth.) This way, the second-order
coefficient $F(y)$ can be obtained over the whole ranges of $y<0.03$
(i.e., the DL and CL limits given in the second column of 
Tab.~\ref{tab_kparam_largey} for $C(y)$).

\begin{table}[htb]
\begin{center}
\begin{tabular}{|c|c|c|c|c|c|}
\hline
\rule[0cm]{0cm}{0cm}
Algorithm & $k_0$ & $k_1$ & $k_2$ & $k_3$ & $k_4$ 
                      \\ \hline\hline
\rule[0cm]{0cm}{0cm}
DL  & $-76.264$ & $4.257$ & $4.108$ & $-0.296$ & $0.333$ \\ 
CL  & $14.275$  & $-27.217$ & $5.522$ & $-0.296$ & $0.333$ \\ 
\hline
\end{tabular}
\caption{Parametrization of the second-order coefficient $F(y)$ in the
average number of jets as a polynomial $\sum_n k_n(\ln(1/y))^n$,
for the DL and CL algorithms. The range of validity is $y<0.03$.}
\label{tab_kparam_smally}
\end{center}
\end{table}

To obtain the final perturbative predictions for the mean number of
jets, we now proceed as in Ref.~\cite{CDFW1}. To next-to-leading
logarithmic (NLL) accuracy, the resummed results are independent of
the version of the algorithm. Therefore the only differences 
between DL and CL come from the matching to
the fixed-order results. We simply subtract the
first- and second-order terms of the NLL resummed result and
substitute the corresponding exact terms. Denoting by $\cN_q$ the
NLL multiplicity in a quark jet, given in \cite{CDFW1}, we obtain
\beq\label{nfin}
\cN(y) = 2\cN_q(y) + C_F \left(\frac{\as}{2\pi} \right) r(y)
           + \left( \frac{\as}{2\pi} \right)^2
             \left( F(y)-2F_q(y) \right)
\eeq
where $F_q$ is the second-order coefficient in $\cN_q$,
given in \cite{CDW2}: 
\begin{eqnarray}
\label{nadd}
F_q(y) & = & C_F \left\{\frac{1}{24}C_A\ln^4 y
- \frac{1}{18}(C_A-N_f)\ln^3 y \right.   \nonumber \\
 & & \left. + \frac{N_f}{9}\left(1-\frac{C_F}{C_A}\right)
\left[\left(4\frac{C_F}{C_A}-1\right)\frac{N_f}{C_A}
-1\right]\ln^2y \right\}.
\end{eqnarray}

We will make practical 
use of the formulae (\ref{nfin})--(\ref{nadd}) later on, in 
Sect.~\ref{subsec_jetrates}.

\section{Non-perturbative comparisons}
\label{sec_nonpertcomparison}

In this Section we will attempt to quantify the effects due to
hadronization for the various algorithms we have been discussing so
far. In doing so we will resort to three among the most widely used
QCD-based MC event generators: \herwig\ \cite{herwig} (version 5.9
\cite{herwig59}), \jetset\ \cite{jetset} (\pythia\ version 6.1
\cite{pythia}, which incorporates \jetset\ 7.4) and \ariadne\ (version
4.10 \cite{ariadne}).  In order to avoid `philosophical' arguments
about what {\sl hadronization} actually is, we give here an {\sl
operational} definition useful for our purposes: {\sl hadronization
corrections} are the `empirical adjustments' applied to the
theoretical perturbative predictions before comparing them with the
experiment. In an event generator, the former is represented by the
partonic state before the hadronization routines are called and the
latter by the state of final particles after hadronization and decays.

However, one must note a difference between the two. In the end there is 
an unambiguous identification between the hadron level of a generator
and experiment, since the former is eventually tuned to reproduce data.
The partonic state, on the other hand, is not a physically well-defined
quantity. We therefore have to cope with the arbitrariness inherent in
generators, which generally implement only enhanced terms of the infrared 
(i.e., soft and collinear) dynamics of quarks and gluons, thus introducing 
unnatural cut-offs and kinematic boundaries into the original QCD evolution. 
As a consequence, the mutilations done to the exact QCD dynamics in its
PS approximation could well give rise to non-perturbative contributions
already at the parton level \cite{camjet}. 

In many studies, one wants to take one step further, and extract an
$\as$ value from the deduced partonic level, based on the same
parton-shower generator. This is more dangerous. For example, it has
been shown in Ref.~\cite{camjet} that \herwig\ at the parton level
overestimates the number of jets, as compared to the pQCD result, if
the same $\as$ is used. Therefore a larger $\as$ has to be used for
the resummed pQCD results than in the \herwig\ shower to obtain
the same parton level. Clearly, this knowledge is important in order
to extract an $\as$ value, but it is irrelevant in a study of
hadronization corrections. (More details on this point will be given
later on, in Sect.~\ref{subsec_jetrates}.)  The only thing we would
like to recall here is that progress in this direction is being made:
e.g., that exact ${\cal O}(\alpha_s^2)$ LO matrix elements have been
dressed with string fragmentation in the \jetset\ modeling since a
long time, that further studies in the same environment involving a
matching of the mentioned MEs to the parton cascade have also been
carried out \cite{andre} and, finally, that an `${\cal O}(\alpha_s^2)$
+ PS + {cluster hadronization}' version of \herwig\ will soon be
publicly available \cite{Bryan}.  
The general problem of how completely to match
the `two parton levels' remains a key issue to be addressed, but that
is evidently far beyond the scope of this paper.

Therefore, for most part of this Section we leave aside the analytical
formalism of the previous one, and only compare the partonic and the
hadronic levels of generators. (We will however come back to it in one
instance, at the very end of Subsect.~\ref{subsec_jetrates}.)  One is
indeed comforted in doing so by what we have already mentioned and
will illustrate below: that we do find agreement among different MC
programs. It is rather straightforward to study hadronization
corrections in this spirit, since the generators provide lists of all
partons and hadrons, event by event. In each case, jets are
reconstructed both from the quarks and gluons at the end of the parton
cascade and from the particles arising after hadronization and
decays\footnote{Note that no simulation of detector acceptance and
  resolution are implemented in the latter case.}. We will perform our
analysis in the following Subsections. Each of these corresponds to a
different phenomenological context. In the first, we study the
three-jet resolution in a simple tube model.  In \ref{subsec_jetrates}
we study hadronic events at LEP1, focusing
our attention to the case of typical multi-jet quantities, in particular
the mean number of jets defined in eq.~(\ref{n_jets}).  In the
following Subsection the emphasis will be on some kinematic properties
of two- and three-jet events at LEP1 energies.  Finally, in
Subsect.~\ref{subsec_WW}, we will study hadronization effects in the
context of the mass reconstruction of $W^\pm$-bosons in four-jet
events at LEP2.

Before proceeding further, for completeness, we briefly recall the
properties of the three mentioned MC event generators.  \herwig\
implements the parton shower by coherent branching of the partons
involved down to a fixed transverse momentum scale $Q_0\sim 1$ GeV,
and then converts these partons into hadrons using a cluster
hadronization model \cite{clu}.  In particular, the branching
algorithm includes angular-ordering and azimuthal correlations of the
emission (due to QCD coherence) along with the retention of gluon
polarization and $p_\perp$ is the $\as$ scale. 
The \jetset\ shower algorithm orders emissions in
decreasing mass, with angular ordering imposed as an additional
constraint, down to a cut-off mass $Q_0\sim 1$ GeV. Azimuthal
anisotropies from coherence and gluon polarization are also included,
and the $\as$ scale is $p_{\perp}^2$.  Hadronization is done according
to the Lund string model \cite{AGIS}.  In \ariadne\ \cite{ariadne},
the ordering of emissions is in the invariant transverse momentum
defined above in eq.\ (\ref{LL:invpt}) for the \Diclus\ algorithm. The
same \tpt\ is used as the scale in $\as$ and the cascade is continued
down to $p_{\perp 0}\approx0.6$ GeV. The coherence, treated by
angular ordering in the \herwig\ and \jetset\ parton showers, is
inherent in the way gluons are emitted as dipole radiation from color
connected pairs of partons. The azimuthal anisotropy due to gluon
polarization is not explicitly taken into account but is reproduced to
some extent by the dipole structure of the emissions.  Hadronization
is handled by the \jetset\ string fragmentation\footnote{An up-to-date review 
of our current understanding of hadronization is found in Ref.~\cite{book}.}.

Above we have seen that the algorithms in part are based on different
considerations. One example is the picture of the perturbative shower
evolution, which can be organized either in terms of decreasing opening
angles of emissions or in terms of decreasing transverse momenta of 
emissions. Either of these two pictures gives a perfectly legitimate
description of nature, but they arrive at different answers for what is 
the `right' way to cluster a set of $m$ partons into $n$ jets.
Even within a given calculational scheme, further uncertainties exist,
such as how to distribute the recoil of an emission, i.e., the details
of how energy and momentum is conserved. Add to this differences 
in the way non-perturbative physics is viewed, e.g., in string vs.
cluster fragmentation models, and it is clear that there is not one
unique view of the world. 
Therefore there is also not a unique criterion for what is the best
possible clustering algorithm. One may then expect to find that different
algorithms have complementary strengths and realize
 that the choice of algorithm should be based on the intended application. 

\subsection{Tube model results}
\label{subsec_tubemodel}

\begin{figure}[htb]
\begin{center}
~\epsfig{file=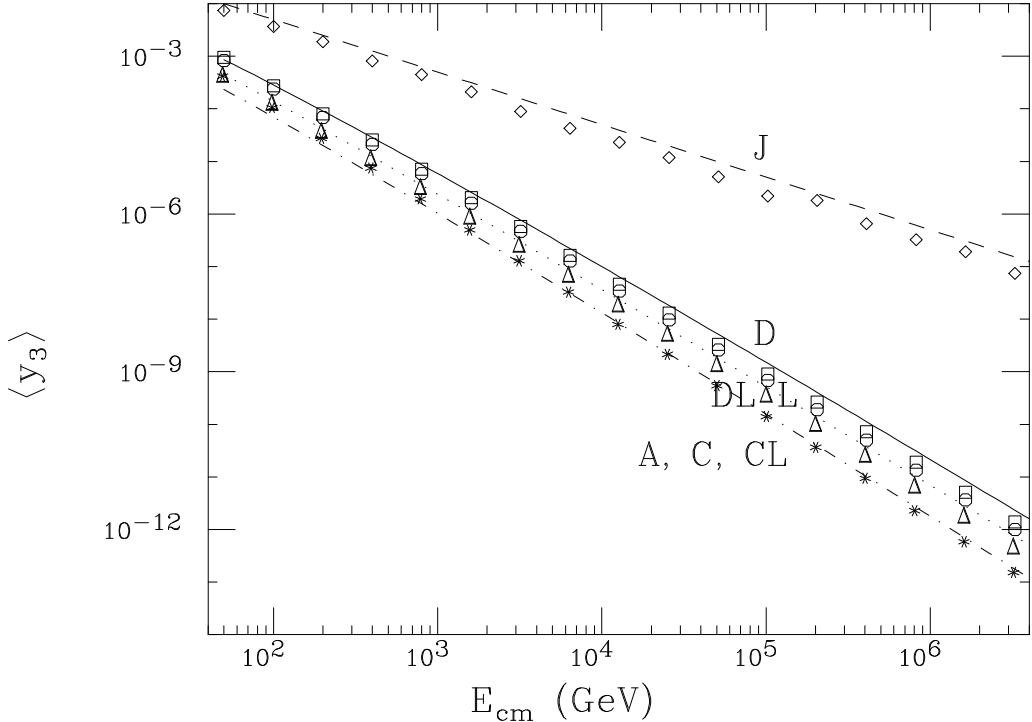,height=16cm,angle=90}
\caption{Mean value of the three-jet resolution $y_3$ in the tube
  hadronization model, for the J scheme (open-diamond symbols), D
  scheme (open-square symbols), AR1 scheme (circle symbols,
  not labelled for reason of space), DL and L schemes
  (open-triangle symbols), A, C and CL schemes (asterisk symbols).
  Note that the DL and L data points visually coincide, so do the A, C
  and CL ones.  The corresponding curves show the approximate formulae
  discussed in the text.}
\label{fig_y3_bis}
\end{center}
\end{figure}

\begin{figure}[htb]
\begin{center}
~\epsfig{file=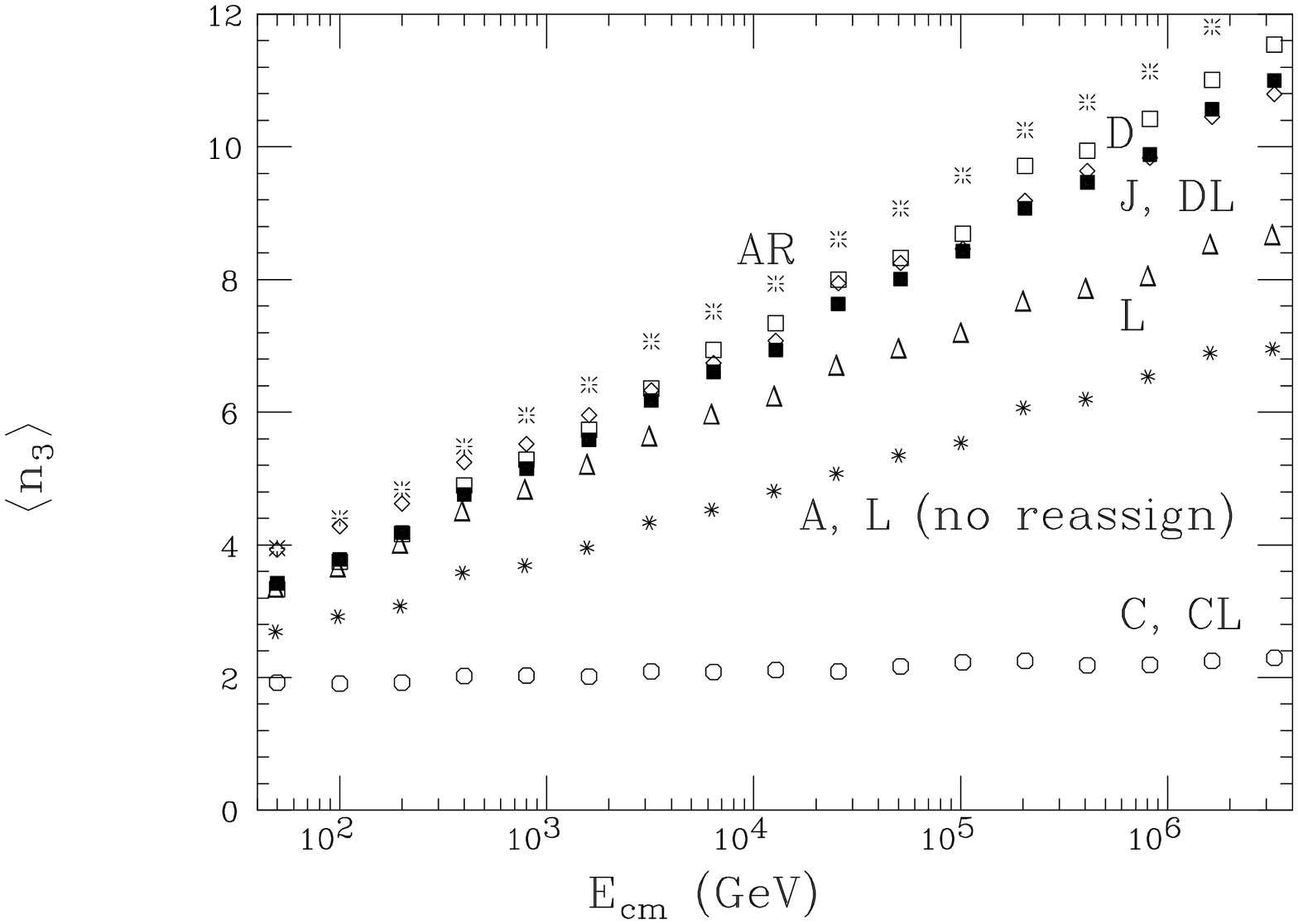,height=16cm,angle=90}
\caption{Mean number of particles in the third jet at $\ycut=y_3$
in the tube hadronization model, 
for the J scheme (open-diamond symbols), D scheme
(open-square symbols), DL scheme (full-square symbols), 
L scheme (open-triangle symbols),
A and L (no reassignment) schemes (asterisks symbols),
C and CL schemes (open-circle symbols),
AR scheme (star symbols). 
Note that the A and L (no reassignment) data points visually coincide, 
so do the C and CL ones.}
\label{fig_n3_bis}
\end{center}
\end{figure}

As a preliminary exercise, we consider the simplest possible hadronization
mechanism \cite{tube}, the so called `longitudinal phase space' or `tube' 
model, in fact a simplified version of the Lund string model.
Here a color-connected   pair of partons produces a jet of
light hadrons over a cylindrical $(y,p_{\perp})$ phase space, where $y$ is
the rapidity and $p_{\perp}$ the transverse momentum 
(note that $y=\frac{1}{2}\ln[(E+p_z)/(E-p_z)]$, with $z$ the cylinder axis
and $(E,p_x,p_y,p_z)$ the four-momentum). In practice, a number $N$ of 
massless four-momenta are generated at random with an exponential 
transverse momentum distribution and a uniform rapidity distribution 
in the interval $-Y<y<Y$. The maximum rapidity $Y$ is given by 
$E_{\mathrm{cm}}=Q=2\lambda\sinh Y$ and the multiplicity $N$ by 
$\lambda = N\VEV{p_{\perp}}/2Y$ (see Ref.~\cite{camjet} for details).
As illustrative values, we have taken $\lambda=0.5$ GeV and 
$\VEV{p_{\perp}}=0.3$ GeV. In Ref.~\cite{camjet}, use was made of this model
in order to illustrate some shortcomings of the \Durham\ algorithm: that
is, junk-jet formation and misclustering.
We resume here those studies for the case of the algorithms that were
not treated there.

As explained in Sect.~\ref{subsec_Angular}, 
by studying the mean value of the scale $\ycut=y_3$ (denoted by $\VEV{y_3}$)
at which a third junk-jet is resolved
in the tube model one can get an insight on the effectiveness of the
modifications to the \Durham\ scheme proposed in Ref.~\cite{camjet}.
In the sense that, the smaller $\VEV{y_3}$ is, the more contained is
the junk-jet phenomenon.
As a matter of fact, both in the A and C algorithms \cite{camjet}
the average value of $y_3$ is much smaller,
as compared to the D scheme, over an enormous range of 
energies $Q\equiv\Ecm$. Fig.~\ref{fig_y3_bis} illustrates this. 

We supplement the results for D, A and C of Ref.~\cite{camjet} by
adding in Fig.~\ref{fig_y3_bis} the corresponding behaviors for the L,
DL and CL binary schemes, and for the AR1 scheme.  Note that here L
refers to the original \Luclus\ scheme implementing both preclustering
and reassignment.

It is curious to realize once more (see Sect.~\ref{subsec_fixedorder})
the beneficial effects in the DL scheme of adopting the \Luclus\ 
measure instead of the \Durham\ one, as the corresponding data points
lay well below those of the original D scheme. In contrast,
preclustering and rearrangement bring no noticeable improvement in
this context, neither separately nor together: notice the overlap of
the DL and L curves.  (For readability of the figures, we have avoided
plotting the cases of the \Luclus\ scheme with only one of
preclustering and reassignment.)  This probably indicates that the
junk-jet formation is dominated by the existence of a single
high-$p_{\perp}$ track or a few very nearby tracks.  However, it is
clear that the further step of angular-ordering is needed even in the
case of the \Luclus\ measure in order to bring the results further
down. Once this is implemented, there is no sizable difference between
the performances of the two different measures (C vs. CL).  In a
sense, the sole adoption of the \Luclus\ $y_{ij}$ helps to alleviate
the original problem, but is not enough to cure it in the same way the
angular-ordering does. In addition, it is evident that the latter
procedure removes any distinction between the two measures.  For
reference, we should also mention that the \AOLuclus\ (which we do not
study further here, see Sect.~\ref{subsec_Angular}) can boast
identical performances to those of the A, C and CL schemes.  The
\Diclus\ algorithm nicely interpolates between the D and L ones,
following the relative behaviour of the measures, similarly to the
case of Fig.~\ref{fig_afunction_bis}.

Thanks to simple kinematic relations valid within the tube model and
depending on the clustering procedures of the various algorithms, it
was shown in Ref.~\cite{camjet} that one can approximate the behaviors
of the binary algorithms
in Fig.~\ref{fig_y3_bis} by means of some analytical formulae, as a
function of the collider CM energy $Q$. We recall here those for the
J, D, A and C schemes:
\beq\label{yJDtube}
\VEV{y_3}^J\sim\frac \lambda Q,\qquad\qquad\qquad 
\VEV{y_3}^D\sim\left(\frac{2\lambda\ln(Q/\lambda)}{\pi Q}\right)^2,
\eeq
\beq\label{yACtube}
\VEV{y_3}^A\approx\VEV{y_3}^C\approx\VEV{y_3}^{CL}\sim
\left(\frac{\lambda}{Q}\ln\ln(Q/\lambda)\right)^2.
\eeq
As evident from the plot, CL coincides with A and C.
It is also easy to show that the corresponding equation for both the DL and
L schemes reads as follows:
\beq\label{yLDLtube}
\VEV{y_3}^L\approx\VEV{y_3}^{DL}\sim
\left(\frac{\lambda\ln(8Q/\lambda)}{\pi Q}\right)^2.
\eeq
The curves in Fig.~\ref{fig_y3_bis} correspond to the above formulae:
the dashed line to the J scheme, the continuous one to the D algorithm,
the dotted one refers to L and DL whereas the dot-dashed one to the
A, C and CL cases.

In order to test the efficiency of the soft-freezing procedure proper
of the C scheme a good quantity to study is the mean number of
particles contained in the the third (softest) jet when $\ycut=y_3$,
which was denoted by $\VEV{n_3}$ in Ref.~\cite{camjet}. Clearly, the
smaller this quantity is on average, the less particles have been
attracted inside the original resolved (soft) cluster (see the
discussion in Sect.~\ref{subsec_Cambridge}), and the misclustering
effect is thus reduced. Fig.~\ref{fig_n3_bis} shows $\VEV{n_3}$ as a
function of the CM energy, over the same $\Ecm$ range as in the
previous plot. The relative behaviors of the J, D, A and C schemes
were already illustrated in detail in Ref.~\cite{camjet}, so we do not
dwell here on them. Rather we emphasize that the adoption of the
\Luclus\ measure is less helpful in this case, as the DL and D curves
almost coincide. Furthermore, it is worth spotting that the
reassignment procedure increases the mean value of $n_3$, as can be
appreciated by comparing the data points labeled `L' and `L (no
reassignment)'.  It is therefore clear that such a procedure, which
does remedy the problem of the misassignment of soft particles (see
Sect.~\ref{subsec_LUCLUS}), is instead inefficient in suppressing
misclustering.  This is evident if one considers that such a step
tends to `balance' the event, by reassigning tracks among clusters so
that the jets show in the end similar multiplicity (thus acting in
contrast to what soft-freezing does).

Curiously, the preclustering seems to work on the same footing as the
angular-ordering: compare `A' and `L (no reassignment)'. However, here
the almost exact agreement is somewhat of a coincidence.  As intimated
in the Introduction, this is a consequence of having adopted the
default $d_{\mathrm{init}}$ value in producing the `L (no
reassignment)' curve. We have verified that in the limit
$d_{\mathrm{init}} \to 0$ the DL results are in fact recovered.  This
makes clear the possible danger of ascribing artifacts of the
algorithm to real physics effects, if a wrong setting of
$d_{\mathrm{init}}$ is adopted.  Finally, like in the case of
$\VEV{y_3}$, the adoption of the soft-freezing procedure wipes out
differences between the \Durham\ and \Luclus\ measures (compare C and
CL). As for the \Diclus\ scheme, we note that it yields the largest
$\VEV{n_3}$, especially at high energies, this witnessing the tendency
of this algorithm of clustering soft resolved particles which are
source of dipoles (see discussion in Subsect.~\ref{subsec_DICLUS}
while commenting Fig.~\ref{fig_soft_resolved}.)

Evidently, the tube model adopted in the previous paragraph should be regarded
as a useful tool in order to test the performances of the various algorithms 
with respect to the misbehaviors responsible for junk-jet formation and
misclustering, which naturally arise in the physics dominion governed by
soft radiation, for all algorithms based on $p_{\perp}$- and $m$-measures
\cite{camjet}. In the very end, however, the benchmark ground to verify
the goodness of an algorithm in terms of hadronization corrections should
be a full MC program, such as those previously mentioned.

\subsection{Jet rates}
\label{subsec_jetrates}

\begin{figure}[!p]
\begin{center}
\epsfig{file=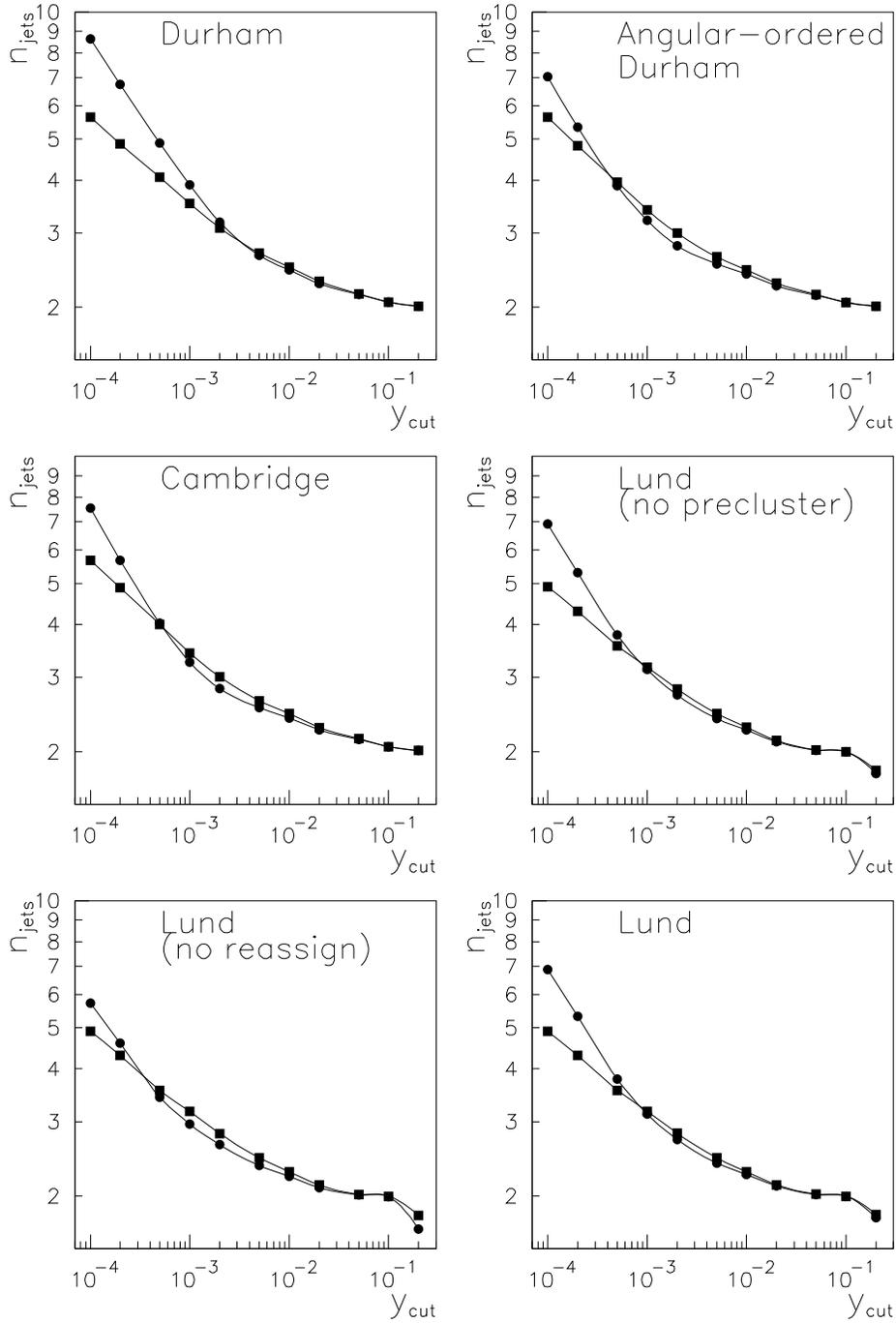, height=18cm}
\caption[Comparison of jet algorithms]{
  Parton level (squares) and hadron level (circles) results from
  \herwig\ on the mean number of jets at $Q=M_Z$ for various jet
  algorithms as labeled, as a function of the jet resolution variable
  $\ycut$. The statistical errors are smaller than the size of the
  points.  Results are very similar for \jetset\ and \ariadne.}
\label{fig_hadronisation_herwig1}
\end{center}
\end{figure}

\begin{figure}[!p]
\begin{center}
\epsfig{file=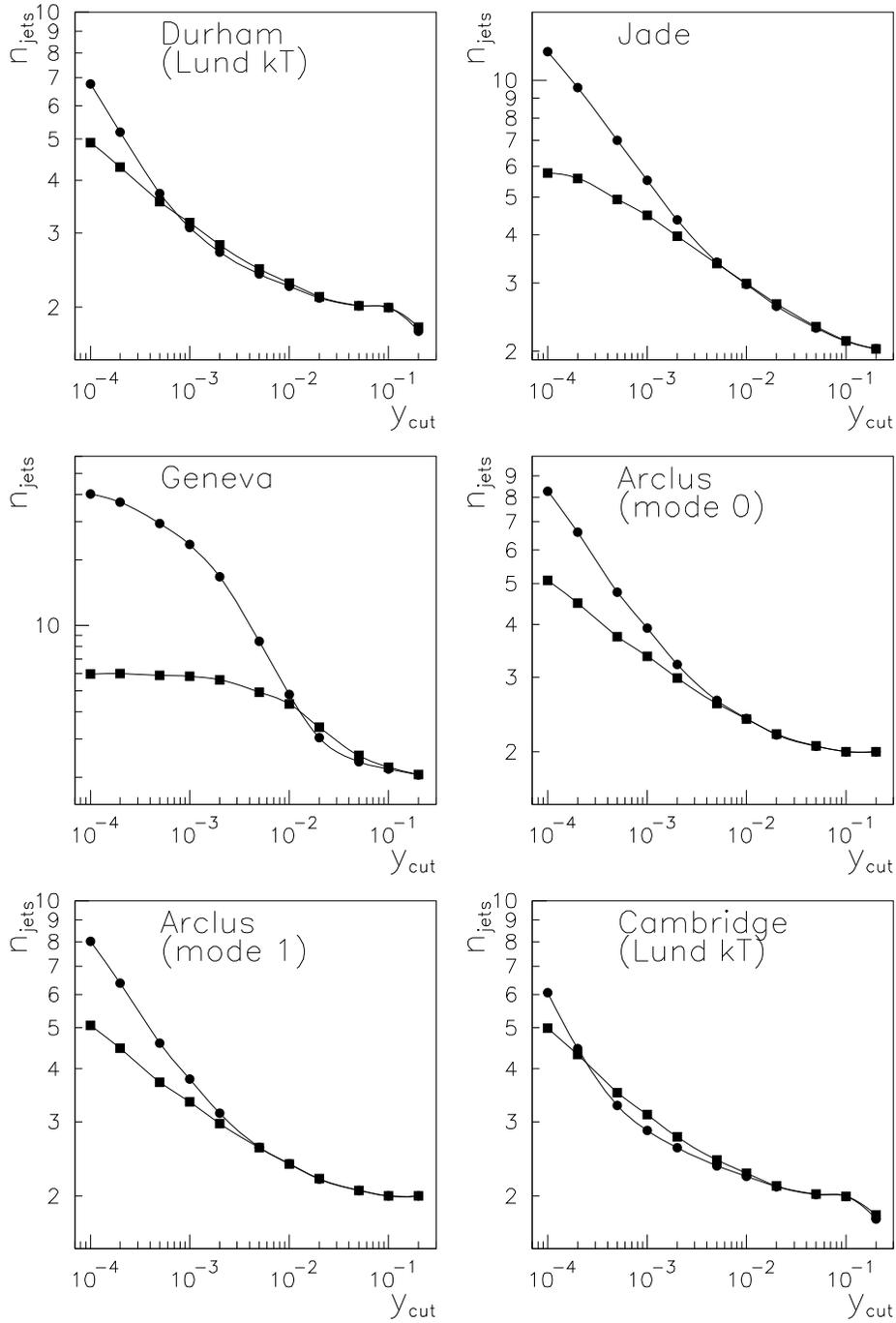, height=18cm}
\caption[Comparison of jet algorithms]{
  Parton level (squares) and hadron level (circles) results from
  \herwig\ on the mean number of jets at $Q=M_Z$ for various jet
  algorithms as labeled, as a function of the jet resolution variable
  $\ycut$. The statistical errors are smaller than the size of the
  points.  Results are very similar for \jetset\ and \ariadne.}
\label{fig_hadronisation_herwig2}
\end{center}
\end{figure}

In Ref.~\cite{camjet} the quantity $n_{\mbox{\tiny jets}}$ was studied,
both at hadron and parton level, as a function of the resolution parameter, 
down to the minimum figure of $\ycut=0.0001$. Such a choice of range was not 
a random
one. Indeed, it is well known that the energy scale at which QCD enters its
non-perturbative phase is around 1 GeV . This is also the typical value 
at which the QCD-based MC programs stop the parton cascade and
turn on the hadronization process. Clearly, if one intends to probe
the interface between perturbative and non-perturbative QCD
by studying jet properties as a function of $\ycut$
at LEP1 energies, where  $Q\approx M_Z$, one should aim for an
algorithm with small hadronization corrections for $\ycut$ values down to 
$(1~\mbox{GeV})^2/Q^2\approx 10^{-4}$. Otherwise it becomes more difficult
to understand the QCD dynamics in such a critical regime, and systematic 
uncertainties (due to different MC modelings) must be accounted for.

The fact that the \Jade\ algorithm 
(see Fig.~\ref{fig_hadronisation_herwig2}) fails to maintain the size
of the hadronization corrections small at low $\ycut$ (as compared
to, e.g., the \Durham\ scheme, see Fig.~\ref{fig_hadronisation_herwig1}) 
is a consequence of the 
adoption of a measure which is an invariant mass one, rather than
a transverse momentum. We have in fact already recalled in the Introduction
that a $p_{\perp}$-distance is better adapted to the 
conventional picture of non-perturbative jet fragmentation and therefore
naturally allows a cleaner separation of perturbative and non-perturbative 
physics \cite{durham,partition,yuri}. We also quantified this 
point in the tube model when
mentioning that the power-suppression in $Q$ on $\VEV{y_3}$ goes 
like  $(\ln Q/Q)^2$ 
in the \Durham\ scheme, whereas the corresponding behavior
in the \Jade\ one is $1/Q$: see eq.~(\ref{yJDtube}) and, for a more
theoretically sound basis, also Ref.~\cite{BPY}.

That a $p_{\perp}$-based measure is indeed a better choice is confirmed not
only by the fact that also the various \Luclus\ algorithms 
can boast a power-suppression similar to that of the \Durham\ 
scheme, see eq.~(\ref{yLDLtube}), but also by observing that
a common feature of 
Figs.~\ref{fig_hadronisation_herwig1}--\ref{fig_hadronisation_herwig2} 
is that {\sl all} the transverse momentum
based schemes (also the \Diclus\ ones) are
better behaved than the \Jade\ one at small $\ycut$'s. 

Comparisons must be done with some care, however, since the horizontal
$y$ scale means different things for many of the algorithms shown in
Figs.~\ref{fig_hadronisation_herwig1}--\ref{fig_hadronisation_herwig2}. 
For a pair of partons/hadrons $i,j$, the definitions give that
$(y_{ij})_{\Jade} > (y_{ij})_{\Durham} > (y_{ij})_{\Luclus}$,
since the three measures share the same angular dependence and differ
only in the energy factors being proportional to $E_i E_j$, 
$\min(E_i^2,E_j^2)$ and $E_i^2 E_j^2/(E_i + E_j)^2$, respectively
(disregarding the difference between $|\bfp_i|$ and $E_i$ for
\Luclus). It then follows that \Jade\ differs significantly 
from the other two when $E_i \ll E_j$, that \Durham\ and
\Luclus\ differ by up to a factor four for $E_i \approx E_j$, and that
\Jade\ and \Luclus\ always differ by more than a factor four.
Even more different is the \Geneva\ scheme, where the energy factor 
on the same scale would be $(4/9) E_i E_j E_{\mathrm{vis}}^2/(E_i + E_j)^2$.
Since normally $E_i + E_j \ll E_{\mathrm{vis}}$, it follows that
$(y_{ij})_{\Geneva} \gg (y_{ij})_{\Jade}$ most of the time.
The transition region between pQCD and non-pQCD is thus no longer around 
$10^{-4}$ in the \Geneva\ scheme, but maybe more like at $10^{-2}$,
judging by the jet rate in Fig.~\ref{fig_hadronisation_herwig2}.

\begin{figure}[htb]
\begin{center}
\centerline{}
\begin{minipage}[b]{.5\linewidth}
\centering\epsfig{file=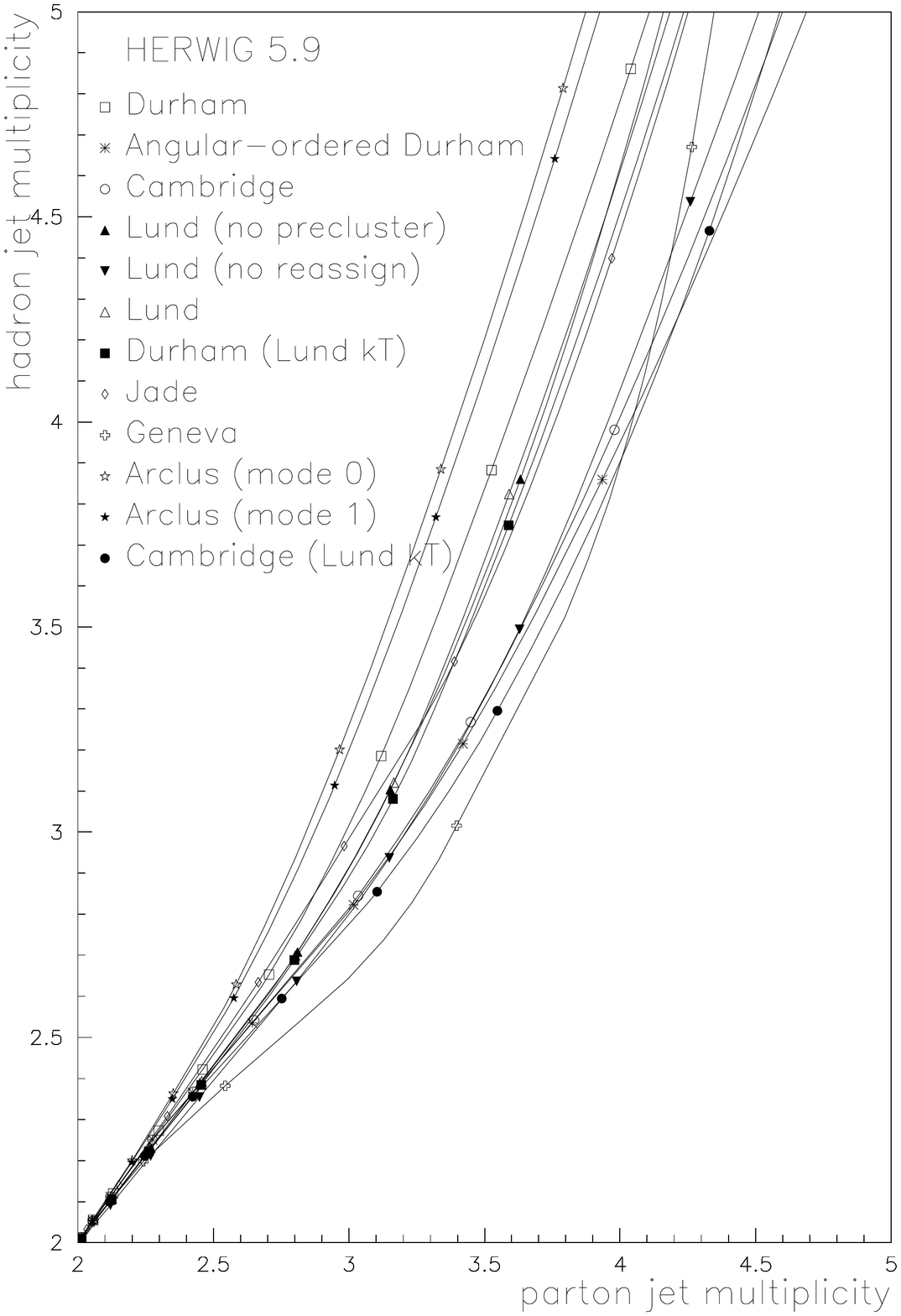,angle=0,height=12cm,width=\linewidth}
\end{minipage}\hfil
\begin{minipage}[b]{.5\linewidth}
\centering\epsfig{file=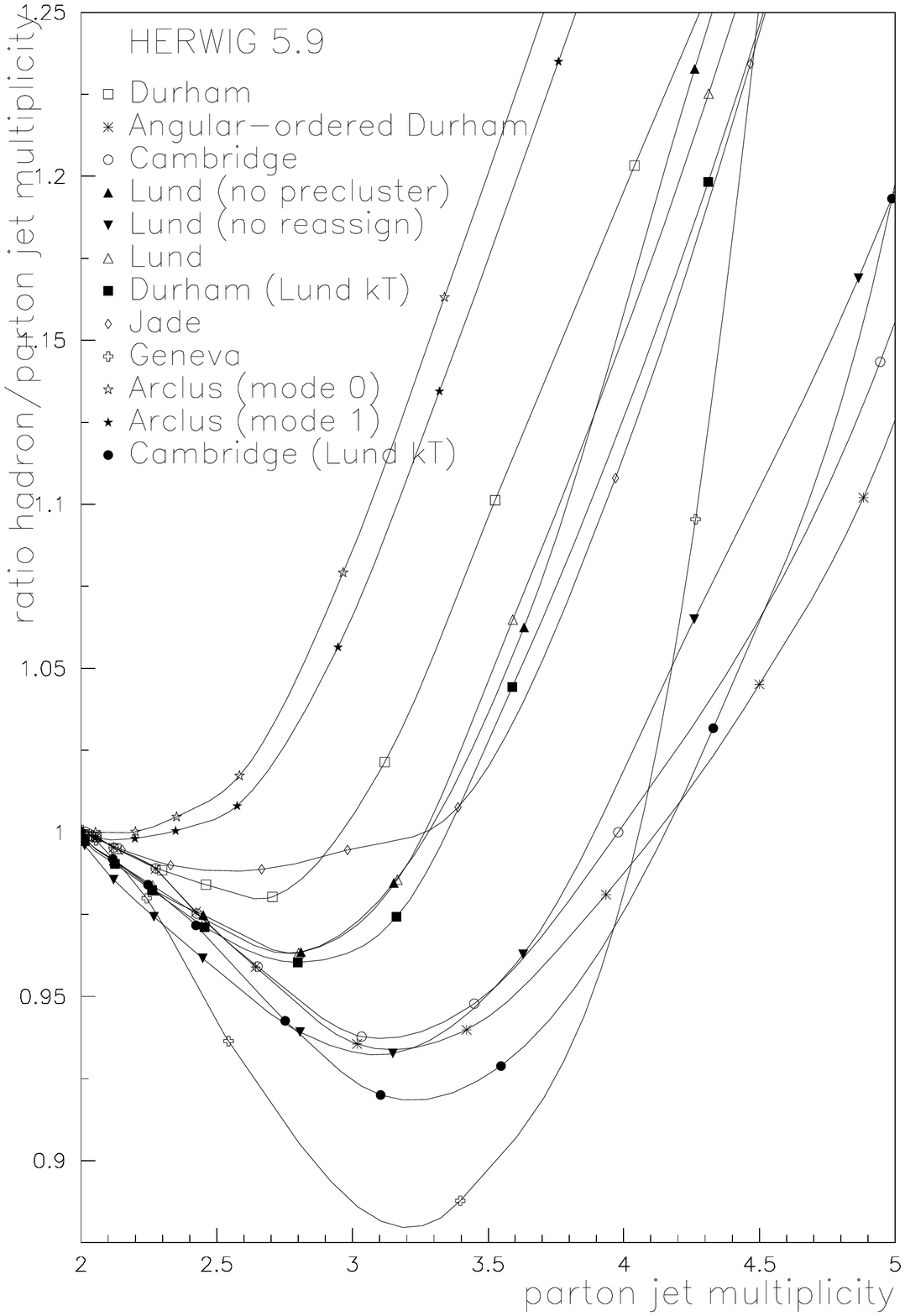,angle=0,height=12cm,width=\linewidth}
\end{minipage}\hfil
\centerline{}
\caption{The average hadron jet multiplicity vs.
 the parton one (left plot) and the ratio of the two (right plot)
at $Q=M_Z$ for various jet algorithms as labeled, using \herwig. 
The statistical errors are smaller than the size of the points.}
\label{fig_parton-hadron_herwig}
\end{center}
\end{figure}

An alternative procedure to compare jet algorithms is
offered by Fig.~\ref{fig_parton-hadron_herwig}, where the 
average hadron jet multiplicity is plotted against the average parton one. 
The $n_{\mbox{\tiny jets}}$ values in the plot are exactly the same as 
defined by the curves in 
Figs.~\ref{fig_hadronisation_herwig1}--\ref{fig_hadronisation_herwig2},
but with the explicit $\ycut$ dependence eliminated in each pair of 
average jet numbers. This is done for the same algorithms analyzed 
in the previous two figures. Both the parton and the hadron jet multiplicity 
increase with a diminishing $\ycut$. The criterion of a good algorithms as 
one with small hadronization corrections thus translates into one with
a curve close to the diagonal in the left-hand side of 
Fig.~\ref{fig_parton-hadron_herwig} or, alternatively, to the
horizontal at unity in the right-hand side of 
Fig.~\ref{fig_parton-hadron_herwig}, where 
the hadron-to-parton jet multiplicity ratio is presented.

If one intends to probe the QCD phase transition at small $\ycut$, then
the relevant region is that at large jet multiplicities. There, we 
notice that four `bunches' of curves distinctively 
separate. Four algorithms remain particularly
close to the diagonal/horizontal line.  Not surprisingly, among these
are the schemes that had been especially designed (in
Ref.~\cite{camjet}) to improve the performances of the \Durham\
scheme: such as the \AODurham\ (asterisks symbols)
and the \Cambridge\ (open-circle symbols), which are the closest
to the diagonal/horizontal line.  To this group also belong the
\Luclus\ scheme with only preclustering implemented (with
$d_{\mathrm{init}}$ kept at its default value 0.25 GeV even at low
$\ycut$, full-down-triangle symbols) and the \Cambridge\ algorithm
using the \Luclus\ measure (full-circle symbols).  This is not
surprising if one recalls Fig.~\ref{fig_n3_bis} in the tube model.
That plot, on the one hand, had already shown the residual effects of
preclustering with the default $d_{\mathrm{init}}$ and, on the other
hand, had also made the point that the \Cambridge\ scheme using
the \Luclus\ measure performs as well as the original one with the
\Durham\ distance.  In addition, always in line with what was assessed
in the tube model, it is clear that the \Luclus\ measure alone is not
enough to improve the hadronization performances of the \Durham\ 
scheme (full- and open-square symbols, respectively).  In fact, the DL
algorithm belongs to another set of curves (along with the \Jade,
\Durham\ and the two other configurations of the \Luclus)
whose hadronization corrections are much larger with increasing
multiplicity (i.e.\ decreasing $\ycut$).  A third group is constituted
by the two \Diclus\ algorithms, which perform very well in the
two-jet-dominated region but rather worse as soon as a third jet is
resolved.  The \Geneva\ algorithm performs worst, since it starts
out with the largest negative hadronization corrections and thereafter
steeply shoots up towards the largest positive ones.

\begin{figure}[htb]
\begin{center}
\centering\epsfig{file=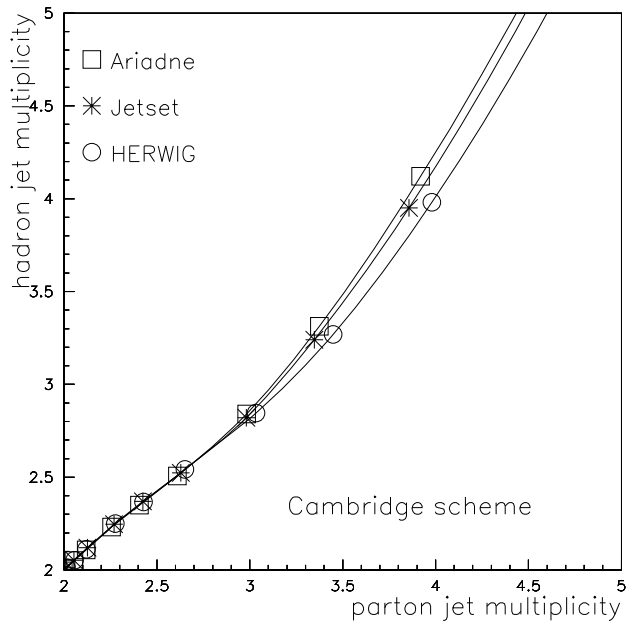,angle=0,height=12cm}
\caption{The average hadron jet multiplicity vs.
 the parton one for the \Cambridge\ scheme
at $Q=M_Z$ for the three event generators \ariadne, \jetset\ and 
\herwig, as labeled. 
The statistical errors are smaller than the size of the points.}
\label{fig_parton-hadron_cambridge}
\end{center}
\end{figure}

Fig.~\ref{fig_parton-hadron_cambridge} reproduces the rates of the
\Cambridge\ algorithm, already given in Fig.~\ref{fig_parton-hadron_herwig} 
for the case of \herwig, now extended to include \jetset\ and
\ariadne\ data points, too. Indeed, the pattern of the hadron/parton 
multiplicity is
very similar among the three programs, though with a more marked tendency of 
the rates
of departing from the diagonal  in the latter two cases. We have verified
(though not shown) that similar consistent behaviors among the
three generators also occur in the case of the other jet schemes. 

A very interesting aspect of 
Figs.~\ref{fig_hadronisation_herwig1}--\ref{fig_parton-hadron_herwig} 
is the `negative hadronization corrections' (see Ref.~\cite{BKSS}
and also Ref.~\cite{negative} for \Jade), i.e., that fewer hadron than
parton jets are reconstructed in the three-jet-dominated region.
This can be easily observed in the case of the \Geneva, 
the \AODurham, the `two' \Cambridge\ and the 
various \Luclus\ schemes. Though less visible there, it also occurs 
for the \Jade\ and \Durham\ algorithms. Not even the two modes of 
{\Diclus}\ are immune from it, though here the effects are small. 

This phenomenon has a very straightforward interpretation in terms of
the well-known string \cite{stringeff} or drag \cite{drageff} effect.
The concept can be illustrated by considering a three-parton event
$q\overline{q}g$. To leading order in $1/N_C^2$, the system separates
into two color dipoles, one $qg$ one where the color of the quark
is compensated by the anticolor of the antiquark, and another similar
$g\overline{q}$ one. Each of these dipoles can act as a source for 
further softer perturbative emission or define the topology of 
a non-perturbatively hadronizing string piece. In the (transverse) rest 
frame of such a dipole the emission of partons and hadrons is isotropic
in azimuth but, when viewed in the CM frame of the event,  
particle production is not symmetric around the three jet axes. Instead
enhanced soft particle production is found in the angular ranges spanned 
by the dipoles, i.e., in the $qg$ and $g\overline{q}$ ranges. There is no 
corresponding $q\overline{q}$ dipole --- in fact, the color-suppressed 
dipole of this kind enters with a negative sign, i.e., provides
destructive interference to the other two. Therefore the $q\overline{q}$
angular range has a much smaller soft particle production than the other 
two ranges. This effect is well established experimentally \cite{stringdata}. 

A corollary of the string/drag effect is that reconstructed jet
directions also can display a systematic bias. For instance, with
respect to the original quark parton direction, softer particles will
predominantly be produced on the side of the gluon jet, and less on
the antiquark side. Therefore a naive clustering of hadrons will find
a quark jet axis somewhat shifted towards the gluon.  The original
parton direction is not known experimentally, of course, but the
effect is visible by harder particles in the quark jet appearing
slightly more on the antiquark side (lined up with the original parton
direction) and softer particles more on the gluon side
\cite{stringasym}.  The antiquark is similarly affected, while the
gluon receives opposite contributions from the two string pieces
attached to it. Simple geometry shows that a dipole is more boosted
and the string/drag effect therefore more developed when two partons
are nearby. If the gluon is closer to the quark than to the antiquark,
say, the gluon and quark reconstructed directions will be shifted
closer to each other, while the antiquark direction is less affected.
(The reconstruction of jet directions is further tested in the next
Section.) Thus a three-jet event becomes more two-jet like, in the
sense that the $y_{3}$ value where the event flips from a two-jet to
a three-jet is lower on the hadron than on the parton level. Hence the
negative hadronization corrections. These arguments are valid both for 
string and cluster hadronization --- the clusters are aligned by the 
same colour topology as the strings and are similarly boosted ---
and both models correctly reproduce the measured string/drag effect.
In the following, to shorten our discussion, we will discuss the
dynamics of hadronization in terms of the string model, but recall
that an equivalent formulation in terms of the cluster one is always
possible.

The magnitude of the above effect depends on the details of the
algorithm used. For instance, the shift inwards of the two jet
directions can be viewed as a kinematical consequence of replacing two
massless partons with two massive jets, while still retaining the same
total invariant mass of the pair. Thus the negative hadronization
corrections should be absent in the \Jade\ E scheme, where the correct
invariant mass is used as distance measure, eq.~(\ref{Jade_E_dist}).
(But, of course, \Jade\ E has its own problems of misclustering, so is
not the solution.) The corrections are still rather small in the
normal \Jade\ scheme. The \Diclus\ algorithm is the only one
deliberately designed to reflect particle production along hyperbolae,
and to reconstruct the directions of the asymptotes of those
hyperbolae, while others implicitly are based on a picture of a jet as
a set of particles extending away from the origin along a fixed
momentum direction.  As we see, \Diclus\ does a pretty good job of
its intended task in the two-jet region. Other things being the same,
in this region \Diclus\ would thus be the preferred choice. 
The other algorithms have differently large
negative hadronization corrections from this effect, reflecting the
details of the distance measure and the clustering scheme. For
instance, using $|\bfp_i|$ rather than $E_i$ emphasizes the importance
of jet masses acquired in the hadronization; thus the DL curve lies
below the D one.

\begin{figure}[htb]
\begin{center}
\centerline{}
\begin{minipage}[b]{.5\linewidth}
\centering\epsfig{file=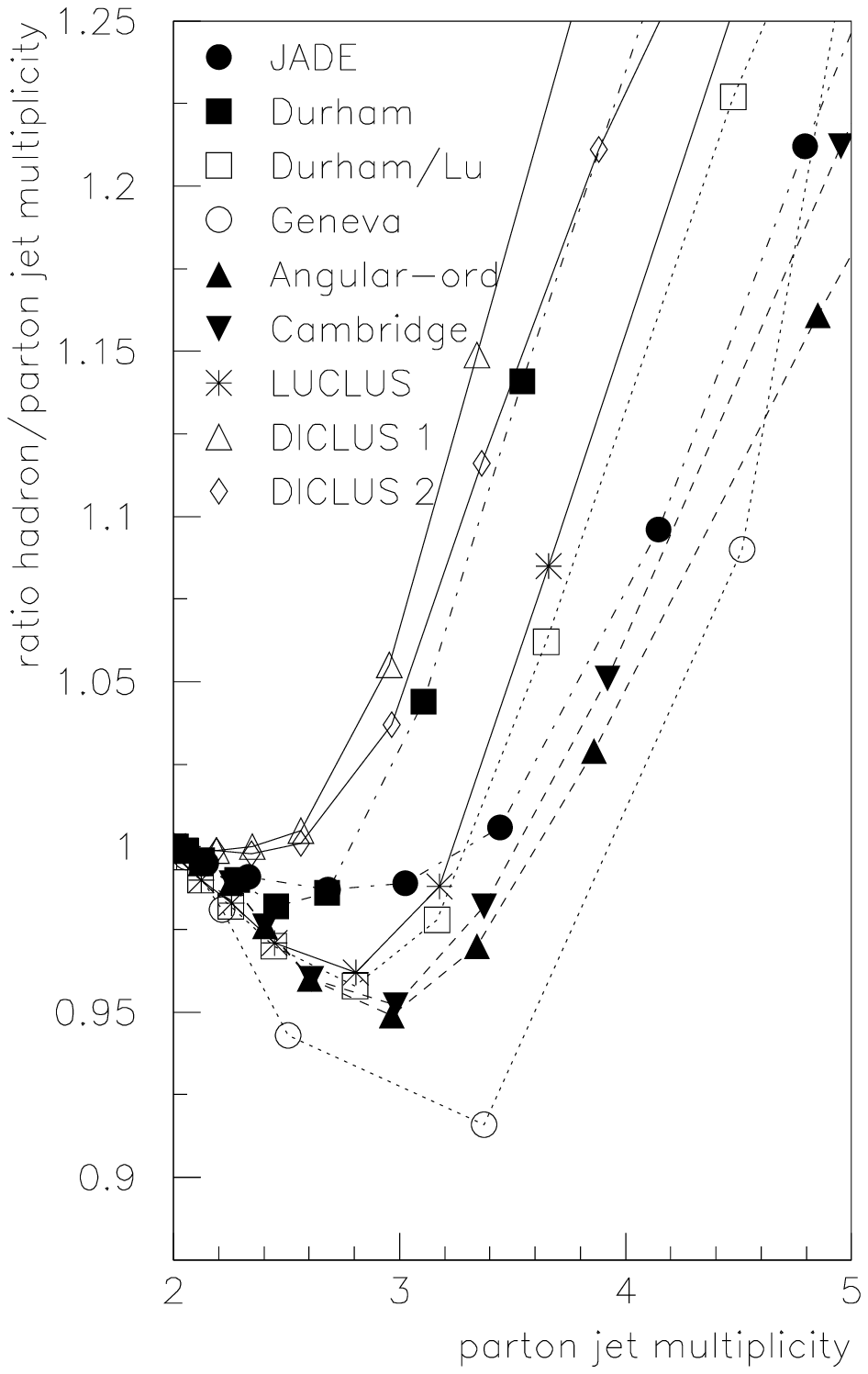,angle=0,height=12cm,width=\linewidth}
\end{minipage}\hfil
\begin{minipage}[b]{.5\linewidth}
\centering\epsfig{file=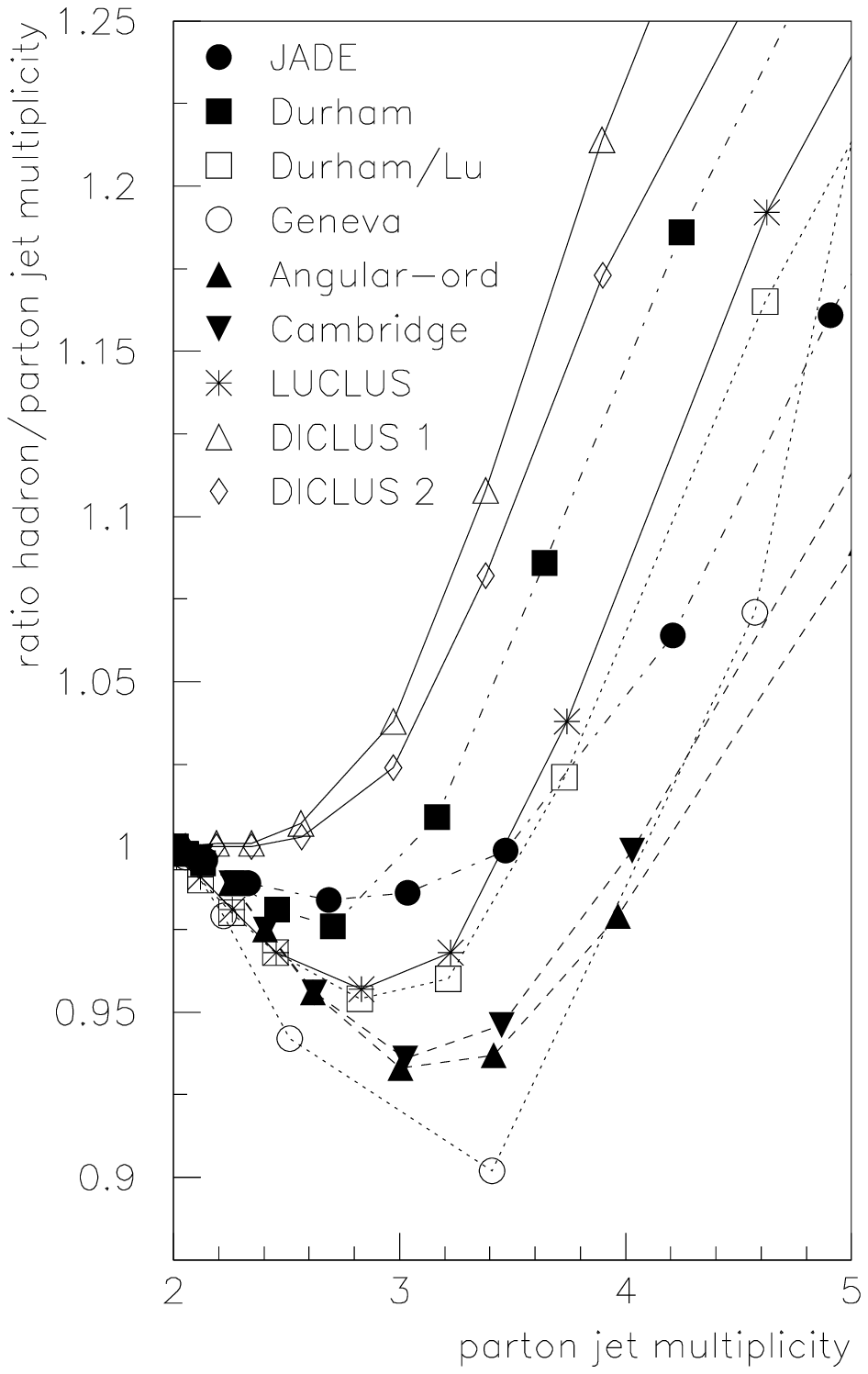,angle=0,height=12cm,width=\linewidth}
\end{minipage}\hfil
\centerline{}
\caption{The ratio of the average hadron and parton jet multiplicities
  as a function of the average parton jet multiplicity, at $Q=M_Z$ for
  various jet algorithms as labeled, using \jetset\ with the
  \ariadne\ dipole cascade. Left plot is for all flavors, while the
  right plot is for $u$ quark events only.}
\label{fig_parton-hadron_jetset}
\end{center}
\end{figure}

What is the most `desirable' behaviour here is not so easy to tell. 
On the one hand, small hadronization corrections would be better for
perturbative studies, on the other hand, the string phenomenon is a
genuinely interesting piece of non-perturbative physics that deserves to
be studied in its own right. And even for an $\as$ determination,  
ultimately what matters is not whether a hadronization correction has to
be applied or not, but how large an error bar has to be assigned to this
correction. Fig.~\ref{fig_parton-hadron_jetset} (left picture)
here illustrates that the 
 event generators we have tried agree very well once again. That is, the 
phenomena described above are of general validity, and seem to be
accountable to a similar extent for most measures.

{}From the figures, it would seem that the above string/drag effects are 
present 
for three-jets but absent for higher jet multiplicities. This is not
fully correct, however. In any hadronic $n$-jet event, the two closest 
jets are likely to correspond to dipole-connected partons. (In the
leading-log picture of shower evolution, the only exception is given by
the rather infrequent $g \to q\overline{q}$ branchings.) Therefore the 
hadron-level jet directions will sit closer than the parton-level ones,
and one is more likely to get an $(n-1)$-jet event on the hadron level than
on the parton level. The turnaround of the curves, and ultimately the
larger hadron than parton average jet multiplicity at small $\ycut$, 
is thus rather a 
reflection of other effects entering and becoming more important.
These can collectively be classified as fluctuations in the hadronization 
process, but can have different origins. One example is the junk-jet
formation discussed repeatedly above, e.g., in Sect.~\ref{subsec_tubemodel}.
Another is the kinematics of particle decays, especially of bottom
and charm hadrons. This latter effect is illustrated in 
Fig.~\ref{fig_parton-hadron_jetset} (right picture), 
where results for the production of
all initial quark flavors are compared with those for $u$ quarks only.

If jet rates are to be used to extract an $\as$ value, one also needs
to understand the relation between the parton level curves given by a
generator and those expected in an exact theory. Indeed, as previously
recalled and as already shown in Ref.~\cite{camjet} for the D, A and C
algorithms, if the same $\as$ produced by the generator is used for
the pQCD leading-log resummed $+$ ${\cal O}(\as^2)$ fixed-order
predictions, then the corresponding parton level curves would fall
below the hadron level over a much larger $\ycut$ spectrum. In fact, a
good matching (for all $\ycut$'s) between the pQCD predictions and the
\herwig\ parton level is obtained if the former use $\as \equiv
\as(M_Z^2) = 0.126$, instead of 0.114, the value obtained from the
generator by interpreting the input parameter {\tt QCDLAM}, with the
default value of 0.18 GeV, as the NLO scale parameter $\lms$.  The
necessity of such `rescaling' should be not surprising, as such an
interpretation of {\tt QCDLAM} is only justified in a small region of
phase space (see \cite{CMW}) which should not be dominant.

\begin{figure}[htb]
\begin{center}
\centerline{}
\begin{minipage}[b]{.5\linewidth}
\centering\epsfig{file=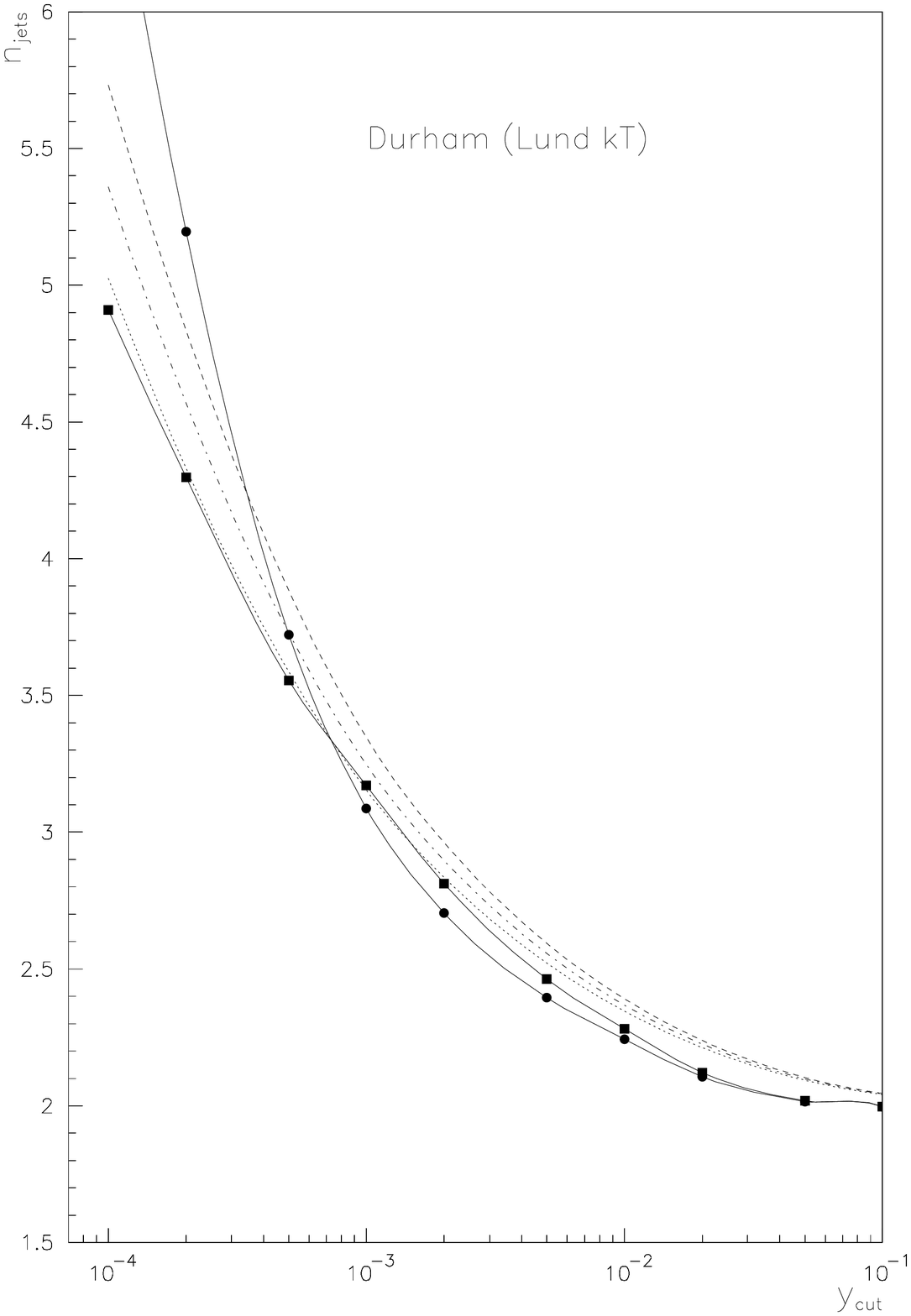,angle=0,height=12cm,width=\linewidth}
\end{minipage}\hfil
\begin{minipage}[b]{.5\linewidth}
\centering\epsfig{file=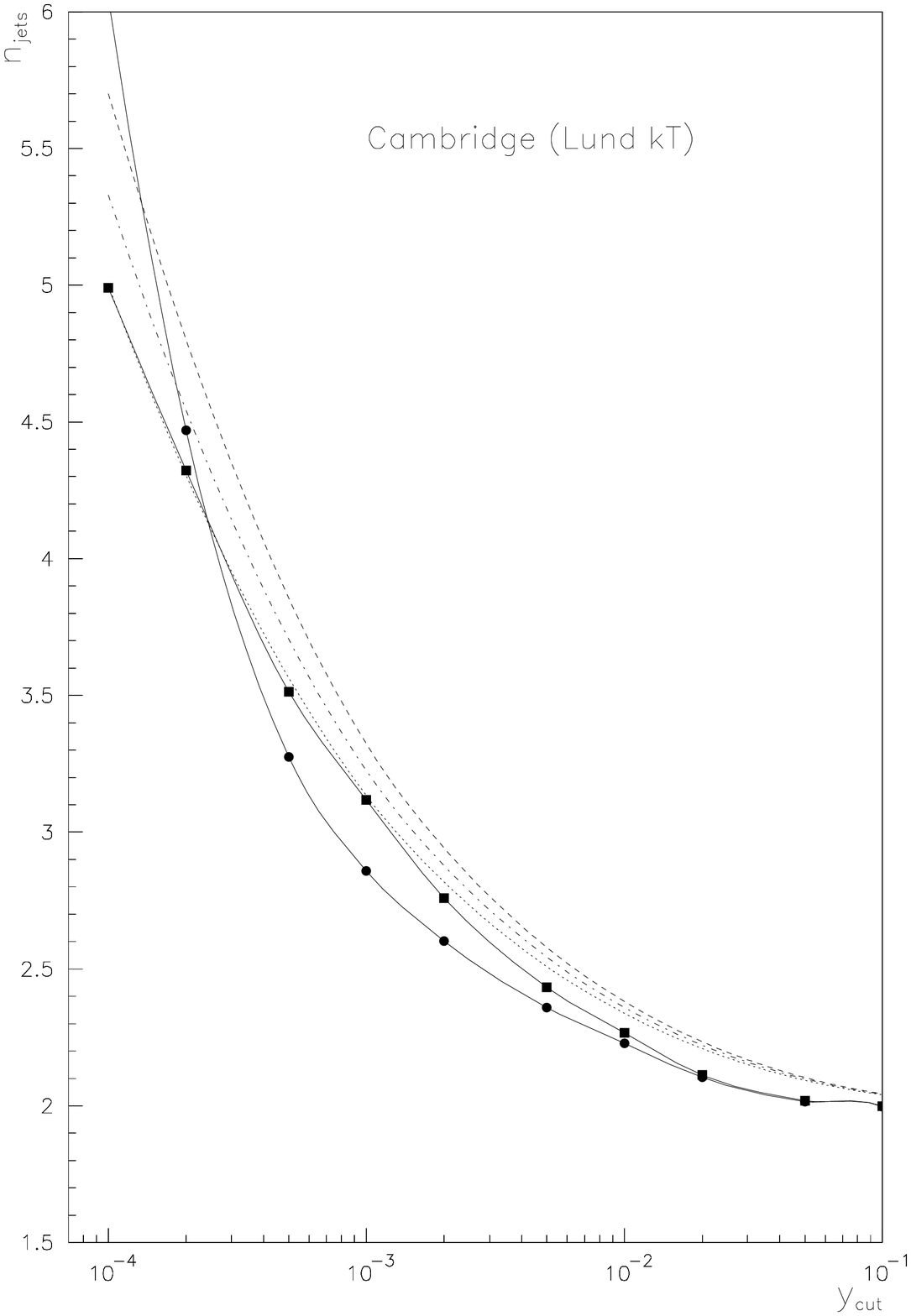,angle=0,height=12cm,width=\linewidth}
\end{minipage}\hfil
\centerline{}
\caption{The mean number of jets at $Q=M_Z$
  for the \Durham\ (left-hand side plot) and \Cambridge\ (right-hand
  side plot) algorithms using the \Luclus\ measure.  \herwig\ 
  predictions: squares, parton level; circles, hadron level.  Resummed
  predictions: dashed, $\as=0.126$; dot-dashed, $\as=0.120$; dotted,
  $\as=0.114$.}
\label{fig_as_herwig}
\end{center}
\end{figure}

Fig.~\ref{fig_as_herwig} plots the rates as obtained from the formulae
(\ref{nfin})--(\ref{nadd}), that is, the resummed predictions for the
DL and CL algorithms, for three values of $\as$, against the \herwig\ 
parton and hadron levels. (They are the counterpart of those for the
D, A and C scheme presented in Figs.~13--15 of Ref.~\cite{camjet}.)
{}From Fig.~\ref{fig_as_herwig} is then clear that, if the \Luclus\ 
measure is used instead of the \Durham\ one, things go the other way
round: the dotted curves, for which $\as=0.114$, are the ones closer
to the MC parton level.  This seems to indicate a further advantage in
using the \Luclus\ $p_{\perp}$: the MC parton level reproduces more
accurately the best perturbative results for the same $\as$. That is,
it appears that this measure is more blind to the differences between
the MC and the perturbative results than the \Durham\ one.
Furthermore, this is particularly true at very small $\ycut$, the
critical regime where not only a reduced size of the hadronization
corrections is required to study the transition between pQCD and
non-pQCD, but possibly also a good matching between the theoretical
and phenomenological `parton levels'. Even more reassuring is the fact
that, of the two schemes, CL is the one doing best in that region (the
dashed curve in the right-hand side plot practically coincides with
the MC parton level), in line with various other results previously
obtained for this new hybrid scheme.
 
As a summary of our hadronization studies in the context 
of multi-jet event rates
at LEP1, we can recapitulate  the following.
\begin{enumerate}
\item There is substantial correspondence between the simple tube
  model and more sophisticated MC programs. The main kinematic
  features recognised in the former reflect onto the latter. Thus, the
  simple hadronization mechanism based on a longitudinal phase space
  represents a good guidance in order to test in first instance the
  performances of new (and old) algorithms.
\item After full hadronization is implemented, four among the 
  clustering schemes studied since the beginning appear to have
  rather contained hadronization corrections at small values of the
  resolution parameter, say, around $10^{-4}$ at LEP1 energies,
  corresponding to partonic multiplicities of five or so. These are
  the \AODurham, the original \Cambridge\ scheme (i.e., that using the
  \Durham\ distance), the one using the \Luclus\ measure and,
  curiously, the original \Luclus\ scheme deprived of the reassignment
  step and implementing the default $d_{\mathrm{init}}$.  All other
  schemes perform significantly worse, particularly the \Geneva\ one,
  which appears to be very unstable. \Diclus\ offers unique advantages
  in having very small hadronization corrections in the
  two-jet-dominated region, but then has larger corrections than other
  algorithms (except \Geneva) at smaller resolution scales.
\item The hadronization corrections come about for several reasons.
  The string/drag effects usually give a negative contribution at
  large $\ycut$, i.e., results in fewer jets on the hadron than on the
  parton level, the size of which depends on the details of the
  algorithm. Fluctuations in the hadronization process, such as charm
  and bottom decays and junk-jet/misclustering effects, give a positive
  contribution, that always wins out at small $\ycut$. That some
  algorithms have a smaller net hadronization correction 
at medium $\ycut$
thus in part is the result of a
cancellation between opposite effects. From this point of view, our
results  for the case of the \Durham\ and \Cambridge\
schemes are in general agreement with those
presented in Ref.~\cite{stan}.   
\item Our results are substantially independent of the MC program
  used, that is, of the model adopted for the hadronization mechanism,
  and for the QCD cascade.  This means that hadronization corrections
  can be estimated rather reliably for many algorithms.  As a
  by-product of this conclusion, we observe that the angular-ordering
  procedure recommended in Ref.~\cite{camjet} as a refinement of the
  \Durham\ algorithm is then not restricted to the `angular
  ordered' emission as implemented in the \herwig\ parton shower
  (the MC program exploited in that reference).
\item A warning should be borne in mind, concerning the comparison of
  parton and hadron level.  It should be recalled that the partonic
  dynamics implemented in event generators is an approximation of the
  actual prediction, as the parton cascade only exploits some
  (logarithmically) enhanced terms of the infrared (soft and
  collinear) emission. Therefore, there is an intrinsic danger in
  interpreting the hadron--parton difference as generated by the MCs
  as a method-independent estimate of non-perturbative effects and
  simply adding it to the resummed predictions.
  
\item Specifically, we did not address here the key issue of how to
  combine the best perturbative QCD predictions, based on resummed
  contributions matched to fixed-order results, with hadronization
  corrections.  However, in this respect, we have shown that the
  \Luclus\ $p_{\perp}$ distance measure seems to offer an interesting
  alternative to the traditional \Durham\ one, in the sense that the
  theory and MC parton levels appear to match better in $\as$,
  especially when implemented along with the clustering sequence of
  the \Cambridge\ algorithm.
\end{enumerate}

\subsection{Jet reconstruction}
\label{subsec_jetrecon}

In this section we study various aspects of how well algorithms
reconstruct jet directions and energies, as well as a few other 
related quantities. The results presented here have been obtained 
with the \jetset\ program, but all essential features come
out very similarly with \herwig\ and \ariadne.

Since some of the studies are based on reconstructing a fixed number
of jets, like three or four, it should be noted that the \AODurham\ 
and \Cambridge\ algorithms do not always allow this, and do not
necessarily provide a unique answer. To understand these points, first
consider simple binary joining algorithms, such as \Jade\ or \Durham.
In these, it is always the cluster pair with smallest distance
$y_{ij}$ that are joined next. Starting from $n$ clusters, there is
thus a unique sequence of joinings giving $n-1, n-2, n-3, \ldots, 3,
2, 1$ clusters. If one wants to obtain three jets, say, one simply has
to perform binary joinings till exactly three clusters remain. For an
exercise like that, the $\ycut$ value need never even be specified.

For use in the continued discussion, let us make an alternative
description. By $\hat{y}_{m(m-1)}$ we may denote the smallest $y_{ij}$ 
value in the $m$-cluster configuration, which thus sets the scale for 
the joining to $m-1$ clusters. Standard distance measures
are constructed such that, in a joining, the joined cluster is always
further away from any third particle than was the nearest of the 
two original clusters, i.e., $y_{\hat{ij}k} > \min (y_{ik},y_{jk})$.
Thus, when joining the pair with smallest $y_{ij}$, the new 
configuration has a larger smallest $y_{ij}$. This way one 
obtains a unique ordered sequence of joining scales 
$\hat{y}_{n(n-1)} < \hat{y}_{(n-1)(n-2)} < \ldots <
\hat{y}_{43} < \hat{y}_{32} < \hat{y}_{21}$. Looking
for three clusters, there is always a non-vanishing range of $\ycut$ 
values $\hat{y}_{43} < \ycut < \hat{y}_{32}$, and all $\ycut$ in this 
range correspond to exactly the same three-cluster configuration. 

In the \AODurham\ scheme, on the other hand, $\ycut$
is used not only to interrupt a sequence of joinings, but also to
influence the sequence itself. We remind that the procedure joins the 
pair with smallest $v_{ij} \equiv 2(1 - \cos\theta_{ij})$ among all those with
$y_{ij} < \ycut$. Change $\ycut$, and you can change which pair is 
joined in the step from $m$ to $m-1$ clusters, and in turn all the 
subsequent joinings. Therefore, even if there should be a range
$\hat{y}_{43} < \ycut < \hat{y}_{32}$, that range may split into
subranges corresponding to different three-jet configurations.
Furthermore, it may be impossible to obtain three clusters, since an
infinitesimal $\ycut$ change may give a flip from one joining
sequence, ending with four clusters, to a completely different one,
ending with two. A further consequence of such flips is that the number 
of clusters need not be a monotonous function of $\ycut$.

The \Cambridge\ algorithm introduces one further task for $\ycut$, on
top of the \AODurham, namely to provide the scale
for sterilization/soft-freezing. This increases the fraction of events
that fail to reconstruct a requested number of jets, but reduces the
number of cases with several different three-jet topologies.

One should not exaggerate the problem, however. Typically only for
0.15\% of LEP1 events it is impossible to find a three-jet
configuration with \AODurham, which increases to 1.3\% in \Cambridge.
Furthermore, 4.7\% give two (or more) different three-cluster
configurations for \AODurham\ and 0.9\% for \Cambridge. The numbers are
somewhat higher for four-jets.  None of the other algorithms failed to
reconstruct the requested number of jets, nor have any ambiguity in
which jets are reconstructed. In the studies below, events which
failed to reconstruct are not considered at all, while the choice
among alternative three-cluster configurations is simply based on
which is found first. (In a search procedure that involves a measure
of randomness, so there should be no special bias\footnote{In 
Ref.~\cite{stan} a special-purpose algorithm was devised in order to
determine the $\ycut$ transition values at which an event flips from an
$n$-jet to an $m$-jet configuration, with $m$ and $n$ not necessarily
consecutive. It was used to study the characteristics of those events
that have two different $n$-jet configurations.}.) 


The parton shower starts out from a back-to-back $q \overline{q}$
pair. The observable event axis is smeared by the parton shower and
hadronization, but one interesting measure is how well the original
axis can be reconstructed. Thus clustering algorithms are requested to
find two jets; alternatively measures such as thrust and sphericity
can be used here. In the first result column of Tab.~\ref{tab_jet_ang}
it is shown that most algorithms do comparably well, including thrust,
while \Diclus\ without reclustering does worse and sphericity has
the largest error. In three-jet events there is no `correct' answer,
and so here the comparison is based on matching the jets clustered on
the parton level with those obtained on the hadron level. Each event
thus gives three angles. Only events with $0.85 < T < 0.95$ on the
parton level have been used to produce the numbers. Again \Diclus\
without reclustering does worse, third column of
Tab.~\ref{tab_jet_ang}, though less dramatically so than above, while
\Luclus\ does somewhat better than the others. The same pattern
holds for four-jets (not shown).

\begin{table}[htb]
\begin{center}
\begin{tabular}{|l|c|c|c|c|c|}
\hline
\rule[0cm]{0cm}{0cm}
Algorithm & \multicolumn{2}{c|}{Two-jet} & 
\multicolumn{3}{c|}{Three-jet}  \\ \hline
\rule[0cm]{0cm}{0cm}
     & $\VEV{\Delta\theta}$ & $\VEV{(p_z)_{\mathrm{back}}}$ & 
$\VEV{\Delta\theta}$ & $\sigma(\Delta E)$ &  
$\VEV{\Delta\theta_{\mathrm{min}}}$ \\
     & ($^{\circ}$) & (GeV) & ($^{\circ}$) & (GeV) & ($^{\circ}$) \\ \hline
\rule[0cm]{0cm}{0cm}
\Jade       & $3.22$ & $0.33$ & $4.01$ & $2.74$ & $-2.4$ \\ 
\Durham     & $3.09$ & $0.11$ & $3.91$ & $2.41$ & $-2.4$ \\ 
\LDurham    & $3.14$ & $0.19$ & $3.86$ & $2.48$ & $-3.0$ \\ 
\Geneva     & $3.05$ & $0.04$ & $4.01$ & $2.45$ & $-2.7$ \\ 
\AODurham   & $3.09$ & $0.10$ & $3.81$ & $2.25$ & $-1.9$ \\ 
\Cambridge  & $3.09$ & $0.10$ & $3.88$ & $2.28$ & $-2.4$ \\ 
\Luclus     & $3.06$ & $0.00$ & $3.52$ & $2.02$ & $-2.3$ \\ 
\Diclus\ 1   & $3.66$ & $0.00$ & $4.43$ & $1.99$ & $-0.3$ \\
\Diclus\ 2   & $3.56$ & $0.00$ & $4.23$ & $1.93$ & $-0.3$ \\
\Diclus\ 2 reclustered   & $3.07$ & $0.00$ & $3.65$ & $2.16$ & $-2.3$ \\
thrust           & $3.23$ & ---    & ---    & ---    & ---     \\
sphericity       & $4.36$ & ---    & ---    & ---    & ---     \\
\hline
\end{tabular}
\caption{Average angular and momentum/energy error on jet reconstruction 
  in two- and three-jet events at LEP1; see text for further details.
  \jetset\ results.}
\label{tab_jet_ang}
\end{center}
\end{table}

The error on the jet axis reconstruction need not be entirely of a
statistical character, however. In Sect.~\ref{subsec_jetrates} above,
we have mentioned the string/drag effect as the reason for the
`negative hadronization corrections'. In a three-jet event, normally
the smallest angle between two jets, $\theta_{\mathrm{min}}$, would be
formed by the gluon jet and a quark/antiquark jet. These are connected
by a dipole and thus should be `pulled closer' by the hadronization.
The last column in Tab.~\ref{tab_jet_ang} shows the average
$\Delta\theta_{\mathrm{min}}$, the difference between the hadron- and
parton-level $\theta_{\mathrm{min}}$ values. We see that indeed there
is the expected systematic bias in all algorithms, although markedly
smaller in \Diclus\ than in the others. \Diclus\ is the only
algorithm intended to correctly account for dipole effects in the
hadronization, and is thus seen to achieve this purpose. To set the
scale of the effect, the width of the $\Delta\theta_{\mathrm{min}}$
distribution is about $10^{\circ}$ in all algorithms, so the
systematic bias is still significantly smaller than the event-to-event
fluctuations. Also the smallest angle in four-jet analyses show a
similar pattern, with \Diclus\ the only one to be almost
bias-free. We note that if the sum of the momenta of the particles
assigned to each jet in \Diclus\ are allowed to redefine the jet
directions, the bias returns and this `reclustered' \Diclus\ behaves
more or less like the standard binary algorithms.

In a study of fixed three-parton configurations at lower energies,
where then the parton-level was known to have a fixed smallest angle 
of about $70^{\circ}$, most algorithms reconstruct an angle around 
$65^{\circ}$, while the \Diclus\ average is around $72^{\circ}$,
second results column of Tab.~\ref{tab_jet_fixede}.
Thus, while still doing best, there are indications that \Diclus\
at times overcompensates for the string/drag effect.

\begin{table}[htb]
\begin{center}
\begin{tabular}{|l|c|c|c|c|c|c|}
\hline
\rule[0cm]{0cm}{0cm}
Algorithm & $\VEV{\Delta\theta}$ & 
$\VEV{\theta_{\mathrm{min}}}$ & $\sigma(\Delta E)$ &
$\VEV{\Delta E_q}$ & $\VEV{\Delta E_{\overline{q}}}$ & 
$\VEV{\Delta E_g}$ \\
   & ($^{\circ}$) & ($^{\circ}$) & (GeV) &
(GeV) & (GeV) & (GeV) \\ \hline
\rule[0cm]{0cm}{0cm}
\Jade       & $5.70$ & $64.9$ & $1.34$ & $-0.16$ & $0.47$ & $-0.39$ \\ 
\Durham     & $5.68$ & $64.8$ & $1.33$ & $-0.03$ & $0.60$ & $-0.58$ \\ 
\LDurham    & $5.67$ & $64.7$ & $1.30$ & $-0.08$ & $0.46$ & $-0.39$ \\ 
\Geneva     & $5.74$ & $64.4$ & $1.44$ & $~0.25$ & $0.73$ & $-0.97$ \\ 
\AODurham   & $5.60$ & $65.0$ & $1.29$ & $-0.01$ & $0.60$ & $-0.61$ \\ 
\Cambridge  & $5.82$ & $63.6$ & $1.43$ & $~0.16$ & $0.66$ & $-0.85$ \\ 
\Luclus     & $5.34$ & $65.0$ & $1.11$ & $-0.05$ & $0.57$ & $-0.53$ \\ 
\Diclus\ 1   & $5.67$ & $72.0$ & $1.06$ & $-0.06$ & $0.71$ & $-0.65$ \\ 
\Diclus\ 2   & $5.38$ & $71.6$ & $1.03$ & $-0.04$ & $0.69$ & $-0.65$ \\ 
\Diclus\ 2 reclustered  & $5.13$ & $66.0$ & $0.96$ & $-0.32$ & $0.59$ & 
$-0.27$ \\ 
\hline
\end{tabular}
\caption{Average angular and energy error on jet reconstruction 
in three-parton events at 30~GeV. All events are $u \overline{u} g$,
to avoid contamination from heavy-flavour effects, and the three-jet
kinematics is fixed by $x_q = 0.9$, $x_{\overline{q}} =x_g = 0.55$;
hence $\theta_{\mathrm{min}} = 70.2^{\circ}$ on the parton level.
The last three columns give the difference between the hadron- and
parton-level numbers. See text for further details. \jetset\ 
results.}
\label{tab_jet_fixede}
\end{center}
\end{table}

Also the jet energy reconstruction can be compared between the hadron
and parton level, fourth column of Tab.~\ref{tab_jet_ang} and third
column of Tab.~\ref{tab_jet_fixede} give the width of the jet energy
difference distribution. Here \Diclus\ and \Luclus\ perform better
than any of the others. The tendency for systematic bias can be
studied in fixed three-parton configurations, last three columns of
Tab.~\ref{tab_jet_fixede}. The energy of the most energetic jet is
usually reconstructed without much bias, whereas there is a tendency
in all algorithms for the other quark to gain energy from the gluon,
reflecting the fact that gluon jets are softer and broader and thus
easily lose particles to the other jets, especially the most nearby
one. This systematic bias is largest in \Geneva, as could be expected
from the way \Geneva\ favours clustering around energetic particles.
\Cambridge\ shows the second largest bias, and the reclustered
\Diclus\ the smallest.

\begin{figure}[htb]
\begin{center}
\centerline{}
\centering\epsfig{file=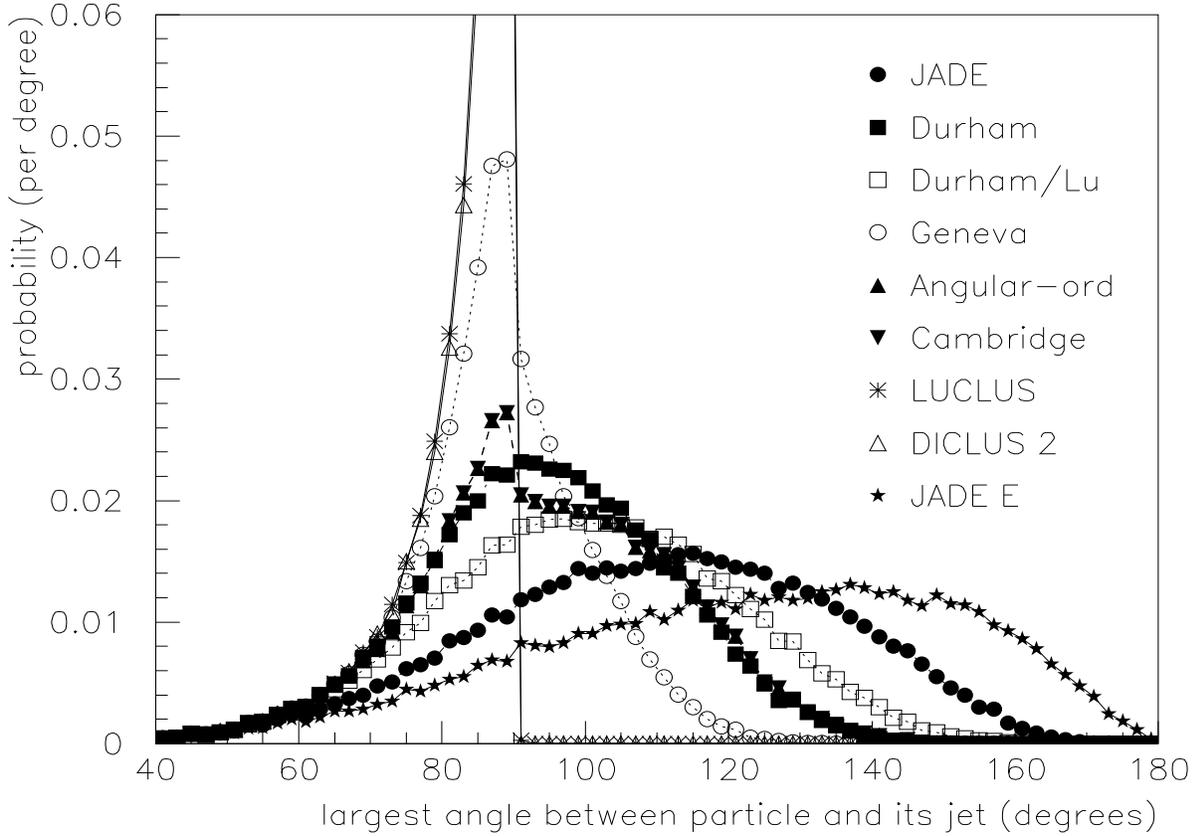,angle=0,height=12cm,width=\linewidth}
\centerline{}
\caption{Largest angle, per event, between a particle and the jet it is 
  assigned to, in two-jet events at LEP1. For clarity the $y$ axis has
  been truncated; the \Luclus\ curve goes up to 0.13, and \Diclus\ to
  0.15 in the last bin before $90^{\circ}$. \jetset\ results.}
\label{fig_largest_angle}
\end{center}
\end{figure}

{}From a practical point of view, a jet is a collection of `nearby'
particles, where `nearby' obviously is a very subjective criterion.
One measure is how far out in angle a jet extends from its core.  For
instance, if two back-to-back jet axes are reconstructed for a LEP1
event, one may expect an optimal subdivision of particles to be by
hemisphere, so that no particle is found more than $90^{\circ}$ from
its jet axis. Fig.~\ref{fig_largest_angle} shows the angle for the
particle furthest away from its assigned jet, on a per-event basis. It
is seen that only \Luclus\ and \Diclus\ respects the $90^{\circ}$
criterion.  The reason is that the standard distance measures allow
two soft particles to be joined, also when they are somewhat away in
angle. In a normal binary joining scheme, they will thereafter
together enter into one of the final jets, even if one of them is much
closer to another jet. The reassignment step of \Luclus\ is
specially devised to overcome this limitation, i.e., to reevaluate
prior joining decisions in the light of the joinings that have been
performed since.

Among the other algorithms, \Jade\ is most likely to have a particle
in the `wrong' hemisphere. In fact, the \Jade\ E scheme, using the
true mass as distance measure, is the very worst of the algorithms
studied. This is the well-known instability problem, already
mentioned. \Durham\ and the other $p_{\perp}$-based algorithms are
better, but note that it is important that the angular dependence is
$2(1 - \cos\theta_{ij})$ rather than the correct $\sin^2\theta_{ij}$,
or else two back-to-back particles would have $p_{\perp} = 0$ and be
joined. \Durham\ with the \Luclus\ measure is slightly worse than
normal \Luclus, since the clustering of two soft particles is somewhat
more favoured than in standard \Durham. The \AODurham\ and \Cambridge\ 
schemes offer no visible improvement. \Geneva\ is the pure binary
joining algorithm with best performance, reflecting that clustering of
two soft particles is disfavored.
 
The same phenomenon obviously carries over when more than three jets are
reconstructed. \Luclus\ always gives much narrower jets than any of 
the other algorithms, and \Jade\ gives the broadest ones. The one
notable change is that for three-jets, \Geneva\ no longer gives
narrower jets than \Durham\ and its relatives, probably indicating
how the \Geneva\ distance measure allows an energetic jet to pick
up particles also fairly close to another softer jet. 
While the wide-angle tracks 
are very important for the visual impression, they normally carry little
momentum. The second column in Tab.~\ref{tab_jet_ang} shows 
$(p_z)_{\mathrm{back}}$, the average amount of longitudinal momentum
carried by particles moving `backwards' with respect to their jet axis.  
Typically this number is only 0.1~GeV per event, rising to 0.3~GeV for
\Jade\ and 0.6~GeV for \Jade\ E. 

\begin{figure}[htb]
\begin{center}
\centerline{}
\begin{minipage}[b]{.5\linewidth}
\centering\epsfig{file=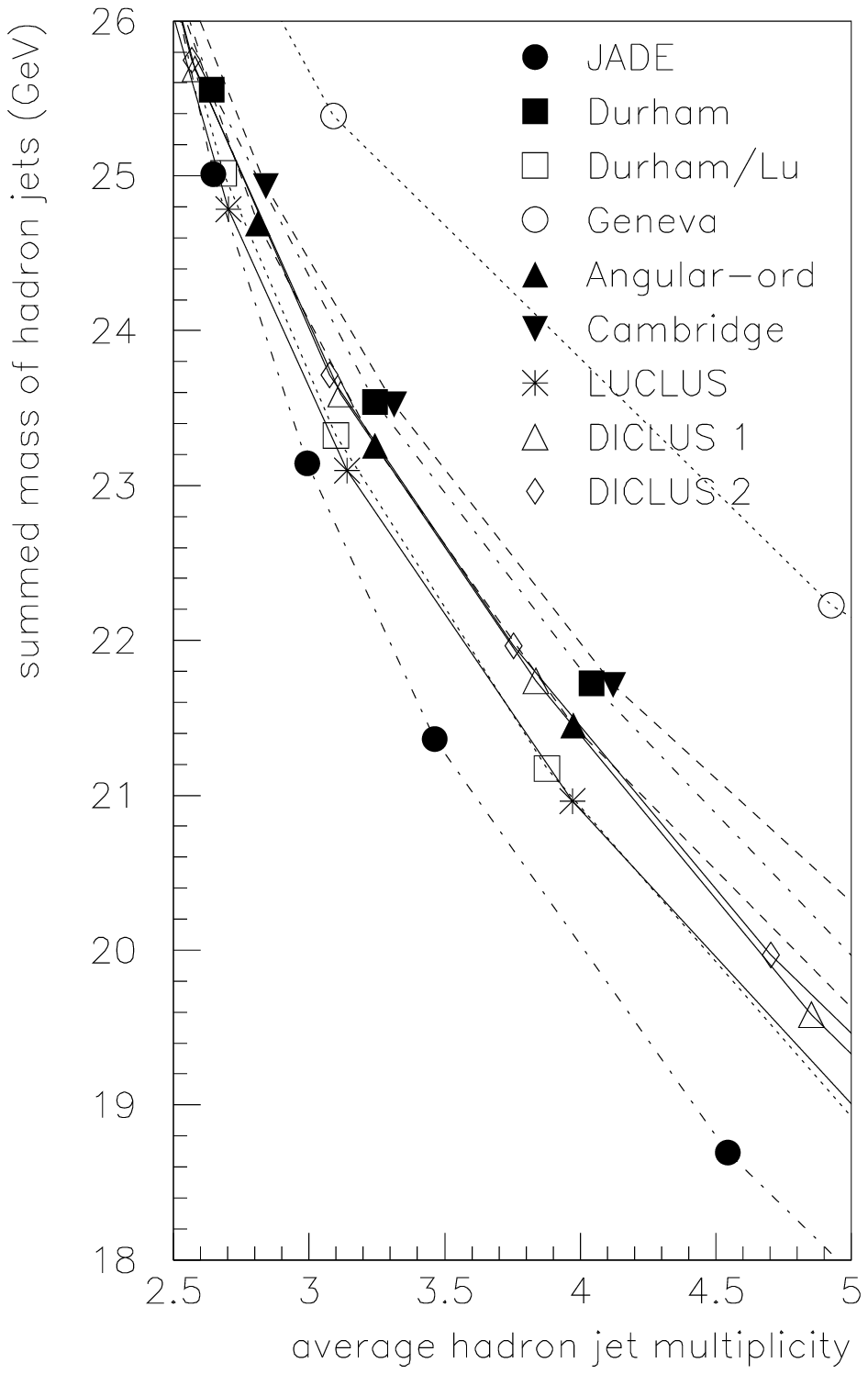,angle=0,height=12cm,width=\linewidth}
\end{minipage}\hfil
\begin{minipage}[b]{.5\linewidth}
\centering\epsfig{file=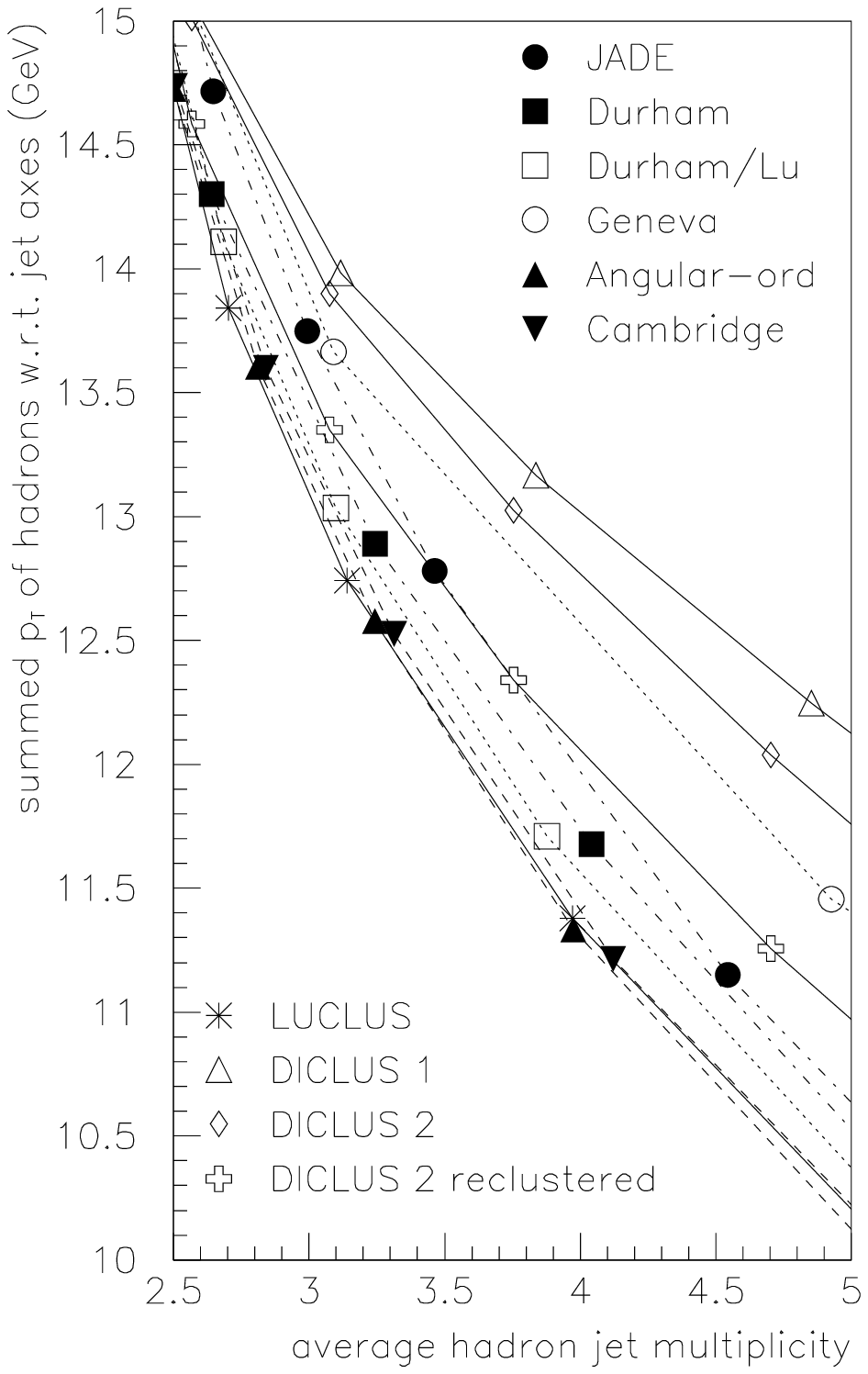,angle=0,height=12cm,width=\linewidth}
\end{minipage}\hfil
\centerline{}
\caption{Sum of jet masses (left) and of particle $p_{\perp}$ as a 
function of the average number of jets (i.e., implicitly as a function 
of $\ycut$) for LEP1 events. Note that the $y$ axis does not start at
0, i.e., differences appear exaggerated. \jetset\ results.}
\label{fig_mass_pT_best}
\end{center}
\end{figure}

Another alternative measure for the narrowness of jets is offered by
the sum of the invariant jet masses. This is studied in
Fig.~\ref{fig_mass_pT_best}. Since the $\ycut$ definition is scheme
dependent, results are plotted as a function of the average number of
jets at the $\ycut$ values studied, as in Sect.~\ref{subsec_jetrates}.
Here the \Jade\ algorithm indeed does best, in line with its distance
measure being intended to minimize jet masses. \Luclus\ and \Durham\ 
with the \Luclus\ measure come next, i.e., here reassignment is not
important. \Geneva\ does markedly worse than other schemes.
 
A third measure is the summed transverse momentum of all particles in
an event, relative to their respective jet axis. This is shown, again
as a function of the average number of jets, in
Fig.~\ref{fig_mass_pT_best}.  The difference between the
$p_{\perp}$-based algorithms is here small, while \Jade\ and
reclustered \Diclus\ are somewhat worse and \Geneva\ together with
the other \Diclus\ modes are the worst. The large difference between
the standard \Diclus\ and the reclustered one is due to that \Diclus\ 
jets typically are asymmetric with most of the particles lying on one
side of the jet direction. The reclustering pulls the jet direction
more to the center of the particles assigned to it, thus reducing the
summed \tpt.

In summary, we draw the following conclusions from the analysis of
two-, three-, and four-jet events (also supported by some studies not 
shown):
\begin{enumerate}
\item The \Diclus\ algorithm generally does best (among the algorithms
  studied) in jet energy reconstruction, and also successfully
  addresses the issue of a systematic hadronization bias in the
  opening angle between two nearby jets, caused by the string/drag
  effect. The price to be paid is that the average error on the
  individual jet direction is larger than in other algorithms.
  Reclustering the jets from \Diclus\ makes it behave more like the
  standard binary algorithms. We also note that using the transverse
  mass in Eq.~(\ref{LL:invmt}) as measure (mode 2) is somewhat
  favoured as compared to using the one in eq.~(\ref{LL:invpt}) (mode
  1).
\item \Luclus\ does almost as well as \Diclus\ in jet energy
  reconstruction, and best in jet angles. (A similar conclusion was
  reached in Ref.~\cite{BKSS}.) The reassignment step means it is the
  only algorithm that does not have stray particles in a jet that are
  visibly much closer to another jet. Since the stray particles
  normally carry rather small momenta, the impact of reassignment on
  momentum-weighted quantities should not be overstressed, however.
\item The \AODurham\ and \Cambridge\ algorithms here offer no
  significant advantages over the basic \Durham\ scheme, nor does a
  use of the \Luclus\ distance measure. All these algorithms therefore
  share a common `average' level of performance.
\item \Jade\ fulfills the intended task of reconstructing small
  cluster masses, but at the price of a larger rate of large-angle
  stray particles.
\item \Geneva\ does better than the average in some quantities, 
and significantly worse in others. Its distance measure means that the 
jet energy determinations show larger systematic biases than with any 
of the other measures used.
\end{enumerate}

\subsection{W mass reconstruction}
\label{subsec_WW}

Above we have studied jet finding in quite general terms. For an
intended application, special further studies may be necessary. 
The criteria for a good algorithm are going to be different in the
determination of an $\as$ value and in the study of angular 
distributions as a test of the three-gluon vertex, to give but two 
examples. Currently, the $W$ mass determination at LEP2 is another such
topic of large interest \cite{Wmass}, representing 
different optimization criteria than the ones illustrated above. We
here focus on the hadronic production channel, where $e^+ e^- \to W^+
W^- \to q \overline{q} Q \overline{Q}$.  Thus the signal is the
presence of four jets, where the two jet pairs ought to have a mass
around $m_W \approx 80$~GeV. There are several complications.
Backgrounds exist, both from the four `wrong' jet pairs in the same
event as the two `right', and from other processes such as the QCD
four-jets $e^+ e^- \to \gamma^*/Z^{*} \to q \overline{q} g g, 
q\overline{q} Q\overline{Q}$. 
The mass distribution is smeared by the intrinsic $W$ width $\Gamma_W
\approx 2$~GeV in combination with the production matrix element
itself, by initial-state QED radiation, by neutrinos that escape
without detection, by cracks in the detector acceptance, by
measurement errors on particle four-momenta and, of course, by
misassignments in the clustering procedure.  A full study can
therefore only be carried out within the context of a complete
detector simulation, which is rather beyond the scope of the current
report. To illustrate some of the clustering issues we have carried
out a rather more modest exercise.

Hadronic $W^+ W^-$ events are generated at 180 GeV CM energy,
but none of the background processes are studied. Detectors are assumed 
perfect, i.e., the correct four-momenta of outgoing particles are
used to reconstruct exactly four jets per event, by the respective
jet algorithm. (Some small number of times \AODurham\ 
and \Cambridge\ fail to find four jets, as explained in
Sect.~\ref{subsec_jetrecon}; such events are left out from the 
statistics of the respective algorithm.) In experimental analyses usually 
some further cuts are imposed, e.g., on the opening angles between
jets and on jet energies. This makes sense, since events where two
jets are very close are not reconstructed so well. However, then the 
retained event sample would differ between clustering algorithms, so we 
have avoided cuts here. Instead all six jet--jet masses in all events 
are found and studied, and the success of an algorithm is reflected in 
how often it can reconstruct sensible $W$ masses.

Some impression of how good the jet reconstruction is can be gleaned by
matching the four jets to the four original partons by minimizing the
sum of jet--parton opening angles. The average value of this sum, as well
as the sum of deviations in the energies between jets and partons, is 
given in the first two 
columns of Tab.~\ref{tab_Wmass_simple}. It generally agrees with
the picture in the previous section: \Luclus\ does good overall,
while \Diclus\ does worse with angles unless reclustering is performed. 
The poor numbers for \Geneva\ are more marked than in previous studies, 
however.

\begin{table}[htb]
\begin{center}
\begin{tabular}{|l|c|c|r|c|}
\hline
\rule[0cm]{0cm}{0cm}
Algorithm & $\VEV{\sum\Delta\theta}$ & $\VEV{\sum|\Delta E|}$ 
& \multicolumn{1}{c|}{$\VEV{\delta}$} & $\sigma(\delta)$ \\
   & ($^{\circ}$) & (GeV) & (GeV) & (GeV) \\ \hline
\rule[0cm]{0cm}{0cm}
Jade     & $41.0$ & $25.3$ & $ 0.06$ & $2.8$  \\ 
\Durham   & $36.5$ & $21.7$ & $-0.03$ & $2.6$  \\ 
\LDurham& $37.0$ & $22.5$ & $ 0.05$ & $2.6$  \\ 
\Geneva   & $46.1$ & $27.3$ & $-0.70$ & $3.4$  \\ 
\AODurham & $37.6$ & $22.0$ & $-0.08$ & $2.8$  \\ 
\Cambridge & $38.2$ & $23.0$ & $-0.13$ & $2.8$  \\ 
\Luclus   & $35.6$ & $19.9$ & $ 0.01$ & $2.6$  \\ 
\Diclus\ 1 & $39.3$ & $19.7$ & $-0.58$ & $3.1$  \\ 
\Diclus\ 2 & $38.8$ & $19.5$ & $-0.57$ & $3.0$  \\ 
\Diclus\ 2 reclustered & $35.6$ & $18.8$ & $0.16$ & $2.5$  \\ 
\hline
\end{tabular}
\caption{Analysis of hadronic $W^+ W^-$ events at LEP2. First two
  columns give angular and energy mismatch between reconstructed jets
  and original partons. Second two give average and spread between
  best reconstructed and true average $W$ mass of event.  \pythia\
  results.}
\label{tab_Wmass_simple}
\end{center}
\end{table}
 
\begin{figure}[htb]
\begin{center}
\centerline{}
\centering\epsfig{file=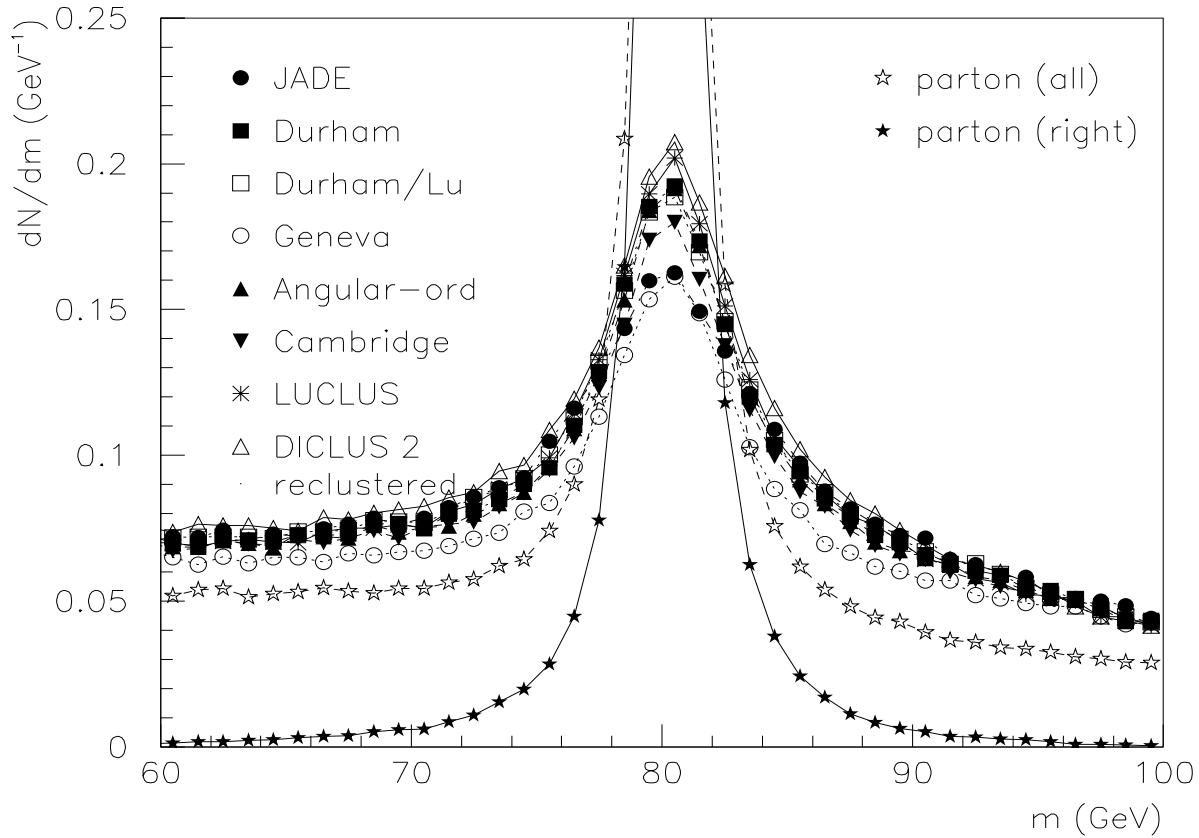,angle=0,height=12cm,width=\linewidth}
\centerline{}
\caption{Jet--jet mass spectrum in hadronic $W^+W^-$ events. 
  Each event is reconstructed to four jets and all six jet--jet
  combinations are included. \pythia\ results.}
\label{fig_jetjetmass}
\end{center}
\end{figure}

The true test is in the jet--jet mass spectrum, illustrated in
Fig.~\ref{fig_jetjetmass}, where one may discern the peaked signal 
from correct combinations of well reconstructed jets, over a 
smoother background of mismeasured jets or incorrect jet combinations.
The 70--90~GeV mass window has been used to produce {\sc Minuit} 
\cite{minuit} fits for a signal plus background shape. The choice
of best fit function is not trivial: the signal Breit-Wigner shape
is combined with misassignment errors in a complicated and
clustering-algorithm-dependent way. For simplicity we have assumed
a Breit-Wigner shape, characterized by a peak height $h$ (given as
the number of events per 0.2 GeV mass bin; this is related to the 
input normalization for \minuit), position $m_W$ and width
$\Gamma_W$. The $h$ and $\Gamma_W$ may be combined to an area $A$
underneath the Breit-Wigner. Normalization is such that 2 should
be the maximum, corresponding to two correctly reconstructed jet pairs
per event. Since the Breit-Wigner has rather large tails, this
ansatz may have a tendency to paint too rosy a picture of how
well algorithms do. An alternative would have been a Gaussian fit,
where the tails are rather strongly dampened, and the bias would go 
in the other direction. The qualitative differences between 
algorithms that we report below are the same, however. Two background
shapes have been used, one a three-term polynomial in mass and another
corresponding to a smeared step function (motivated by the 
kinematical-limit shoulder at large masses). Results with these two 
backgrounds come rather close, thus in Tab.~\ref{tab_Wmass_fit}
\pythia\ numbers are for the former and \herwig\ numbers
for the latter background. The fits described here correspond to the 
`Individual $W$ mass' columns. The `parton (right)' row makes used of the
two correct $W$ masses, and thus represents the best possible answer 
for algorithms, while `parton (all)' contains all six possible
combinations of the four original partons. The fact that fitted areas 
above 2 are obtained illustrate imperfections in the fitting ansatz. 

\begin{table}[htbp]
\begin{center}
\begin{tabular}{|l|r|r|r|r|r|r|r|r|}
\hline
\rule[0cm]{0cm}{0cm}
Algorithm & \multicolumn{4}{c|}{Individual $W$ mass} &
\multicolumn{4}{c|}{Average $W$ mass} \\ \hline
\rule[0cm]{0cm}{0cm}
   & \multicolumn{1}{c|}{$h$} & $m_W$ & $\Gamma_W$ & $A$ &
     \multicolumn{1}{c|}{$h$} & $m_W$ & $\Gamma_W$ & $A$ \\
   &  & (GeV) & (GeV) &  &  & (GeV) & (GeV) &  \\ \hline
\rule[0cm]{0cm}{0cm}
 & \multicolumn{8}{c|}{\pythia\ results} \\ \hline
\rule[0cm]{0cm}{0cm}
\Jade     & 200 & 80.321 & 8.000 & 1.26 & 163 & 80.719 & 3.697 & 0.47 \\ 
\Durham   & 260 & 80.354 & 5.663 & 1.16 & 195 & 80.482 & 3.296 & 0.51 \\ 
\LDurham & 247 & 80.333 & 5.800 & 1.12 & 200 & 80.537 & 3.288 & 0.52 \\ 
\Geneva   & 226 & 80.376 & 6.206 & 1.10 & 190 & 80.180 & 3.237 & 0.48 \\ 
\AODurhamshort & 260 & 80.396 & 5.721 & 1.17 & 227 & 80.454 & 3.216 & 0.57 \\ 
\Cambridge & 238 & 80.376 & 5.871 & 1.10 & 240 & 80.396 & 3.249 & 0.61 \\ 
\Luclus   & 268 & 80.387 & 5.447 & 1.14 & 190 & 80.492 & 3.395 & 0.51 \\ 
\Diclus\ 1 & 182 & 80.008 & 6.883 & 0.98 & 136 & 79.671 & 3.923 & 0.42 \\ 
\Diclus\ 2 & 188 & 80.044 & 6.732 & 0.99 & 138 & 79.657 & 3.758 & 0.41 \\ 
\Diclus\ 2 reclustered & 270 & 80.486 & 5.709 & 1.21 & 184 & 80.615 & 3.375 
& 0.49 \\ 
parton (all)   &1267 & 80.320 & 2.088 & 2.08 & 656 & 80.329 & 2.118 & 1.09 \\ 
parton (right) &1270 & 80.324 & 2.076 & 2.07 & 661 & 80.325 & 2.053 & 1.07 \\ 
\hline
\rule[0cm]{0cm}{0cm}
 & \multicolumn{8}{c|}{\herwig\ results} \\ \hline
\rule[0cm]{0cm}{0cm}
\Jade     & 235 & 80.218 & 6.553 & 1.212 & 220 & 80.491 & 3.337 & 0.576 \\ 
\Durham   & 315 & 80.326 & 4.893 & 1.211 & 295 & 80.268 & 2.999 & 0.694 \\ 
\LDurham & 299 & 80.310 & 5.203 & 1.222 & 238 & 80.293 & 3.216 & 0.601 \\ 
\Geneva   & 260 & 80.359 & 5.287 & 1.081 & 180 & 80.010 & 3.248 & 0.460 \\ 
\AODurhamshort & 311 & 80.345 & 4.944 & 1.209 & 320 & 80.284 & 2.863 & 0.719 \\ 
\Cambridge & 281 & 80.376 & 5.288 & 1.168 & 319 & 80.252 & 2.895 & 0.725 \\ 
\LCambridge    & 280 & 80.387 & 5.261 & 1.155 & 299 & 80.250 & 2.879 & 0.677 \\ 
\Luclus   & 324 & 80.368 & 5.247 & 1.335 & 241 & 80.291 & 3.212 & 0.608 \\ 
\Luclus\ (no pre)  & 324 & 80.371 & 5.249 & 1.334 & 239 & 80.286 & 3.200 & 
0.601 \\ 
\Luclus\ (no reas) & 177 & 79.984 &10.000 & 1.392 & 236 & 80.602 & 3.383 &
0.626 \\ 
\Diclus\ 0 & 193 & 79.920 & 9.904 & 1.498 & 179 & 79.494 & 4.600 & 0.646 \\ 
\Diclus\ 1 & 244 & 79.896 & 7.521 & 1.440 & 271 & 79.561 & 3.528 & 0.750 \\ 
parton (all)   &1312 & 80.422 & 1.995 & 2.056 & 680 & 80.408 & 1.986 & 1.060 \\ 
parton (right) &1319 & 80.427 & 1.988 & 2.059 & 694 & 80.418 & 1.911 & 1.041 \\ 
\hline
\end{tabular}
\caption{Fits to the $W$ mass spectrum in hadronic $W^+ W^-$ events at LEP2.  
First four columns for the each of the two $W$'s in an event, last four for
the average $W$ mass of an event. $h$ is peak height (normalization based on
event sample used), $m_W$ and $\Gamma_W$ fitted $W$ mass and width of
a Breit-Wigner shape, and area $A$ the number of combinations per event under 
the fitted Breit-Wigner. First part \pythia\ results fitted with a 
polynomial background, second part \herwig\ results fitted with a 
smeared step background. Note that \pythia\ and \herwig\ use
different input masses and widths; the last row for each program sets the 
standard of optimal performance.}
\label{tab_Wmass_fit}
\end{center}
\end{table}

Comparing algorithms, several aspects should be kept in mind. 
A larger area $A$ implies a larger efficiency for sensible jet finding, 
i.e., fewer misassignments that completely kill the signal. A smaller 
width $\Gamma_W$ is a sign of good performance for those jet pairs 
that are still correctly combined, i.e., fewer misassignments of a less 
disastrous character. For good $m_W$ determination in an experiment one
should thus have both a large $A$ and a small $\Gamma_W$. As a third 
criterion one could imagine the systematic offset between the 
reconstructed $W$ mass and the parton-level one. However, so long as
such an offset is not too large and can be well modelled, it is not so
important. One anyway has to make other corrections, e.g., the input
$m_W$ parameter does not coincide  with the average generated $m_W$ 
because of the convolution with matrix-element  and phase-space factors.
Unfortunately, while \pythia\ and \herwig\ results largely 
agree, there are some discrepancies that we do not fully understand,
and that thus should act as a warning not to take these studies as the
definite word.

One possible conclusion from the numbers in Tab.~\ref{tab_Wmass_fit}
is that many of the algorithms perform comparably well. In particular,
the correlation between sophistication and performance is weak or
non-existent, moving, e.g., from \Durham\ to \AODurham\ to \Cambridge.
It appears that \Luclus\ consistently reconstructs the largest area,
i.e., does fewest severe misassignments, but has a rather standard
peak width. The difference between \Durham\ and \Luclus\ $p_{\perp}$
measures is small; if anything the latter gives a wider peak and thus
is worse. Whether preclustering is performed or not in \Luclus\ is
irrelevant so long as reassignment is allowed, but without
reassignment the preclustering is disastrous --- the peak is so
broadened that $\Gamma_W$ hits the upper bound allowed in the fit.
Thus, to the extent that \Luclus\ does somewhat better than \Durham,
the reason is the reassignment step. The \Diclus\ fits without
reclustering give problems with the $\Gamma_W$ or $A$ values, but also
displays a large systematic bias in the estimated $m_W$. With
reclustering, \Diclus\ again does fairly well. One reason for the
problems could be that \Diclus\ is designed with QCD events in mind,
where two nearby partons are connected by a color dipole. Here two
nearby jets would come from different $W$'s and not share a dipole (we
did not include the possibility of color rearrangement
\cite{LEP2QCDgen,col_rearr}).

\begin{figure}[htb]
\begin{center}
\centerline{}
\centering\epsfig{file=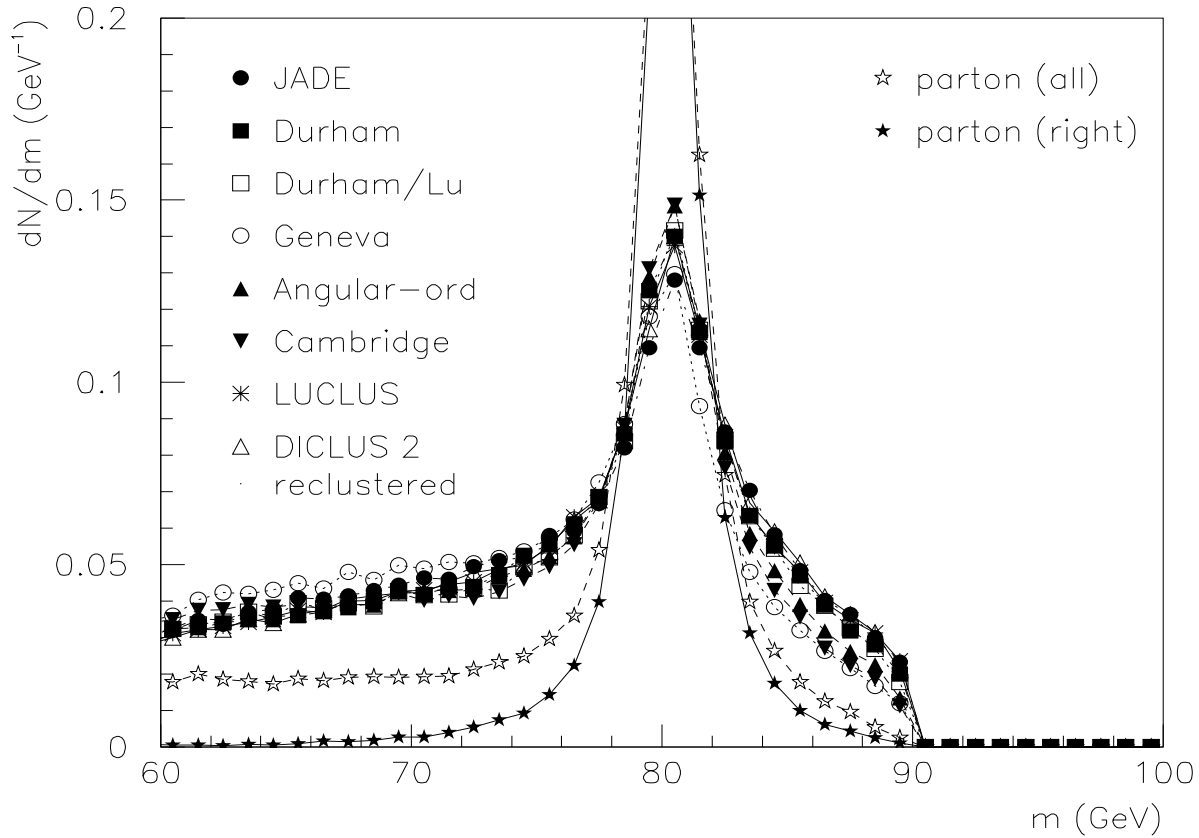,angle=0,height=12cm,width=\linewidth}
\centerline{}
\caption{Average jet--jet mass in hadronic $W^+W^-$ events. 
  Each event is reconstructed to four jets, which can be paired three
  different ways. The pairing with largest energy difference between
  the pairs is omitted and the average masses of the other two
  pairings are plotted. \pythia\ results.}
\label{fig_avgWmass}
\end{center}
\end{figure}

In experiments, it is advantageous to study the average $W$ mass of an
event rather than the two individual ones. There are several reasons
for this, but of interest here is that misassignments of particles 
in part cancel, in that a reassignment of one particle from one $W$ to 
the other reduces the first mass and increases the second, leaving the 
average less affected than each separately. Per event there are thus 
three possible jet pairings, each giving one potential average $W$ mass.
Of these three, we exclude the one where the two most energetic jets 
are paired with each other, since kinematically this is seldom the
right combination. The remaining two combinations give mass distributions
as illustrated in Fig.~\ref{fig_avgWmass}. Note that indeed the signal 
peak is much more narrow, and that there now is an absolute kinematic 
limit at 90~GeV. \minuit\ fits have been performed, as before, with
results as shown in the `Average $W$ mass' columns of 
Tab.~\ref{tab_Wmass_fit}. Normalization is such that an ideal fit would 
give $A =1$.
 
It is notable that the relative performance of algorithms changes
rather drastically compared with above. The two best ones now are
\Cambridge\ (the original one, not the one employing the \Luclus\ 
measure, which is not shown in the plots) and \AODurham, whereas
\Luclus\ falls below the average. This could indicate that the
particles that get misassigned are somewhat different in the former
two and in the latter algorithm.  That is, in the former two, the
errors on the two individual $W$ masses tend to cancel better in the
average. \Diclus\ still gives a larger $\Gamma_W$ than other
algorithms. Geneva has a reasonable width but a small area $A$.

The right two columns of Tab.~\ref{tab_Wmass_simple} shows that the
pattern between models is not so easy to understand. Here the average
mass is evaluated for all three possible jet pairings and compared
with the correct average $W$ mass of the same event. The pairing which
agrees best is retained, and $\delta$ denotes the average mass
difference between the reconstructed and the true average $W$ mass. As
we see, \Geneva\ and \Diclus\ without reclustering here show a
significant bias in the negative direction, and also give a larger
width $\sigma$ of the $\delta$ distribution. \Luclus\ does quite
well in these `behind-the-scene' numbers, so the poor \Luclus\
numbers above do not seem to have a simple explanation.

As possible conclusions for the $W^+W^-$ analysis, we attempt the following.
\begin {enumerate}
\item The choice of the algorithm is in general not so trivial in such
  a context. However, there are some that cannot be recommended,
  notably \Diclus\ without reclustering and \Geneva, and also \Jade.
\item If the `Individual $W$ mass' distribution is preferred in the
  selection procedure, that \Luclus\ performs slightly better than
  the other algorithms.
\item If, instead, one resorts to the `Average $W$ mass' spectrum,
  then \AODurham\ and the original \Cambridge\ (i.e., that with the
  \Durham\ measure) come out best.
\item However, differences between the three algorithms that excel
do not show a simple pattern, so that, in the end, a definite 
decision between these 
latter could probably only be made in the context of some specific
detector simulation and mass extraction procedure.
\end{enumerate}

\subsection{Speed}
\label{subsec_speed}

All binary clustering algorithms are comparably fast. Starting from 
an initial configuration of a large number $n$ of partons and hadrons,
a small number of jets is to be found. Therefore $O(n)$ binary
joinings have to be performed. Each in principle requires 
$O(n^2)$ distances to be evaluated to find the smallest one. 
In practice, distances can be kept in a table that is only updated
for those entries affected by a binary joining. Therefore execution 
time scales more like $O(n^2)$ than the expected $O(n^3)$.
At LEP1 energies, the clustering time is about two thirds of
the time it takes to generate an event (with \jetset\ and \herwig). 

The \Luclus\ preclustering time roughly scales like $O(n^2)$:
each particle can be the seed of a precluster and all particles
have to be tested whether they belong to the precluster.
If $m$ preclusters are formed, normally with $m$ much smaller than
$n$, then subsequent joinings take $O(m^3)$, since the reassignment
steps means one cannot reuse older numbers. The reassignment step
after each joining requires assigning $n$ particles to $m$ clusters, 
i.e., a total $O(m^2n)$ for $O(m)$ joinings. In practice, scaling 
of the total time is about like $O(n^2)$. 
At LEP1 the algorithm is somewhat faster than the binary joining ones, 
but at most by a factor of two. If the trick of pretabulation is 
not used in the binary routines, the difference is more like a factor 
five.
 
The basic step of the \Diclus\ algorithm is the joining of three
clusters into two. Therefore $O(n^3)$ distances have to be evaluated
to find the smallest one. Again, keeping a table of distances allows
the total time to scale more like $O(n^3)$ than like the $O(n^4)$
that might have been expected. Still, there is a significant price to
be paid, and at LEP1 energies \Diclus\ is about a factor fifty
slower than the other algorithms.  

\section{Summary and conclusions}
\label{sec_summary}

Jet clustering algorithms are an expression of {\sl time} and {\sl place}.
The time evolves with the calculational methods developed and these can 
in turn be limited by the computing power available.
The place is circumscribed by the experimental contexts where algorithms 
are needed and the tasks that they are asked to accomplish.
In ten, fifteen years from now, the two will both have changed.
Specifically, if the theoretical methods adopted (e.g., in determining 
the higher-order and exponentiation properties of pQCD, the parton-shower
evolution and/or the non-perturbative dynamics of hadronization)
in some years time would be different, so should clustering algorithms be. 

Inevitably then, our study can claim no prerogative to being definitive.
We have undertaken it for the present era and for the imminent 
phenomenology. The aim was to survey the many and different jet finding 
algorithms for electron-positron  events available on 
the market nowadays and study which algorithm to use 
where, if at all possible. As anticipated in the Introduction, we have not
found {\sl one} single best choice that prevails in {\sl all} 
cases we have addressed. However, as the reader should have agreed upon
by now, there need not exist such a one. 
Nonetheless, in several instances it has been possible to recognize, if not
the {\sl most suitable} algorithm to use, at least the {\sl attractiveness} 
of some of its basic components. In this Section, we attempt to summarize
our findings.

As a first example we have considered the realm of pQCD,
by studying jet fractions at parton level and
resorting to the most advanced techniques of perturbation theory: that is, 
exact next-to-leading fixed order results 
combined with resummed predictions 
to next-to-leading-logarithmic accuracy.
Such a choice was not made by chance, as it was dictated by the crucial r\^ole 
that jet rates
 play, e.g., in the determination of the strong coupling constant 
$\as$ and of its running with energy.
In this respect, we should point out that the main features illustrated in 
this paper for the case of LEP1 energies, survive unaltered for the case of 
LEP2.

By studying the three-jet fraction in pQCD we have taken for granted
the well-established result that a jet measure based on some relative
transverse momentum of the clusters involved is the most appropriate
to use, thus neglecting consideration of jet finders based on other
quantities (such as the invariant mass). Under these circumstances,
one historically recognizes three different such measures. Namely, the
so-called \Luclus, \Durham\ and \Diclus\ ones. The first two cluster
two particles into one whereas the last one merges three into two.
Neglecting imperceptible differences (we used `massless' partons)
between energy and momentum, they can geometrically be viewed as
follows.  The first represents the transverse momentum of either
particles with respect to the sum of the momenta of the two.  The
second is the transverse momentum of the lower-energy cluster with
respect to the higher-energy one.  The third is the transverse
momentum of one cluster with respect to the other two.

Among the three our preference would go to the \Luclus\ measure. In fact,
algorithms based on the latter display a reduced  (renormalization)
scale dependence of the three-jet fraction at NLO, as
compared to the cases of the \Durham\ and \Diclus\ expressions. 
The stability of the perturbative results in higher order against 
variations of such a scale is a measure of the smallness of even higher terms
in the perturbative expansion, this ultimately reflecting a better degree 
of convergence of the corresponding power series. As $\as$ measurements
are unavoidably biased by a theoretical error, and since this is 
assessed in no other way than the range in $\as$ spanned by the QCD 
predictions for different choices of the above scale, in our opinion, 
the \Luclus\ measure comes to be a recommendable choice in this context.
We have hypothized its improved behaviour,
with respect to the \Durham\ one, as due to their respective definitions:
whereas the former is a continuous function of the energies of the
two clusters the latter is not.  As a matter of 
fact, the presence of discontinuities at the
edge of the phase space of an observable has recently been advocated to act as a
source of misbehaviors in higher order perturbation theory. 

An additional neat attribute of the \Luclus\ transverse momentum
appears while combining the fixed-order with the resummed perturbative
predictions, for example in computing the average number of jets
produced in electron-positron annihilation events. 
Such a quantity can be predicted reliably from QCD
over a wide range in $\ycut$ and, furthermore, it is also
particularly sensitive to the actual value
of $\Lambda^{(5)}_{\overline{\mbox{\tiny{MS}}}}$. These two aspects render
it then a particularly good variable for the determination of $\as$.
The advantage of using the \Luclus\ measure in this case 
is that the parton level of the theory matches more naturally the parton
level produced by the Monte Carlo generator, as no rescaling of $\as$
is needed to find an adequate agreement
between the two (contrary to the case of the \Durham\ measure).

The difference between the parton level and the hadron level
as generated by a phenomenological Monte Carlo program is customarily
used as an estimate of the hadronization corrections. However,
one should notice that even in presence
of a good agreement between exact parton level from the theory and
the approximate one from the  Monte Carlo, there is a danger in 
interpreting the hadron-parton difference in the phenomenological generator
as an estimate of non-perturbative effects and simply adding it to
the matched prediction. In fact, the presence of unnatural 
cut-offs and kinematic 
boundaries in the parton shower could well induce non-perturbative
contributions already at the parton level. Thus, we
have refrained here from doing so. Instead, we have compared the partonic 
and the hadronic
outputs as they come from the generator, without any attempts to correct
the former. 

The non-perturbative hadronization is clearly a genuine physics
process, but for it we do not have at present a well established theory.
Rather, our knowledge is based on the
phenomenological experience and is implemented in the above-mentioned programs. 
Although the agreement between the latter is remarkable, and these
in turn reproduce well real data,
there are systematic dissimilarities in their implementation
of the non-pQCD dynamics that must be accounted for.
In other terms, the differences in the predictions of the Monte Carlo
programs contribute to build up our systematic uncertainties on the actual
measurements. These so-called hadronization
 corrections turn out to be algorithm dependent, thus 
to design one for which these are \sensibly\ reduced would represent a clear 
improvement: the smaller those are, the more under control would
the differences between generators be.
This is  of particular relevance at very small values of the resolution
parameter $\ycut$, where
the interface between perturbative and non-perturbative QCD occurs.

In order to to reduce the size of the non-perturbative corrections in
multi-jet rates, the implementation of the angular-ordering and
soft-freezing procedures has proven to be decisive, particularly at
low $\ycut$.  The first one consists in distinguishing between the
variable used to decide which pair of objects to test first and that
to be compared with the resolution parameter.  The second one
corresponds to eliminating from the sequence of clustering the less
energetic one in a resolved pair of particles.  These two steps help to 
heal two of the unwanted phenomena occurring in the dominion of soft
physics, that is, `junk-jet' formation and `misclustering',
respectively. The first takes place because of the tendency of soft
`unresolved' particles of acquiring momenta from particles at low
transverse momentum and forming spurious jets from these whereas the
second happens because of the bias of soft `resolved' particles of
attracting wide-angle radiation.

These two remedies are however effective {\sl only} if inserted into
$p_{\perp}$-based jet finders. In fact, although these two steps were
originally implemented as part of the \Cambridge\ algorithm, we
have assessed their efficiency also in presence of the \Luclus\
measure while reminding the reader of their inadequacy if the \Jade\ one is 
used instead. 
If one then combines this result with what we have already
mentioned for the fixed-order and resummed predictions, it is evident
that the hybrid scheme that we had originally introduced for purpose
of comparison, based on the \Luclus\ transverse momentum and the 
\Cambridge\ clustering sequence, performs better than any other
tested, so to deserve the status of new algorithm.  In our opinion, it
has come to set the standard as far as the dominion of soft physics in
multi-jet events is concerned.

Before proceeding further, we should mention that
the overall features obtained with respect to the size of the hadronization 
corrections are in part the result of the fortuitous 
cancellations between opposite 
tendencies. On the one hand, junk-jet formation and misclustering (and heavy
quark decays as well) induce
positive corrections. On the other hand, the well-known string or drag effect
(i.e., the pulling closer of the
two nearest jet directions by the hadronization mechanism) 
produces negative contributions.
The increased size of the `negative hadronization 
corrections' for some algorithms at medium values of $\ycut$ is then the 
consequence of having reduced the former while leaving untouched the latter 
effect. Therefore, as $\ycut$ grows larger, to diminish 
the extent of the corrections  becomes more and more
matter of finding a delicate balance between the two.
At the upper extreme of the $\ycut$ range, that is, 
in the two-jet limit, the \Diclus\ algorithm admirably
contains the size of the hadronization effects.

If one abandons the subject of QCD studies in multi-jets, that is,
the dominion of soft 
physics and global quantities (such as jet fractions, shape variables, 
etc.), and enter, for example, the territory of the search for mass resonances,
the criteria that define a good algorithm are going to be rather different.
In the new context, as it is now for the mass determination
of the $W$ boson at LEP2, kinematical quantities 
such as energies and angles (which build up the definition of
invariant mass) are of main concern. Also in this case,
although we have not carried out
a sophisticated analysis of four-jet events at LEP2, including detector
effect and background simulations, we believe to have achieved interesting
results.

In hadronic decays of $W^+W^-$ pairs, the four partons emerging
from the unstable resonances
are naturally energetic and far apart. QCD radiation from the two
$W$ decays does not interfere till the next-to-next-to-leading order
in the strong coupling constant. In other terms,
the soft dynamics that determines to a large extent the phenomenology
of jet rates is of little concern here. Instead, in this case,
it is how well an algorithm is able to reconstruct
at hadron level the original partonic energy and direction, and
ultimately the shape of the mass resonance, that sets the target
of a good jet clustering performance.

Therefore, the next step of our analysis has been to quantify the ability
of the various clustering algorithms in minimizing the average
angular and energy error in the jet reconstruction. As a preliminary
exercise, to allow for an understanding of the typical biases, we have 
addressed the 
simplified case of the kinematics of two-, three- and four-parton events, for 
some fixed phase space configuration. The procedure has been eventually
generalized to include all final state jet multiplicities, by studying the
sum of the invariant jet masses as well as of the transverse momentum of all
particles of an event.

After these tests, two out of our list of clustering algorithms excel
above all others, which share an ordinary degree of performance. They
are the \Luclus\ and \Diclus\ schemes. The former is undoubtedly the
best in reconstructing angles and it is second in case of energy only
to the latter, which is however very modest with angular quantities.
The ability of {\Luclus} in reconstructing angles and energies can be
attributed to the reassignment procedure, which it is the only to implement.
In other schemes, it is not uncommon with stray particles at the edge
of a jet that, by any distance criterion, are closer to another jet. 
The poor performances of \Diclus\ in angles are the price paid for an 
implementation especially designed to remedy systematic biases in the 
hadronization, notably the mentioned string or drag effect.

Studies in energies and angles similar to those above have been
carried out also for the case of $W^+W^-$ into four-jet events at LEP2. The
general picture for these two quantities separately is similar to that
outlined above, with \Luclus\ best overall. One would then expect this
algorithm to come first also when energies and angles are combined to
reconstruct the $W$ mass invariant spectrum. This is however true only
if one plots in the corresponding histogram {\sl all} individual
jet-jet masses (six in total). The majority of \Luclus\ events are in
fact concentrated around the $W$ mass, whereas misassignments take
place more often for other algorithms, whose spectra can be
significantly more spread out.

Surprisingly enough, if one plots instead the mass distribution formed
from the two average masses which can be obtained from the two
possible pairings that most likely reconstruct better the $W$ mass
(those in which the two most energetic jets are not paired together),
then the original \Cambridge\ algorithm (the one employing the
\Durham\ measure) comes out best (ahead of the \AODurham).  The
reasons for this are not entirely understood. On the one hand, the use
of the `average masses' rather then the `individual masses' is
generally dictated by the fact that misassignments of particles
partially cancel, on the other hand, our studies of jet angle and
energy reconstruction did not furnish us with an obvious explanation
why angular-ordering and/or soft-freezing should be beneficial to the
four-jet decays of $W^+W^-$ pairs.  (In addition, notice that in the
context of energy, angle and mass reconstruction, there is no
intrinsic advantage in using the \Luclus\ transverse momentum
rather than the \Durham\ one. Indeed, in the average $W$ mass
distribution the adoption of the former worsen the good performances
obtained with the latter.)

Since in high-statistic Monte Carlo simulations the actual speed
of the program is not a secondary issue (hundreds of hadrons
are typically involved), we have studied the performances 
of the various algorithms in this respect. In general, all binary clustering
algorithms are equally fast, whereas \Diclus\ is slower by more than
one order of magnitude.

Finally, three different Monte Carlo event generators have been used to carry
out all aspects of our analysis. We have never found any significative
difference among them. 

\section*{Acknowledgements}

SM is grateful to the UK PPARC for financial support and to the
Theoretical Physics Group in Lund for their kind hospitality during
his visit in Sweden, which has been partially supported by the Italian
Institute of Culture `C.M.~Lerici' (Stockholm) under the grant Prot.
I/B1 690, 1997.  SM finally acknowledges useful discussions with James
Stirling and Bryan Webber as well as various numerical comparisons
with Garth Leder. Finally, we all thank Yuri Dokshitzer, Mike Seymour
and Bryan Webber for carefully reading the manuscript version of this 
paper.

\end{document}